\documentclass{aa}  
\usepackage{times,txfonts,latexsym,graphics,graphicx,amssymb,hyperref}
\hypersetup{colorlinks,linkcolor=red,citecolor=blue,urlcolor=blue}
\usepackage{longtable}
\usepackage{natbib}
\bibpunct{(}{)}{;}{a}{}{,}

\newcommand{\Teff}{$T_{\rm eff}$}
\newcommand{\Fbol}{$F_{\rm bol}$}
\newcommand{\Ang}{$\theta_{\rm LD}$}
\newcommand{\logg}{$\log g$}
\newcommand{\ulogg}{cm~s$^{-2}$}

\newcommand{\FeH}{[Fe/H]}

\newcommand{\aFe}{[$\alpha$/Fe]}

\newcommand{\Dnu}{$\Delta\nu$}
\newcommand{\numax}{$\nu_{\rm max}$}

\newcommand{\alfCen}{$\alpha$~Cen}
\newcommand{\alfCet}{$\alpha$~Cet}
\newcommand{\alfTau}{$\alpha$~Tau}
\newcommand{\betAra}{$\beta$~Ara}
\newcommand{\betGem}{$\beta$~Gem}
\newcommand{\betHyi}{$\beta$~Hyi}
\newcommand{\betVir}{$\beta$~Vir}
\newcommand{\delEri}{$\delta$~Eri}
\newcommand{\epsEri}{$\epsilon$~Eri}
\newcommand{\epsFor}{$\epsilon$~For}
\newcommand{\epsVir}{$\epsilon$~Vir}
\newcommand{\etaBoo}{$\eta$~Boo}
\newcommand{\gamSge}{$\gamma$~Sge}
\newcommand{\ksiHya}{$\xi$~Hya}
\newcommand{\muAra}{$\mu$~Ara}
\newcommand{\muCas}{$\mu$~Cas}
\newcommand{\muLeo}{$\mu$~Leo}
\newcommand{\psiPhe}{$\psi$~Phe}
\newcommand{\tauCet}{$\tau$~Cet}
\newcommand{\Cyg}{61~Cyg}

\newcommand{\tab}[1]{Table~\ref{#1}}
\newcommand{\tabs}[1]{Tables~\ref{#1}}
\newcommand{\fig}[1]{Fig.~\ref{#1}}

\newcommand{\figu}[1]{Figure~\ref{#1}}
\newcommand{\figus}[1]{Figures~\ref{#1}}
\newcommand{\sect}[1]{Sect.~\ref{#1}}
\newcommand{\sects}[1]{Sects.~\ref{#1}}
\newcommand{\gbs}{\emph{Gaia} FGK benchmark stars}

\usepackage{color}

\begin{document}    
  
\title{\emph{Gaia} FGK benchmark stars: Effective temperatures and surface gravities
} 
 
\author{U. Heiter\inst{\ref{inst:Uppsala}}
\and P. Jofr\'e\inst{\ref{inst:Cambridge}}
\and B. Gustafsson\inst{\ref{inst:Uppsala},\ref{inst:Nordita}}
\and A.~J. Korn\inst{\ref{inst:Uppsala}}
\and C. Soubiran\inst{\ref{inst:Bordeaux}}
\and F. Th{\'e}venin\inst{\ref{inst:Nice}}
}

\institute{
           Institutionen f\"or fysik och astronomi, Uppsala universitet, Box 516, 751\,20 Uppsala, Sweden\\
              \email{ulrike.heiter@physics.uu.se}
              \label{inst:Uppsala}
\and       Institute of Astronomy, University of Cambridge, Madingley Rd, Cambridge, CB3 0HA, U.K.
              \label{inst:Cambridge}
\and       Univ. Bordeaux, CNRS, LAB, UMR 5804, 33270, Floirac, France
              \label{inst:Bordeaux}
\and       Nordita, Roslagstullsbacken 23, 106\,91, Stockholm, Sweden
              \label{inst:Nordita}
\and       Universit\'{e} de Nice Sophia Antipolis, CNRS (UMR7293), Observatoire de la C\^{o}te d'Azur, CS 34229, 06304, Nice Cedex 4, France
              \label{inst:Nice}
           }  

\date{Received / Accepted}  

\authorrunning{Heiter et al.}
\titlerunning{\gbs}  

\abstract
{ 
In the era of large Galactic stellar surveys, carefully calibrating and validating the data sets has become an important and integral part of the data analysis.
Moreover, new generations of stellar atmosphere models and spectral line formation computations need to be subjected to benchmark tests to assess any progress in predicting stellar properties.
}
{
We focus on cool stars and aim at establishing a sample of 34 \gbs\ with a range of different metallicities.
The goal was to determine the effective temperature and the surface gravity independently of spectroscopy and atmospheric models as far as possible.
Most of the selected stars have been subjected to frequent spectroscopic investigations in the past, and almost all of them have previously been used as reference, calibration, or test objects.
}
{
Fundamental determinations of \Teff\ and \logg\ were obtained in a systematic way from a compilation of angular diameter measurements and bolometric fluxes and from a homogeneous mass determination based on stellar evolution models. The derived parameters were compared to recent spectroscopic and photometric determinations and to gravity estimates based on seismic data.
}
{
Most of the adopted diameter measurements have formal uncertainties around 1\%,
which translate into uncertainties in effective temperature of 0.5\%.
The measurements of bolometric flux seem to be accurate to 5\% or better, 
which contributes about 1\% or less to the uncertainties in effective temperature.
The comparisons of parameter determinations with the literature in general show good agreements with a few exceptions, most notably for the coolest stars and for metal-poor stars.
}
{
The sample consists of 29 FGK-type stars and 5 M giants. Among the FGK stars, 21 have reliable parameters suitable for testing, validation, or calibration purposes.
For four stars, future adjustments of the fundamental \Teff\ are required, and for five stars the \logg\ determination needs to be improved.
Future extensions of the sample of \gbs\ are required to fill gaps in parameter space, and we include a list of suggested candidates.
}


\maketitle

\section{Introduction}
\label{sect:introduction}

%
%

We are about to revolutionise our knowledge about the Milky Way Galaxy thanks to the recently launched \emph{Gaia} mission \citep[e.g.][]{2001A&A...369..339P,2005ESASP.576.....T,2008IAUS..248..217L}. With its exquisite precision of parallaxes and proper motions for millions of stars of our Galaxy, complemented with spectrophotometric and spectroscopic observations, a much improved picture of the kinematics and chemical composition of the Galactic populations will become available. This will allow us to understand the formation history of the Milky Way and its components in more detail. Achieving this goal requires proper calibration of the data sets.

In this context it is important to note that the system developed for the astrophysical parameter (AP) determination of \emph{Gaia} sources will be almost entirely based on model spectra \citep[Apsis\footnote{\emph{Gaia} astrophysical parameters inference system},][]{2013A&A...559A..74B}.
Apsis uses methods that perform supervised classification, involving a comparison of the observed data with a set of templates. 
For the purpose of estimating stellar atmospheric parameters one may use either observed or synthetic spectra as templates.
Observed spectra are expected to represent individual \emph{Gaia} spectrophotometric and spectroscopic observations better, but they do not cover the whole parameter range with sufficient density because of the small number of stars with accurately known APs.
Synthetic templates allow one to deal with the latter issue, and also to attempt to classify sources that are very rare and possibly as yet unobserved.
Furthermore, the effects of observational noise and interstellar extinction can be considered in a systematic way.
However, synthetic spectra are based on simplified descriptions of the complex physical processes taking place in real stars.
This causes a possible mismatch between synthetic and observed spectra, which may introduce large external errors in derived APs.
Thus, the algorithms of Apsis using synthetic templates will need to be calibrated to account for deficiencies in the physics of the models used.

The calibration can be implemented by applying corrections either to the synthetic spectra before they are used or to the APs produced by each algorithm.
For both approaches we need a set of reference objects, which will be observed by \emph{Gaia}, but for which accurate APs have been determined by independent methods from ground-based observations (e.g. higher resolution spectra).
For those algorithms that estimate \Teff, \logg, and \FeH,\ the reference objects will be divided into two goal levels.
At the first level, we identify a small number of well-known bright benchmark stars.
These will not necessarily be observed by \emph{Gaia} owing to its bright magnitude limit, but their properties are being investigated in detail by ground-based observations.
The second level consists of a large number (several hundred) of AP reference stars to ensure dense coverage of the AP space and to have magnitudes within the reach of \emph{Gaia}.
These stars will be characterised in an automated way using ground-based high-resolution spectroscopy with a calibration linking their APs to those of the benchmark stars.

Another approach to mitigating the spectrum mismatch problem is to improve the modelling of the synthetic spectra.
For example, the synthetic stellar libraries currently in use for classification of FGK-type stars in Apsis \citep[see Table~2 in][]{2013A&A...559A..74B} are mostly based on 1D hydrostatic model atmospheres where local thermodynamic equilibrium is assumed. More realistic 3D radiation-hydrodynamics (RHD) simulations and results of detailed statistical equilibrium calculations are expected to be implemented in future versions of the libraries. 
It will be important to test the actual degree of improvement achieved with the new generations of atmospheric models.
The benchmark stars may serve as test objects, for which synthetic observables generated from atmospheric models using different assumptions for the input physics can be compared with the observed data. 

A set of stars with atmospheric parameters determined independently from spectroscopy will also be suitable to calibrate methods and input data for spectroscopic analyses, such as those based on the characterisation of thousands of spectra from ground-based surveys complementary to \emph{Gaia}. Examples of such surveys are the \emph{Gaia}-ESO Public Spectroscopic Survey \citep{2012Msngr.147...25G,2013Msngr.154...47R}, RAVE \citep{2006AJ....132.1645S}, APOGEE \citep{2008AN....329.1018A}, and GALAH \citep{2015MNRAS.449.2604D}.
To test that a method yields robust results, it should be applied to a set of spectra with known parameters to verify that the results agree.

However, after several decades of Galactic science using spectroscopic surveys, there is still no set of reference stars common to different data sets. In fact, the Sun has been used by most studies to calibrate methods of analysing the spectra of solar-type stars. 
Arcturus has also been widely used as a reference giant, although with varying stellar parameters. Some recent publications using Arcturus as a reference are \citet{2014A&A...564A.119M}, \citet{2014A&A...564A.109S}, \citet{2013AJ....146..133M}, \citet{2013A&A...557A..70M}, \citet{2013MNRAS.429..126R}, \citet{2012A&A...543A.160T}, \citet{2012A&A...542A..48W}, \citet{2011AJ....142..193B}, \citet{2011A&A...531A.165P}. The \Teff\ values used in these publications range from 4230 to 4340~K, the \logg\ values from 1.5 to 1.9~dex, and the overall metallicities from $-0.60$ to $-0.47$~dex. Each of these publications used a different source for the parameters of Arcturus.

%
%
In this article, we aim at defining a sample of benchmark stars covering the range of F, G, and K spectral types at different metallicities, which is representative of a significant part of the stellar populations in our Galaxy. Stars of these spectral types will be the most numerous among the objects observed by \emph{Gaia} -- 86\% of all stars with $V<20$ according to estimations by \citet{2005ESASP.576...83R}. 
The benchmark stars are selected to have as much information as possible available to determine their effective temperature and surface gravity independently from spectroscopy.
We refer to these stars as the \gbs, as they will be used as pillars for the calibration of the \emph{Gaia} parameter determination.

This is the first in a series of articles about the work on \gbs, introducing the sample and describing the determination of effective temperature and surface gravity from fundamental relations.
The second article of this series \citep[hereafter Paper~II]{2014A&A...566A..98B} 
describes our libraries of observed high-resolution optical spectra for \gbs\ and includes an assessment of the data quality.
In the third article \citep[hereafter Paper~III]{2014A&A...564A.133J}, 
the fundamental \Teff\ and \logg\ and the spectral libraries were used for metallicity determinations, based on the analysis of Fe lines by several methods.
The three articles provide a full documentation, which is also summarised in \citet{2014ASInC..11..159J}.
In addition, an abundance analysis for more than 20 chemical elements accessible in the spectral libraries is in progress, and we are working on a future extension of the sample to improve the coverage of parameter space (cf. Appendix~\ref{app:suggested}). 

The remainder of the article is organised as follows.
In \sect{sect:sample} we describe the selection of the current set of 34 \gbs.
In \sects{sect:teff} and \ref{sect:logg} we describe the determination of effective temperatures and surface gravities, respectively, based on angular diameters, bolometric fluxes, masses, and distances.
In \sect{sect:results} we present our results and compare our set of parameters with spectroscopic and photometric determinations extracted from the literature. \sect{sect:lastsection} contains a star-by-star summary of parameter quality and concludes the article.

\section{Sample selection}
\label{sect:sample}

\begin{table*}
   \caption{General information for current sample of \gbs.}
   \label{tab:general}
   \centering
   \begin{tabular}{lrlllrrr}
      \hline\hline
      Name & HD   & RA (J2000) & DEC (J2000) & Spectral Type & $V$mag & \FeH\tablefootmark{$\dagger$} & $u$(\FeH)\tablefootmark{$\ddagger$} \\
      \hline
      {\bf F dwarfs} & & & & & & & \\
Procyon&61421&07 39 18.119& +05 13 29.96  &F5IV-V         &0.4&0.01&0.08\\
HD 84937&84937&09 48 56.098& +13 44 39.32  &sdF5           &8.3&-2.03&0.08\\
HD 49933&49933&06 50 49.832& -00 32 27.17  &F2V            &5.8&-0.41&0.08\\
\hline
{\bf FGK subgiants} & & & & & & & \\
\delEri &23249&03 43 14.901& -09 45 48.21  &K1III-IV       &3.5&0.06&0.05\\
HD 140283&140283&15 43 03.097& -10 56 00.60  &sdF3           &7.2&-2.36&0.10\\
\epsFor&18907&03 01 37.637& -28 05 29.60  &K2VFe-1.3CH-0.8&5.9&-0.60&0.10\\
\etaBoo&121370&13 54 41.079& +18 23 51.79  &G0IV           &2.7&0.32&0.08\\
\betHyi&2151&00 25 45.070& -77 15 15.29  &G0V            &2.8&-0.04&0.06\\
\hline
{\bf G dwarfs} & & & & & & & \\
\alfCen A&128620&14 39 36.494& -60 50 02.37  &G2V            &0.0&0.26&0.08\\
HD 22879&22879&03 40 22.064& -03 13 01.12  &F9V            &6.7&-0.86&0.05\\
Sun& &&&& &0.03&0.05\\
\muCas&6582&01 08 16.395& +54 55 13.23  &G5Vb           &5.2&-0.81&0.03\\
\tauCet&10700&01 44 04.083& -15 56 14.93  &G8.5V          &3.5&-0.49&0.03\\
\alfCen B&128621&14 39 35.063& -60 50 15.10  &K1V            &1.4&0.22&0.10\\
18 Sco&146233&16 15 37.269& -08 22 09.99  &G2Va           &5.5&0.03&0.03\\
\muAra&160691&17 44 08.701& -51 50 02.59  &G3IV-V         &5.1&0.35&0.13\\
\betVir&102870&11 50 41.718& +01 45 52.99  &F9V            &3.6&0.24&0.07\\
\hline
{\bf FGK giants} & & & & & & & \\
Arcturus&124897&14 15 39.672& +19 10 56.67  &K1.5III        &-0.1&-0.52&0.08\\
HD 122563&122563&14 02 31.845& +09 41 09.95  &F8IV           &6.2&-2.64&0.22\\
\muLeo&85503&09 52 45.817& +26 00 25.03  &K2III          &3.9&0.25&0.15\\
\betGem&62509&07 45 18.950& +28 01 34.32  &K0IIIb         &1.1&0.13&0.16\\
\epsVir&113226&13 02 10.598& +10 57 32.94  &G8III          &2.8&0.15&0.16\\
\ksiHya&100407&11 33 00.115& -31 51 27.44  &G7III          &3.5&0.16&0.20\\
HD 107328&107328&12 20 20.981& +03 18 45.26  &K0IIIb         &5.0&-0.33&0.16\\
HD 220009&220009&23 20 20.583& +05 22 52.70  &K2III          &5.0&-0.74&0.13\\
\hline
{\bf M giants} & & & & & & & \\
\alfTau&29139&04 35 55.239& +16 30 33.49  &K5III          &0.9&-0.37&0.17\\
\alfCet&18884&03 02 16.773& +04 05 23.06  &M1.5IIIa       &2.5&-0.45&0.47\\
\betAra&157244&17 25 17.988& -55 31 47.59  &K3Ib-II        &2.8&-0.05&0.39\\
\gamSge&189319&19 58 45.429& +19 29 31.73  &M0III          &3.5&-0.17&0.39\\
\psiPhe&11695&01 53 38.741& -46 18 09.60  &M4III          &4.4&-1.24&0.39\\
\hline
{\bf K dwarfs} & & & & & & & \\
\epsEri&22049&03 32 55.845& -09 27 29.73  &K2Vk:          &3.7&-0.09&0.06\\
Gmb 1830&103095&11 52 58.769& +37 43 07.23  &G8Vp           &6.4&-1.46&0.39\\
61 Cyg A&201091&21 06 53.952& +38 44 57.99  &K5V            &5.2&-0.33&0.38\\
61 Cyg B&201092&21 06 55.264& +38 44 31.40  &K7V            &6.0&-0.38&0.03\\

      \hline\hline
   \end{tabular}
   \tablefoot{Coordinates and spectral types are extracted from the SIMBAD database. The $V$ magnitudes are mean values extracted from the General Catalogue of Photometric Data \citep[GCPD\footnote{\url{http://obswww.unige.ch/gcpd/gcpd.html}},][]{1997A&AS..124..349M}. 
   \tablefoottext{$\dagger$}{NLTE [Fe/H] value from Table~3 of Paper~III.}
   \tablefoottext{$\ddagger$}{[Fe/H] uncertainty obtained by quadratically summing all $\sigma$ and $\Delta$ columns in Table~3 of Paper~III.}
   }
\end{table*}

The current sample of \gbs\ was selected to cover the range of effective temperatures between about 4000~K and 6500~K. This region of the HR diagram was divided into the categories of F~dwarfs, FGK subgiants, G~dwarfs, FGK giants, and K~dwarfs. For each of these categories we aimed at selecting at least two stars, one with solar and one with sub-solar metallicity.
We also added a few M~giants with \Teff\,$\lesssim$\,4000~K to enable an assessment of models and methods at the parameter range limits.

The main selection criteria are related to the availability of information required to determine the effective temperature and the surface gravity from fundamental relations (see \sects{sect:teff} and \ref{sect:logg}).
Angular diameter measurements should be available or expected in the near future, and bolometric fluxes should be reliably determined, in order to calculate the effective temperature.
Accurate parallax measurements are a prerequisite for a fundamental \logg\ determination.
In addition, we considered stars that are members of visual or eclipsing binary systems and stars for which asteroseismic data are available or which are targets of asteroseismology missions or campaigns, such as \emph{CoRoT} or \emph{Kepler}.
Finally, the stars were primarily chosen among those subjected to frequent spectroscopic investigations in the past.

\tab{tab:general} lists the identifiers and basic information for the current sample of \gbs. The $V$ magnitudes range from 0 to 8. About two thirds of the stars are fainter than 3~mag and will be included in the magnitude range observed by \emph{Gaia} (see \citealt{2014SPIE.9143E..0YM} for a discussion of the detection of bright stars with \emph{Gaia}). 
The stars are evenly distributed along the celestial equator, and all but ten stars have declinations within $\pm30$ degrees. Thus, a large part of the sample can be observed with telescopes on both hemispheres.
The sample contains several visual binary stars. These are discussed in \sect{sect:othermass}.
A few active stars are also included, in case activity diagnostics need to be evaluated.
Procyon, \epsEri, and \Cyg~A were classified as active stars by \citet{2003AJ....126.2048G} 
based on Ca~II H and K line emission, while \muCas, 18~Sco, Gmb~1830, and \Cyg~B were found to be inactive.
Surprisingly, HD~49933, the main asteroseismic target of the \emph{CoRoT} mission, has been discovered as strongly active with surface spots \citep{2005A&A...431L..13M}. 
This type of activity can modify the emergent stellar flux, leading to an increased uncertainty in \Teff.
\citet{1993ApJ...417..157B} 
measured an activity index based on Ca~II H and K line emission for \ksiHya.
For ten stars, an X-ray luminosity was measured by ROSAT \citep{1998A&AS..132..155H}, which is given relative to bolometric luminosity in \sect{sect:mass}.

Almost all stars in the current sample fulfil the main criteria, with two exceptions.
We included two dwarf stars with extreme metallicities, even though their angular diameters do not lie within reach of current interferometers.
HD~84937 is the faintest star ($V$=8), and it has a predicted angular diameter of 0.15~mas (O.~Creevey, priv. comm.). Thus, it is too small and too faint to be measured with currently existing interferometers.
At a distance of $\sim$70~pc, it is, however, the closest dwarf star with a metallicity around $-2$~dex, and one of the most studied halo dwarfs \citep[e.g.][]{2012MNRAS.427...27B,2008A&A...478..529M}. It remains an important candidate benchmark star, awaiting future improvements in capabilities of interferometers, e.g the Navy Precision Optical Interferometer \citep[NPOI,][]{2013JAI.....240002A}.
The most metal-rich star in the sample, \muAra, would be feasible for being measured with an instrument such as the CHARA array at Mount Wilson Observatory (California), but is not accessible with the latter owing to its position in the sky.
This star has been the target of an asteroseismology campaign \citep[and \sect{sect:othermass}]{2005A&A...440..609B}.
The sample should be complemented in the future with a northern dwarf star of similar metallicity, magnitude, and angular size as \muAra\ (see Appendix~\ref{app:suggested}).

%
%

Almost all of the stars in our sample have previously been used as reference, calibration, or test objects.
In the following, we give some examples. 
%
%
Similar to Arcturus (see \sect{sect:introduction}), Procyon has been widely used as a standard star in various studies.
Some recent publications using Procyon as a calibration or test object are 
\citet{2014A&A...563A..52P}, \citet{2013MNRAS.428.3164D}, \citet{2012A&A...546A..90B}, \citet{2012A&A...542A..48W}, \citet{2011CaJPh..89..395L}, \citet{2011A&A...528A..87M}, \citet{2010ApJ...725.2176Q}, \citet{2010MNRAS.405.1907B}, \citet{2010A&A...523A..71G}.
The \Teff\ values used in these publications range from 6490~K to 6650~K, and the \logg\ values from 3.96~dex\ to 4.05~dex.
Each of these publications used a different source for the parameters.


%
%
Apart from Arcturus, several giants have been used as reference objects in differential abundance analyses, illustrated in the following three examples.
In an abundance study of barium stars \citet{2007A&A...468..679S} 
adopted the solar-metallicity giant \epsVir\ as the standard star for a differential analysis\footnote{The history of \epsVir\ as a standard star goes much further back in time. For example, it was used as a reference star in a differential curve-of-growth analysis by \citet{1975A&A....43..127C}.}. The stellar parameters of the reference star were determined from photometric calibrations 
and stellar evolution models. 
%
\citet{2011A&A...534A..80H} 
studied the formation of the Galactic bulge by determining metallicities for clump giants in Baade's window relative to the metal-rich giant \muLeo.
The effective temperature adopted for the reference star was obtained by the infrared flux method, and the \logg\ value was estimated from various independent methods \citep{2007A&A...465..799L}. 
%
\citet{2000ApJ...530..783W} 
studied nucleosynthesis signatures of neutron-capture elements through a line-by-line differential abundance comparison of an r-process enriched low-metallicity giant (HD 115444) with the metal-poor giant HD~122563.
The atmospheric parameters of both stars were determined by demanding excitation and ionization equilibrium of Fe line abundances.

%
%
Arcturus, \alfCen~A, \alfCet, and \alfTau\ are among the standard sources that were used for the flux calibration of the Short-Wavelength Spectrometer on the Infrared Space Observatory by means of synthetic spectra. The stellar spectra and calibration issues were discussed by \citet{2003A&A...400..679D}, for example, 
who used atmospheric parameters from the literature for \alfCen~A and determined parameters from the infrared spectra for the other stars.
These four stars, as well as \betGem, are also included in the sample of 12 standard stars used for calibration purposes in a procedure for creating template spectral energy distributions in the infrared for arbitrary effective temperatures \citep[\emph{autoshape} procedure, 1--35~$\mu$m,][]{2006AJ....132.1445E}. 
%
Procyon and \epsEri\ were used by \citet{2011CaJPh..89..395L} 
to calibrate oscillator strengths of over 900 lines of neutral species in the wavelength range 400 to 680~nm.
%
The star 18~Sco is one of six fundamental stars that are being used for calibrating atomic and molecular line lists for producing theoretical stellar libraries \citep{2014MNRAS.442.1294M}\footnote{\url{http://archive.eso.org/wdb/wdb/eso/sched_rep_arc/query?progid=087.B-0308(A)}}. 
Procyon, \betVir, \epsEri, \etaBoo, \muCas, and \tauCet\ were used among other stars by \citet[][cf. their Table~2]{2011A&A...531A..83C} 
for a calibration between effective temperatures derived from H$\alpha$ line-profile fitting and direct effective temperatures based on interferometric angular diameters and bolometric fluxes. 

%
%
Subsets of the \gbs\ have been used in numerous works for tests of various modelling and analysis techniques.
%
%
\citet{2010MNRAS.405.1907B} 
compared direct and indirect methods to determine astrophysical parameters 
for Procyon, \alfCen~A and B, \betHyi, \betVir, \delEri, \etaBoo, \ksiHya, \muAra, and \tauCet.
Their direct parameters were based on interferometric and asteroseismic data.
HD~22879 was used by \citet{2010A&A...517A..57J} 
as part of a sample of stars with high-resolution spectra 
to validate the MA$\chi$ tool, which had been applied to low-resolution spectra of over 17\,000 metal-poor dwarfs from the SEGUE survey to estimate atmospheric parameters.
Procyon, \alfCen~A and B, \betHyi, \etaBoo, and \tauCet\ appear in a test of automatic determination of stellar parameters from asteroseismic data \citep{2010ApJ...725.2176Q}, 
where various sources were used for the reference stellar parameters.
%

Eleven of the current \gbs\ are included in one of the stellar samples that were used to validate the SDSS/SEGUE stellar parameter pipeline by \citet{2011AJ....141...90L}. 
This sample 
was selected from the ELODIE spectral library \citep{2001A&A...369.1048P,2004astro.ph..9214P,2004PASP..116..693M}. 
These \gbs\ are Procyon, \betGem, \betVir, \epsEri, \muCas, \tauCet, 18~Sco, Gmb~1830, HD~22879, HD~84937, and HD~140283.
The reference parameters were those supplied with the ELODIE library.
Arcturus and \muLeo\ were used together with two other nearby red giants by \citet{2013AJ....146..133M} 
as standard stars to test the results of the APOGEE Stellar Parameters and Chemical Abundances Pipeline.
Arcturus, \alfCen~A, \betVir, \gamSge, \ksiHya, \muCas, and \muLeo\ were used by \citet{2013EPJWC..4303006V} 
to validate the GAUFRE tool for measuring atmospheric parameters from spectra by comparing the results of the tool with average spectroscopic parameters from the literature.
In a similar way, Arcturus and HD~140283 were used by \citet{2014A&A...564A.109S} 
to validate the MyGIsFOS code for spectroscopic derivation of atmospheric parameters, using various sources for the reference parameters.

%
%
Procyon and the metal-poor stars Gmb~1830, HD~84937, HD~122563, and HD~140283 were used (in different combinations) as test objects in a series of investigations of non-LTE line formation for several chemical elements
(Ca~I and Ca~II in \citealt{2007A&A...461..261M}; 
eight elements in \citealt{2008A&A...478..529M}; 
Mn~I in \citealt{2008A&A...492..823B}; 
Co~I and Co~II in \citealt{2010MNRAS.401.1334B}; 
Mg I in \citealt{2013A&A...550A..28M}). 
In most cases, the reference \Teff\ values were based on Balmer line-profile fitting, and the \logg\ values were determined with the parallax method (cf. \sect{sect:logg_other}).
In \citet{2012MNRAS.427...27B} 
non-LTE line formation of Fe was studied in 1D hydrostatic and averaged 3D hydrodynamic model atmospheres, using photometric parameters for the metal-poor test objects.
The analysis of Ba by \citet{2006A&A...456..313M} 
included HD~22879 as well, and 
\betVir\ and \tauCet\ were added in the study of Fe~I and Fe~II by \citet{2011A&A...528A..87M}.
In the latter work, \Teff\ and \logg\ were determined from methods that are largely independent of the model atmosphere.

%
%
The M~giants \alfTau\ and \alfCet\ were used as test objects by \citet{2012A&A...547A.108L} 
in a comparative study of stellar spectrum modelling involving 11 different analysis methods.
%
%
Arcturus, \alfTau, \muLeo, Gmb~1830, HD~122563, and HD~140283 are included in the catalogue of observed and synthetic hydrogen line profiles (P$\delta$, P$\gamma$, H$\alpha$) by \citet{2012A&A...547A..62H}. 
The stellar parameters were taken from various sources, but mainly from the spectroscopic study by \citet{2000AJ....120.1841F}. 
%
Arcturus, Procyon, \epsEri, and \tauCet\ were used by \citet{2010ApJ...710.1003G} to analyse the shapes of spectral line bisectors and to measure basic properties of stellar granulation.
%
The local metal-poor dwarf Gmb~1830 was used by \citet{2000ApJ...532..430V} 
and follow-up publications to validate predictions of stellar models for old stars.

%
%
Furthermore, 18 of the \gbs\ are included in the MILES library of medium-resolution spectra.
The atmospheric parameters for most of the 985 stars in the MILES library were determined by \citet{2011A&A...531A.165P} using an automatic routine. However, for some stars, parameters from the literature were adopted, which the authors judged to be more credible than the results of the automatic analysis. Among the \gbs\, this was the case for Arcturus, 
HD~122563, 
and HD~140283. 
Procyon and \betGem\ are included as two of eight iconic cool stars in the Advanced Spectral Library \citep[ASTRAL,][]{2013AN....334..105A}, 
a library of high-quality UV atlases obtained with HST/STIS. The star
\epsFor\ is included in the X-shooter spectral library \citep{2014A&A...565A.117C}.

As can be seen from these examples, the stars included in our sample of \gbs\ have been widely used in the literature. However, different authors considered different values for the fundamental parameters \Teff\ and \logg. The aim of this work is to provide a set of recommended values for calibration purposes, based on interferometric, photometric, and seismic data.

\section{Effective temperature}
\label{sect:teff}
We employed the fundamental relation $L = 4\pi R^2 \sigma $\Teff$^4$, where $L$ is the luminosity, $R$ the radius, and $\sigma$ the Stefan-Boltzmann constant, to determine the effective temperature.
In fact, the Sun is the only object for which we used the relation in this form.
For the other stars, we used the bolometric flux\footnote{The term ``bolometric flux'' refers to the total radiative flux from the star received at the Earth.} \Fbol\ and the limb-darkened angular diameter \Ang\ instead of $L$ and $R$, and calculated \Teff\ directly from these two measured quantities:
\begin{equation}
 T_{\rm eff} = \left(\frac{F_{\rm bol}}{\sigma}\right)^{0.25} (0.5\,\theta_{\rm LD})^{-0.5}.
\end{equation}
The input data for \Teff\ were compiled from the sources described in the following sections.

When possible, interferometric measurements of \Ang\ and \Fbol\ determinations based on integrations of absolute flux measurements across the stellar spectrum were extracted from the literature.
We refer to these as \emph{\emph{direct}} data.
For the cases where direct data are not yet available, we resorted to indirect determinations based on various calibrations available in the literature.

\subsection{The Sun}
\label{sect:sun}

\begin{table*}
   \caption{Fundamental parameters and data for the Sun, and physical constants.}
   \label{tab:Sun}
   \centering
   \begin{tabular}{lcclr}
\hline\hline
Description & Value & Uncertainty & Unit & Ref. \\
\hline
Solar radius                      & 6.9577$\cdot 10^{+08}$        & 1.4$\cdot 10^{+05}$ & m                        & (1) \\
Solar constant                    & 1360.8                        & 0.5                 & Wm$^{-2}$                & (2) \\
Solar mass parameter $GM_{\odot}$ & 1.3271244210$\cdot 10^{+20}$  & 1$\cdot 10^{+10}$   & m$^{3}$s$^{-2}$          & (3) \\
Astronomical unit of length       & 1.49597870700$\cdot 10^{+11}$ &                     & m                        & (4) \\
Stefan-Boltzmann constant         & 5.670373$\cdot 10^{-08}$      & 2.1$\cdot 10^{-13}$ & Wm$^{-2}$K$^{-4}$        & (5) \\
Newtonian constant of gravitation & 6.67384$\cdot 10^{-11}$       & 8.0$\cdot 10^{-15}$ & m$^{3}$kg$^{-1}$s$^{-2}$ & (5) \\
Zero point $L_0$ for bolometric flux & 3.055$\cdot 10^{+28}$      &                     & W                        & (6) \\
\hline\hline
   \end{tabular}
   \tablefoot{References: (1) \citet{2014A&A...569A..60M,2008ApJ...675L..53H}, (2) \citet{2011GeoRL..38.1706K}, (3) \citet{2011Icar..211..401K}, and Resolution B3 adopted at IAU General Assembly 2006, \url{http://www.iau.org/static/resolutions/IAU2006_Resol3.pdf}, (4) Resolution B2 adopted at IAU General Assembly 2012, \url{http://www.iau.org/static/resolutions/IAU2012_English.pdf}, (5) 2010 CODATA recommended values, \url{http://physics.nist.gov/cuu/Constants/bibliography.html}, (6) \citet{1999IAUTB..23.....A}. 
   }
\end{table*}

The fundamental parameters used for calculating the effective temperature of the Sun are given in \tab{tab:Sun}.
The solar radius is based on the recent measurements by \citet{2014A&A...569A..60M}. 
These authors used a ground-based facility to measure the position of the inflection point of the solar intensity at the limb from 2011 to 2013. The measurements were corrected for the effects of refraction and turbulence in the Earth's atmosphere. We used their result of $959.78\pm0.19$~ arcsec for the angular radius, obtained at a wavelength of 535.7~nm. The radius defined by the inflection point position is somewhat larger than the radius where the optical depth in the photosphere is equal to 1, which is the definition commonly used for stellar (atmosphere) modelling \citep[e.g.][]{Gust:08}. The correction needs to be derived from radiative-transfer modelling of the solar limb, and we used a value of $-340\pm10$~km based on the calculations by \citet{2008ApJ...675L..53H}. 
This corresponds to a difference in effective temperature of about 2~K.

The solar luminosity was computed from the solar constant (total solar irradiance) and the astronomical unit.
A recent measurement of the solar constant from the Total Irradiance Monitor on NASA's Solar Radiation and Climate Experiment during the 2008 solar minimum period is given by
\citet{2011GeoRL..38.1706K}\footnote{The same value is obtained by calculating the median of the 4080 measurements taken by the same instrument between 2003-02-25 and 2015-02-18, available at \url{http://lasp.colorado.edu/lisird/sorce/sorce_tsi/index.html}.}.
For a discussion of solar luminosity, mass, and radius, see also \citet{2011PASP..123..976H}.

\subsection{Angular diameters}
\label{sect:ang}
\paragraph{Direct measurements}
The catalogues of \citet{2001A&A...367..521P} 
and \citet{2005A&A...431..773R} 
and the literature were searched for angular diameter determinations, and the data were extracted from the original references. The compiled \Ang\ values include the effects of limb darkening.
For 27 stars we found \Ang\ values, which are given in \tab{tab:parameters} together with their references.
In the case of multiple measurements by different authors, we adopted the most recent one or the one with the lowest formal uncertainty.
The measurements were done with several different instruments, mainly in the infrared, but in a few cases in optical wavelength regions.
Various approaches to account for limb darkening were used. The most common one was to apply limb-darkening coefficients taken from \citet{2000A&A...363.1081C} or \citet{1995A&AS..114..247C}.

The measurements for \betGem\ and \epsVir\ were done by \citet{2003AJ....126.2502M} at four optical wavelengths between 450 and 800~nm with the Mark~III Stellar Interferometer on Mount Wilson \citep[California,][]{1988A&A...193..357S}. 
$R$-band (700~nm) observations were obtained with the Sydney University Stellar Interferometer (SUSI, New South Wales, Australia, \citealt{1999MNRAS.303..773D}) by \citet{2007MNRAS.380L..80N,2009MNRAS.393..245N} for \betHyi\ and \betVir.
In that work, limb-darkening corrections from \citet{2000MNRAS.318..387D} based on Kurucz stellar atmosphere models were applied.
The angular diameters of the two dwarf stars 18~Sco and HD~49933, the subgiant HD~140283, and the giant HD~220009 were measured in the near-infrared with the CHARA Array at Mount Wilson Observatory \citep[California,][]{2005ApJ...628..453T} and either the PAVO or the VEGA instruments by \citet[][PAVO]{2011A&A...526L...4B}, \citet[][VEGA]{2011A&A...534L...3B}, \citet[][VEGA]{2015A&A...575A..26C}, and Th\'evenin et al. (in prep., VEGA), respectively. \citet{2011A&A...534L...3B} used 3D-RHD atmospheric models to determine the limb-darkening effect.

The most recent diameter of Arcturus was obtained by \citet{2008A&A...485..561L} in the $H$ band (1.5 to 1.8~$\mu$m) with the IOTA (Infrared-Optical Telescope Array) interferometer \citep[Arizona,][]{2003SPIE.4838...45T}.
The diameters of the remaining stars were measured in the $K$ band (2.2~$\mu$m) with either the VLT Interferometer \citep[VLTI, VINCI instrument,][]{2003SPIE.4838..858K} or the CHARA Array using the Classic or FLUOR instrument, and in two cases, data from the Palomar Testbed Interferometer \citep[PTI, California,][]{1999ApJ...510..505C} were used (for HD~122563 by \citealt{2012A&A...545A..17C} and for \etaBoo\ by \citealt{2007ApJ...657.1058V}).
For the \Cyg\ components we used the VINCI measurement by \citet{2008A&A...488..667K}. The system was also observed with the PTI by \citet{2009ApJ...694.1085V}, who determined \Ang=1.63$\pm$0.05~mas for \Cyg~A and \Ang=1.67$\pm$0.05~mas for \Cyg~B. These values are 8\% lower and 5\% higher, respectively, than the values we used. We prefer the values by \citet{2008A&A...488..667K} because \citet{2009ApJ...694.1085V} derived a significant extinction for the K7V component ($A_V$=0.23$\pm$0.01~mag, while $A_V$=0 for the K5V component as expected for this nearby system), and their angular diameter for the K7V component is larger than that of the K5V component (although they are equal within the uncertainties).
Limb darkening was taken into account using individual atmospheric models in the cases of \psiPhe, \alfCet, and \gamSge\ \citep{2004A&A...413..711W,2006A&A...460..855W,2006A&A...460..843W}, for which spherical PHOENIX models \citep[][version 13]{1999JCoAM.109...41H} were computed.
In the case of Procyon, \alfCen~B, and HD~122563, 3D-RHD simulations were applied by \citet{2012A&A...540A...5C}, 
\citet{2006A&A...446..635B}, 
and \citet{2012A&A...545A..17C}, 
respectively.

\begin{table*}
   \caption{Broad-band photometry for \gbs\ used for indirect determinations of \Ang\ and/or \Fbol.}
   \label{tab:phot}
   \centering
   \begin{tabular}{lrrrrrr}
\hline\hline
Name       & $V$   &$\sigma(V)$& BC$_V$ & $(B-V)$ & $J$ & $K$  \\
\hline
HD 84937   &8.324  &0.021 &        &0.389    &  7.359  & 7.062 \\
\epsFor    &5.883  &0.005 & -0.25  &0.790    &  4.364  & 3.824 \\
HD 22879   &6.689  &0.009 &        &0.540    &  5.588  & 5.179 \\
\muAra     &5.131  &0.012 &        &0.700    &  4.158  & 3.683 \\
HD 122563  &6.200  &0.009 & -0.47  &         &         &       \\
\muLeo     &3.880  &0.005 & -0.51  &         &         &       \\
HD 107328  &4.970  &0.012 & -0.49  &         &         &       \\
HD 220009  &5.047  &0.008 & -0.64  &         &         &       \\
\betAra    &2.842  &0.004 & -0.75  &         &         &       \\
\hline\hline
   \end{tabular}
   \tablefoot{The $V$ magnitudes and $B-V$ colour indices are mean values and $\sigma(V)$ the standard deviation extracted from the GCPD \citep{1997A&AS..124..349M}. 
   The bolometric corrections BC$_V$ were calculated as described in the text, \sect{sect:fbol}.
   The $J$ and $K$ magnitudes are taken from the 2MASS catalogue \citep{2003yCat.2246....0C}.
   }
\end{table*}

\paragraph{Indirect determinations}
For two giants in our sample, \citet{1999AJ....117.1864C} 
determined angular diameters by scaling spectra of reference stars with absolute flux calibration in the infrared region (1.2 to 35~$\mu$m) to infrared photometry of the target stars with similar type (\muLeo\ with reference Arcturus, and HD~107328 with reference \betGem). The diameters of the target stars were obtained by multiplying the diameters of the reference stars by the square root of the scale factors. This method was applied to a large sample of giant stars and verified to be accurate to within 2\% by comparison to direct measurements (see their Fig.~8).

For four of the dwarf and subgiant stars (HD~22879, HD~84937, \epsFor, \muAra), we used the surface-brightness relations by \citet{2004A&A...426..297K} to estimate the angular diameters.
We used the average of the results from the relations for the $(B-K)$ and $(V-K)$ colour indices (Eqs.~22 and 23 in \citealt{2004A&A...426..297K}), because those present the smallest dispersions ($\le$1\% in Table~4 of \citealt{2004A&A...426..297K}).
Interstellar extinction was assumed to be negligible for these four stars.
\citet{2014A&A...561A..91L} 
derived maps of the local interstellar medium (ISM) from inversion of individual colour excess measurements.
These show that interstellar reddening is negligible within about 40~pc, applicable to three of the stars. Also the fourth star, HD~84937, seems to be located within a local cavity with negligible reddening.
The photometry is given in \tab{tab:phot}.
The $K$ magnitudes from 2MASS were transformed to the Johnson system using Eqs.~12 and 14 from \citet{2001AJ....121.2851C}, 
followed by Eqs.~13 and 14, as well as Eqs.~6 and 7, from \citet{1994A&AS..107..365A}. 
For the uncertainties of the angular diameters we adopted 2\% in the case of \epsFor\ and \muAra\ and 3\% for HD~22879 and HD~84937. The last two are metal-poor dwarf stars with diameter values of less than 0.4~mas, which is well below the lowest value (0.7~mas) measured for the sample of stars that \citet{2004A&A...426..297K} used for calibrating their relations. Furthermore, the calibration sample was biased against metal-poor stars (only three of the 20 dwarf star calibrators had [Fe/H] below $-0.5$~dex, with a lower limit of $-1.4$~dex), and there are indications that the predicted diameters might be slightly underestimated for metal-poor stars \citep{2012A&A...545A..17C,2015A&A...575A..26C}.
The indirectly determined angular diameters are given in \tab{tab:parameters}.

\subsection{Bolometric flux}
\label{sect:fbol}
\paragraph{Direct measurements}
\Fbol\ values based on integrations of the observed spectral energy distribution (SED) were found in the literature for 24 stars.
These are given in \tab{tab:parameters} together with their references.
For about half of these stars, the data were taken from \citet{1998A&AS..129..505B}, 
after adopting uncertainties of 1.1\% for dwarfs and 2.6\% for giants and subgiants following \citet[Sect.~3.2, p.~863]{1998A&A...339..858D}. 
For Procyon, the \Fbol\ value was taken from \citet{2005ApJ...633..424A}, 
and for Arcturus from \citet{1999AJ....117.2998G}, 
adopting an uncertainty of 2.6\% as derived by \citet{2008AJ....135.1551K}. 
For \alfCen~A and B, \delEri, and 18~Sco, we adopted the measurements by \citet{2013ApJ...771...40B}, 
and for \Cyg~A and B those by \citet{2013ApJ...779..188M}. 
For the remaining stars, the \Fbol\ determinations were taken from the publications for the interferometric angular diameters discussed in \sect{sect:ang}.

\paragraph{Calibrations}
The K dwarf \epsEri\ and the metal-poor subgiant HD~140283 are included in Table~4 of \citet{1996AaAS..117..227A}, 
which lists bolometric fluxes derived with the calibration of \citet{1995AaA...297..197A}, and we adopted these values with uncertainties of 2\%.
For the dwarf star \muAra,\ we used the calibration of \citet[their Eqs.~8, 9, and 10]{1995AaA...297..197A} 
of the bolometric flux as a function of the $K$ magnitude, $(V-K)$ and \FeH\footnote{Note that the original Eq.~8 must be divided by 10 to obtain \Fbol\ in units of mWm$^{-2}$.}, with uncertainties of 2\%.
The photometry can be found in \tab{tab:phot}.
The $K$ magnitudes from 2MASS were transformed to the Johnson system in the same way as for the surface-brightness relations described above.
The metallicity was set to $+0.2$~dex, which corresponds to the upper validity limit of the calibration.

For several giants and subgiants, 
we computed the bolometric flux from the $V$ magnitudes and the bolometric corrections BC$_V$\footnote{\Fbol$=\frac{1}{4\pi (10{\rm pc})^2} \cdot L_0 \cdot 10^{-0.4(V+{\rm BC}_V)}$, where $L_0$ is the zero-point luminosity, given in \tab{tab:Sun}.}.
The latter were derived from the calibration of \citet{1999A&AS..140..261A} 
of BC$_V$ as a function of \Teff\ and \FeH.
Since \Fbol\ is used to compute \Teff, we iterated until we obtained consistent values for \Teff\ and BC$_V$.
For the uncertainty in BC$_V$, we adopted a value of 0.05~mag.
The photometry can be found in \tab{tab:phot}.
The metallicities were those derived in Paper~III of this series.
The bolometric fluxes determined from calibrations are given in \tab{tab:parameters}.

\begin{table*}
   \caption{Measured or calibrated angular diameters and bolometric fluxes, and their uncertainties (absolute, $u$, and in percent, $\%u$) for \gbs. See text for description, and table notes for references.}
   \label{tab:parameters}
   \centering
   \begin{tabular}{lrrrllrrrl}
      \hline\hline
      Name & \Ang\ [mas] & $u$(\Ang) & $\%u$(\Ang) & Band & Ref(\Ang) & \Fbol\ [$10^{-9}$Wm$^{-2}$] & $u$(\Fbol) & $\%u$(\Fbol) & Ref(\Fbol)\\
      \hline
      {\bf F dwarfs} & & & & & & & & & \\
Procyon&5.390&0.030&0.6&K&Ch&17.8600&0.8900&5.0&A\\
HD 84937&0.153&0.005&3.0&--&K04*&0.0127&0.0001&1.1&B98\\
HD 49933&0.445&0.012&2.7&735nm&B11&0.1279&0.0014&1.1&B98\\
\hline
{\bf FGK subgiants} & & & & & & & & & \\
\delEri &2.394&0.029&1.2&K&T&1.1500&0.0008&0.1&Bo13\\
HD 140283&0.353&0.013&3.7&720nm&C15&0.0386&0.0008&2.0&A96*\\
\epsFor&0.788&0.016&2.0&--&K04*&0.1425&0.0066&4.6&H*\\
\etaBoo&2.189&0.014&0.6&K&vB&2.2100&0.0282&1.3&vB\\
\betHyi&2.257&0.019&0.8&700nm&N07&2.0190&0.0525&2.6&B98\\
\hline
{\bf G dwarfs} & & & & & & & & & \\
\alfCen A&8.511&0.020&0.2&K&K&27.1600&0.2670&1.0&Bo13\\
HD 22879&0.382&0.011&3.0&--&K04*&0.0577&0.0006&1.1&B98\\
\muCas&0.973&0.009&0.9&K'&Bo08&0.2504&0.0028&1.1&B98\\
\tauCet&2.015&0.011&0.5&K'&D&1.1620&0.0128&1.1&B98\\
\alfCen B&6.000&0.021&0.4&K&B06&8.9800&0.1220&1.4&Bo13\\
18 Sco&0.676&0.006&0.9&700nm&Ba&0.1734&0.0090&5.2&Bo13\\
\muAra&0.763&0.015&2.0&--&K04*&0.2354&0.0047&2.0&A95*\\
\betVir&1.450&0.018&1.2&700nm&N09&0.9590&0.0105&1.1&B98\\
\hline
{\bf FGK giants} & & & & & & & & & \\
Arcturus&21.050&0.210&1.0&H&L&49.8000&1.2948&2.6&G\\
HD 122563&0.940&0.011&1.2&K&C12&0.1303&0.0061&4.7&H*\\
\muLeo&2.930&0.040&1.4&--&C99*&1.1458&0.0530&4.6&H*\\
\betGem&7.980&0.080&1.0&opt&M&11.8200&0.5319&4.5&M\\
\epsVir&3.280&0.030&0.9&opt&M&2.2100&0.0994&4.5&M\\
\ksiHya&2.386&0.021&0.9&K&T&1.2280&0.0319&2.6&B98\\
HD 107328&1.740&0.020&1.1&--&C99*&0.4122&0.0195&4.7&H*\\
HD 220009&2.045&0.034&1.7&800nm&Tp&0.4409&0.0206&4.7&H*\\
\hline
{\bf M giants} & & & & & & & & & \\
\alfTau&20.580&0.030&0.1&K&RR&33.5700&1.3500&4.0&RR\\
\alfCet&12.200&0.040&0.3&K&W4&10.3000&0.7000&6.8&W4\\
\betAra&5.997&0.037&0.6&K&Tp&3.7179&0.1718&4.6&H*\\
\gamSge&6.060&0.020&0.3&K&W3&2.5700&0.1300&5.1&W3\\
\psiPhe&8.130&0.200&2.5&K&W2&3.2000&0.3000&9.4&W2\\
\hline
{\bf K dwarfs} & & & & & & & & & \\
\epsEri&2.126&0.014&0.7&K'&D&1.0000&0.0200&2.0&A96*\\
Gmb 1830&0.679&0.015&2.2&K&C12&0.0834&0.0009&1.1&B98\\
61 Cyg A&1.775&0.013&0.7&K&K08&0.3844&0.0051&1.3&M13\\
61 Cyg B&1.581&0.022&1.4&K&K08&0.2228&0.0032&1.4&M13\\

      \hline\hline
   \end{tabular}
   \tablefoot{
An asterisk indicates indirect determinations of \Ang\ and \Fbol\ values determined from calibrations.\\
The column headed ``Band'' indicates the wavelength band of the interferometric observations -- ``opt'' for several optical wavelengths, H for several wavelengths from 1.5 to 1.8~$\mu$m, K or K' for $K$-band (about 2~$\mu$m), or wavelength in nm.\\
References for \Ang:
B06 ... \citet{2006A&A...446..635B};  
B11 ... \citet{2011A&A...534L...3B};  
Ba ... \citet{2011A&A...526L...4B};   
Bo08 ... \citet{2008ApJ...683..424B}; 
C12 ... \citet{2012A&A...545A..17C};  
C15 ... \citet{2015A&A...575A..26C}   
C99 ... \citet{1999AJ....117.1864C};  
Ch ... \citet{2012A&A...540A...5C};   
D ... \citet{2007A&A...475..243D};    
K ... \citet{2003A&A...404.1087K};    
K08 ... \citet{2008A&A...488..667K};  
K04 ... indirect, using surface-brightness relations by \citet{2004A&A...426..297K}; 
M ... \citet{2003AJ....126.2502M};    
L ... \citet[][$\tau_{\rm Ross}=1$ diameter]{2008A&A...485..561L}; 
N07 ... \citet{2007MNRAS.380L..80N};  
N09 ... \citet{2009MNRAS.393..245N};  
RR ... \citet{2005A&A...433..305R};   
T ... \citet{2005AaA...436..253T};    
Tp ... Th\'evenin et al. in prep., VLTI/AMBER measurement for \betAra, CHARA/VEGA for HD~220009; 
vB ... \citet{2007ApJ...657.1058V};   
W2/W3/W4 ... \citet[][$\tau_{\rm Ross}=1$ diameter]{2004A&A...413..711W,2006A&A...460..843W,2006A&A...460..855W}\\ 
Additional references for \Fbol:
A ... \citet{2005ApJ...633..424A}; 
A95 ... \citet{1995AaA...297..197A} calibration for dwarf stars; 
A96 ... \citet[their Table~4]{1996AaAS..117..227A}; 
B98 ... \citet{1998A&AS..129..505B}, 
uncertainties of 1.1\% for dwarfs and 2.6\% for giants and subgiants following \citet[Sect.~3.2, p.~863]{1998A&A...339..858D}; 
Bo13 ... \citet{2013ApJ...771...40B}, 
G ... \citet{1999AJ....117.2998G}, 
uncertainty of 2.6\% from \citet{2008AJ....135.1551K}; 
H ... from the $V$ magnitude and the bolometric correction BC$_V$ derived from the calibration of \citet{1999A&AS..140..261A}; 
M13 ... \citet{2013ApJ...779..188M} 
   }
\end{table*}

\subsection{Additional determinations of angular diameter and bolometric flux}
\label{sect:ang_fbol_comparison}

\begin{table*}
   \caption{Other interferometric measurements of angular diameter, which are compared to the adopted values in \fig{fig:ang_fbol_comparison}.}
   \label{tab:ang_comparison}
   \centering
   \begin{tabular}{lccrll}
      \hline\hline
      Name & \Ang\ [mas] & $u$(\Ang) & $\%u$(\Ang) & Band & Reference \\
      \hline
      \alfCet&      13.2&       0.3&       1.9&opt&\citet{2003AJ....126.2502M}\\
\alfCet&      11.9&       0.4&       3.5&K&\citet{1998AJ....116..981D}\\
\gamSge&      6.22&      0.06&       1.0&opt&\citet{2003AJ....126.2502M}\\
\alfTau&      21.1&       0.2&       1.0&opt&\citet{2003AJ....126.2502M}\\
\alfTau&      20.2&       0.3&       1.5&K&\citet{1987AaA...188..114D}\\
Arcturus&      21.4&       0.2&       1.2&opt&\citet{2003AJ....126.2502M}\\
Arcturus&      21.0&       0.2&       1.0&K&\citet{1986AaA...166..204D}\\
Arcturus&      19.5&        1.&       5.1&K&\citet{1996AJ....111.1705D}\\
Arcturus&      20.9&      0.08&       0.4&K&\citet{1998AaA...331..619P}\\
Arcturus&      21.2&       0.2&       1.0&K&\citet{2005AaA...435..289V}\\
Gmb1830&     0.696&     0.005&       0.7&K'&\citet{2012ApJ...746..101B}\\
\betGem&      7.97&       0.1&       1.4&800&\citet{2001AJ....122.2707N}\\
\betGem&      7.90&       0.3&       3.9&K&\citet{1987AaA...188..114D}\\
\betGem&      7.95&      0.09&       1.1&740&\citet{2001AJ....122.2707N}\\
\epsVir&      3.28&      0.05&       1.5&800&\citet{2001AJ....122.2707N}\\
\epsVir&      3.23&      0.05&       1.5&740&\citet{2001AJ....122.2707N}\\
\epsEri&      2.15&      0.03&       1.4&K&\citet{2004AaA...426..601D}\\
\tauCet&      2.07&      0.01&       0.5&opt&\citet{2014ApJ...781...90B}\\
\tauCet&      2.08&      0.03&       1.5&K&\citet{2004AaA...426..601D}\\
18Sco&     0.780&      0.02&       2.2&K'&\citet{2012ApJ...746..101B}\\
\etaBoo&      2.20&      0.03&       1.4&K&\citet{2005AaA...436..253T}\\
\etaBoo&      2.28&      0.07&       3.1&740&\citet{2001AJ....122.2707N}\\
\etaBoo&      2.13&      0.01&       0.6&opt&\citet{2014ApJ...781...90B}\\
\etaBoo&      2.27&      0.03&       1.1&opt&\citet{2003AJ....126.2502M}\\
\etaBoo&      2.17&      0.03&       1.4&800&\citet{2001AJ....122.2707N}\\
Procyon&      5.46&      0.08&       1.5&800&\citet{2001AJ....122.2707N}\\
Procyon&      5.40&      0.03&       0.6&K&\citet{2005ApJ...633..424A}\\
Procyon&      5.45&      0.05&       1.0&K&\citet{2004AaA...413..251K}\\
Procyon&      5.45&      0.05&       1.0&opt&\citet{2003AJ....126.2502M}\\
Procyon&      5.43&      0.07&       1.3&740&\citet{2001AJ....122.2707N}\\

      \hline\hline
   \end{tabular}
   \tablefoot{The column headed ``Band'' indicates the wavelength band of the interferometric observations -- ``opt'' stands for several optical wavelengths, K or K' for $K$-band (about 2~$\mu$m), numbers give the wavelength in nm.}
\end{table*}

\begin{table*}
   \caption{Other measurements of bolometric flux, which are compared to the adopted values in \fig{fig:ang_fbol_comparison}.}
   \label{tab:fbol_comparison}
   \centering
   \begin{tabular}{lccrl}
      \hline\hline
      Name & \Fbol\ [$10^{-9}$Wm$^{-2}$] & $u$(\Fbol) & $\%u$(\Fbol) & Reference \\
      \hline
      \alfCet&      9.57&       0.4&       4.5&\citet{2003AJ....126.2502M}\\
\gamSge&      2.86&       0.1&       4.5&\citet{2003AJ....126.2502M}\\
61Cyg~B&     0.223&     0.004&       2.0&\citet{1996AaAS..117..227A}\\
61Cyg~B&     0.203&     0.001&       0.5&\citet{2012ApJ...757..112B}\\
Arcturus&      48.6&        2.&       4.5&\citet{2003AJ....126.2502M}\\
61Cyg~A&     0.377&     0.002&       0.5&\citet{2012ApJ...757..112B}\\
61Cyg~A&     0.372&     0.007&       2.0&\citet{1996AaAS..117..227A}\\
Gmb1830&    0.0838&     0.002&       2.0&\citet{1996AaAS..117..227A}\\
Gmb1830&    0.0827&    0.0008&       1.0&\citet{2012ApJ...746..101B}\\
\delEri&      1.18&      0.05&       3.9&\citet{2010MNRAS.405.1907B}\\
\ksiHya&      1.17&      0.05&       3.9&\citet{2010MNRAS.405.1907B}\\
\alfCen~B&      8.37&       0.3&       4.2&\citet{2010MNRAS.405.1907B}\\
\muCas&     0.253&     0.005&       2.0&\citet{1996AaAS..117..227A}\\
\tauCet&      1.12&    0.0007&       0.1&\citet{2013ApJ...771...40B}\\
\tauCet&      1.16&      0.02&       2.0&\citet{1996AaAS..117..227A}\\
\tauCet&      1.14&      0.04&       3.3&\citet{2010MNRAS.405.1907B}\\
\tauCet&      1.13&     0.003&       0.3&\citet{2014ApJ...781...90B}\\
\alfCen~A&      26.3&       0.9&       3.4&\citet{2010MNRAS.405.1907B}\\
18Sco&     0.165&     0.003&       2.0&\citet{1995AaA...297..197A}\\
HD22879&    0.0590&     0.001&       2.0&\citet{1996AaAS..117..227A}\\
\betHyi&      1.97&      0.07&       3.7&\citet{2010MNRAS.405.1907B}\\
\betVir&     0.942&      0.02&       2.0&\citet{1996AaAS..117..227A}\\
\betVir&     0.915&      0.03&       3.5&\citet{2010MNRAS.405.1907B}\\
\betVir&     0.944&      0.02&       2.1&\citet{2009MNRAS.393..245N}\\
\etaBoo&      2.17&      0.01&       0.5&\citet{2003AJ....126.2502M}\\
\etaBoo&      2.14&      0.06&       2.9&\citet{2010MNRAS.405.1907B}\\
\etaBoo&      2.14&     0.007&       0.3&\citet{2014ApJ...781...90B}\\
HD84937&    0.0136&    0.0003&       2.0&\citet{1996AaAS..117..227A}\\
Procyon&      18.4&       0.4&       2.0&\citet{1996AaAS..117..227A}\\
Procyon&      18.2&       0.8&       4.5&\citet{2003AJ....126.2502M}\\
Procyon&      17.6&       0.5&       2.7&\citet{2010MNRAS.405.1907B}\\
Procyon&      18.3&      0.02&       0.1&\citet{2013ApJ...771...40B}\\
HD49933&     0.127&    0.0008&       0.6&\citet{2013ApJ...771...40B}\\
HD49933&     0.144&     0.003&       2.0&\citet{1996AaAS..117..227A}\\

      \hline\hline
   \end{tabular}
   \tablefoot{The values with reference \citet{1995AaA...297..197A} or \citet{1996AaAS..117..227A} are based on calibrations using broad-band photometry. All others are determined by integrating over SEDs.}
\end{table*}

\begin{figure}[ht]
   \begin{center}
      \resizebox{\hsize}{!}{\includegraphics{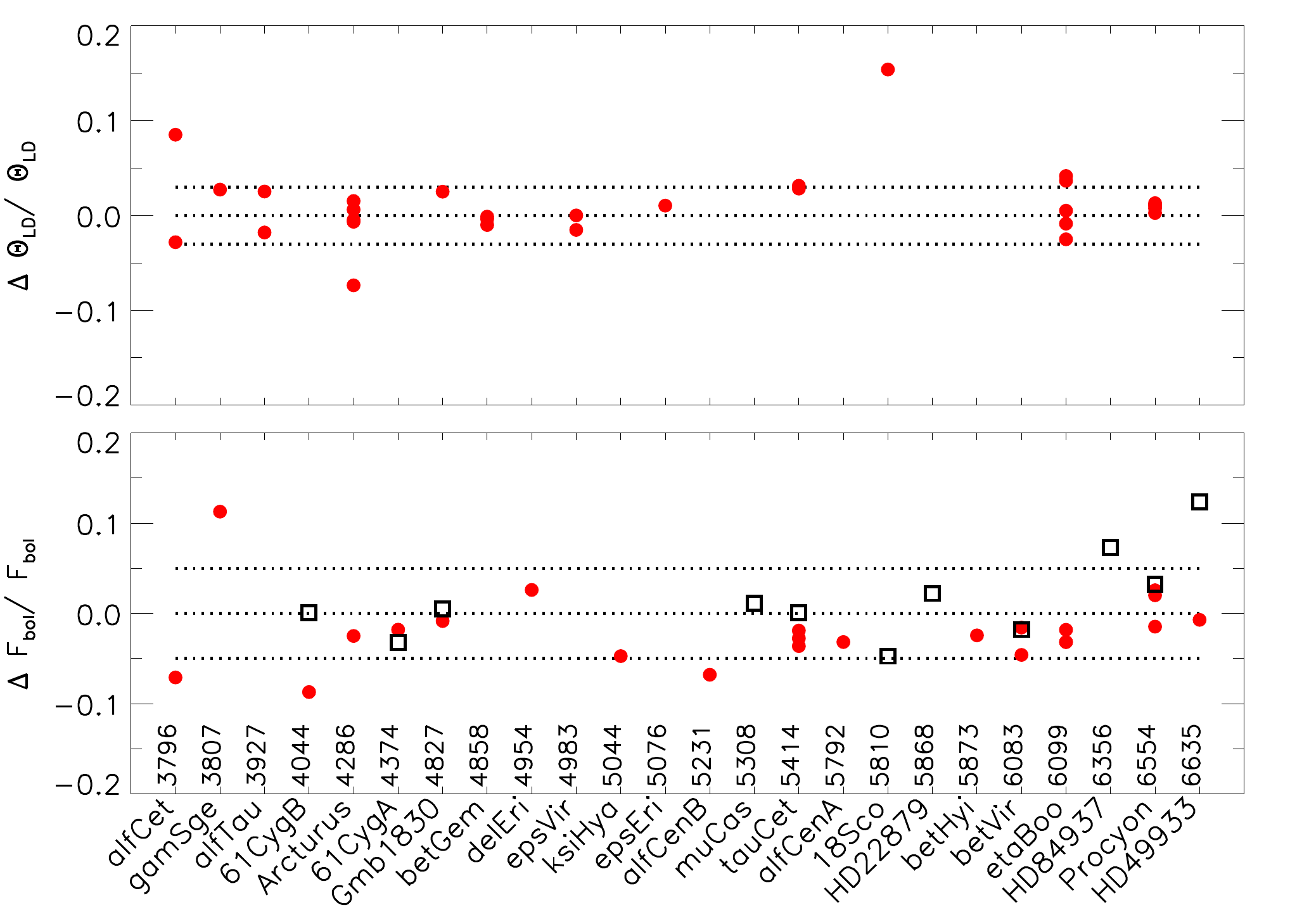}}
   \end{center}
   \caption{Relative difference between additional and adopted values of angular diameter (upper panel) and bolometric flux (lower panel), for \gbs\ with multiple measurements, sorted by \Teff. Red full circles are direct measurements, black open squares are calibrated values (see \tabs{tab:ang_comparison} and \ref{tab:fbol_comparison}). Dotted lines indicate differences of $\pm3$\% (upper panel) and $\pm5$\% (lower panel).}
   \label{fig:ang_fbol_comparison}
\end{figure}

\begin{figure}[ht]
   \begin{center}
      \resizebox{\hsize}{!}{\includegraphics{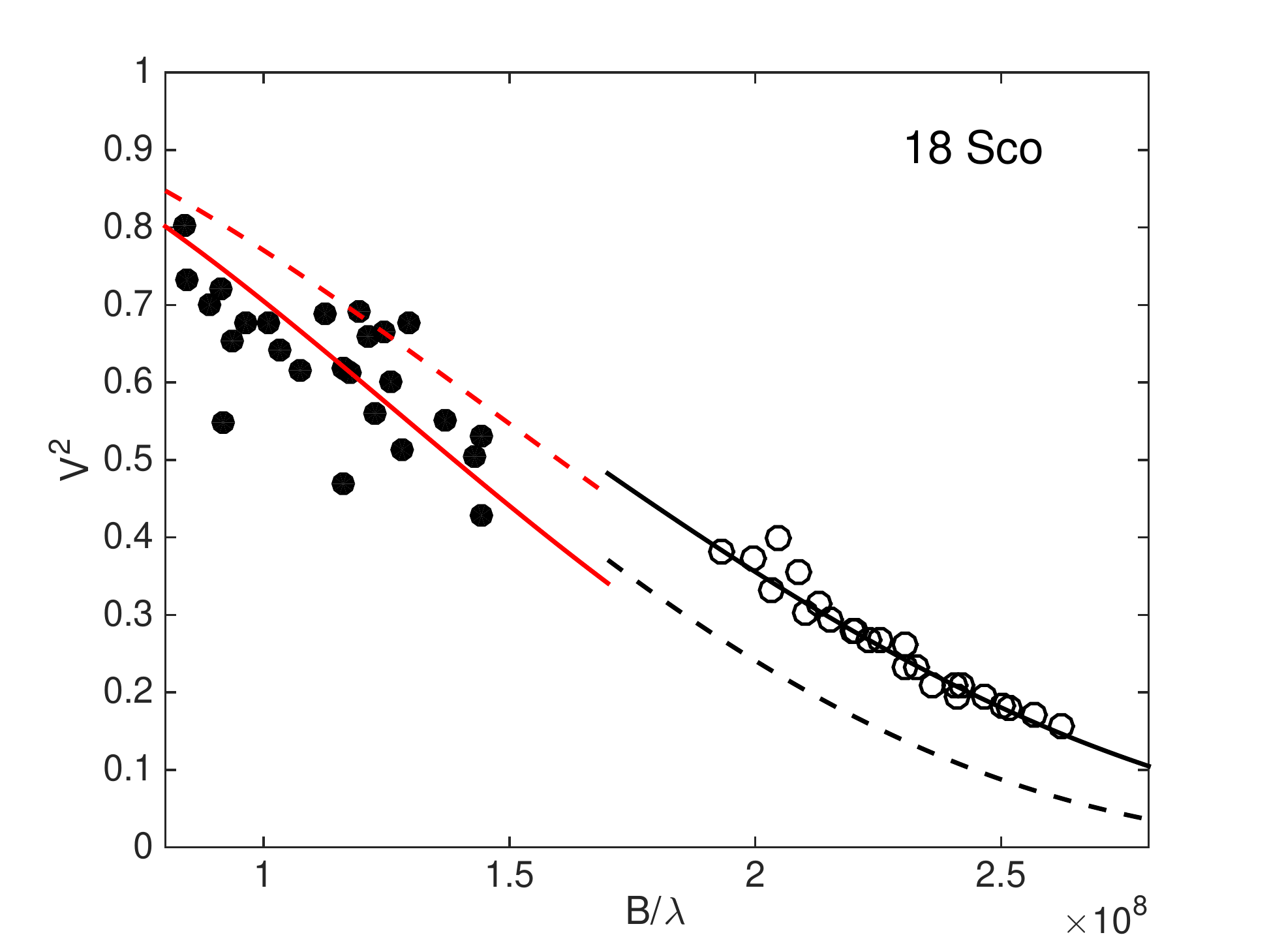}}
   \end{center}
   \caption{18~Sco: Squared visibility as a function of the ratio of projected baseline and wavelength. Open circles: measurements from \citet{2011A&A...526L...4B}; full circles: measurements from \citet{2012ApJ...746..101B}. Lines are calculations using the same model as in the two references. Solid lines: using the author's determination of \Ang; dashed lines: swapped \Ang\ values. Red and black lines: using limb-darkening coefficients for $K$ and $R$ filters, respectively.
}
   \label{fig:ang_18Sco}
\end{figure}

We compiled additional measurements of angular diameter and bolometric flux in \tabs{tab:ang_comparison} and \ref{tab:fbol_comparison}.
The differences from the adopted values are less than 3\% for angular diameter and less than 5\% for bolometric flux for most stars, as shown in \fig{fig:ang_fbol_comparison}.
For \Ang, the deviations are comparable to the uncertainties quoted for the adopted values (\tab{tab:parameters}), with three outliers discussed below.
Uncertainties in \Ang\ of 3\% translate into uncertainties in effective temperature of 1.5\%. 
Also in the case of \Fbol, the differences shown in \fig{fig:ang_fbol_comparison} are comparable to the relative uncertainties quoted in \tab{tab:parameters}. The two stars with the largest differences are discussed below.
Uncertainties in \Fbol\ of 5\% translate into uncertainties in effective temperature of about 1\%. 

One of the outliers in the angular diameter comparison is 18~Sco with a difference of 15\% between the $K$-band measurement of \citet{2012ApJ...746..101B} and the $R$-band measurement of \citet[][adopted here]{2011A&A...526L...4B}, corresponding to about five times the combined uncertainty\footnote{With ``combined uncertainty'' we refer to the uncertainties added in quadrature.}.
We compare the visibility measurements by the two authors in \fig{fig:ang_18Sco}. It is obvious that these measurements are affected by systematic differences, which cannot be explained by the uncertainties in the linear limb-darkening coefficients used by both authors.
A more complex modelling of limb darkening might diminish the discrepancy, since the deviation of realistic limb darkening from 1D approximations could be wavelength dependent.
For example, \citet{2006A&A...446..635B} find that for $K$-band observations of the somewhat cooler star \alfCen~B, the diameter using 1D models for limb darkening was only 0.3\% larger than when using 3D-RHD models.
On the other hand, for $R$-band observations of the somewhat hotter star HD~49933, \citet{2011A&A...534L...3B} derived a 2\% larger diameter from 1D compared to 3D models.
Most of the difference in the measurements may, however, be due to calibration errors (see Sect.~2.2 in \citealt{2013ApJ...771...40B}). 
\citet{2011A&A...526L...4B} and \citet{2012ApJ...746..101B} used three and two calibrator stars, respectively, with one star in common. For that star (HD~145607), they assumed slightly different angular diameters, but they agree within the uncertainties.
In any case, the measurements by \citet{2011A&A...526L...4B} show a much smaller scatter, and the model curve using their \Ang\ value passes through a few of the points by \citet{2012ApJ...746..101B}, while the opposite case is not true.

The two other outliers in the angular diameter comparison (\fig{fig:ang_fbol_comparison}, upper panel) are Arcturus and \alfCet, with the largest and third largest angular diameters. For each star, one of the measurements shows a deviation of 7--8\% from the adopted value (\citealt{1996AJ....111.1705D} and \citealt{2003AJ....126.2502M}, respectively). For Arcturus, the difference is less than two times the combined uncertainties, while the difference is at the 3-$\sigma$ level for \alfCet.
For the latter star, the deviating observation was obtained at optical wavelengths as compared to the infrared. However, there are five other stars with angular diameter determinations at optical wavelengths by the same author \citep{2003AJ....126.2502M} with good agreement with determinations in the infrared.

Regarding the \Fbol\ comparison, the largest differences from the adopted values are found for \gamSge\ and \Cyg~B.
For \gamSge\ \citet{2003AJ....126.2502M} obtained a bolometric flux that is 11\% higher (or twice the combined uncertainty) than \citet[][adopted here]{2006A&A...460..843W}.
\citet{2003AJ....126.2502M} integrated over the SED derived from broad-band photometry alone, while \citet{2006A&A...460..843W} used narrow-band spectrophotometry in the optical/near-infrared region complemented by broad-band photometry in other regions. Thus, the flux estimated by the latter group of authors should be more realistic.

For \Cyg~B \citet{2012ApJ...757..112B} obtained a bolometric flux 9\% lower (or six times the combined uncertainty) than \citet[][adopted here]{2013ApJ...779..188M}.
\citet{2013ApJ...779..188M} measured \Fbol\ for about 20 K- and M-type dwarfs in common with \citet{2012ApJ...757..112B} and obtained 4\% higher values on average. Both authors used a similar method of integrating over stellar SEDs. The difference is that \citet{2013ApJ...779..188M} had spectra observed for each star, while \citet{2012ApJ...757..112B} used spectral templates from an empirical library. \citet{2013ApJ...779..188M} show that the use of templates results in an underestimation of the infrared flux level and argue that this explains the discrepancy for most of the stars. For the extreme cases, such as \Cyg~B, the different input photometry used to determine the absolute flux scale of the spectra might also play a role.

\figu{fig:ang_fbol_comparison} shows that bolometric fluxes determined with the calibration by \citet{1995AaA...297..197A,1996AaAS..117..227A} for dwarf stars agree to within 5\% with those determined from SED integration, except for the hottest stars (\Teff$\gtrsim$6300~K). This justifies using these calibrations for some of the stars in our sample without ``direct'' determinations.

\section{Surface gravity}
\label{sect:logg}

We used the fundamental relation $g = GM/R^2$, where $M$ is the stellar mass, $R$ the stellar radius, and $G$ the Newtonian constant of gravitation (\tab{tab:Sun}), to determine the surface gravity.
The linear radius $R$ was calculated for each star from the angular diameter (see \sect{sect:ang}) and the parallax.
The parallaxes (see \tab{tab:Mass}) were taken from \citet{2007ASSL..350.....V}, 
from \citet{1999A&A...341..121S} for the \alfCen\ system, 
and from \citet{2014ApJ...792..110V} 
for the halo stars HD~84937 and HD~140283.

For the Sun, the surface gravity was calculated from the solar mass parameter given in \tab{tab:Sun} and the radius.
The parameter $GM_{\odot}$ is measured in the process of planetary ephemeris determination with a considerably higher relative accuracy ($\approx10^{-11}$) than that of the constant of gravitation $G$ ($10^{-4}$). The latter is the limiting factor for the achievable accuracy of the solar mass. 
A recent measurement of $GM_{\odot}$ based on tracking data 
from Mars-orbiting spacecraft is given in \citet{2011Icar..211..401K}\footnote{The value given in \citet{2011Icar..211..401K} is compatible with the TDB time scale (Barycentric Dynamical Time) and was converted to the TCB-compatible value (Barycentric Coordinate Time) using the linear transformation adopted in IAU 2006 Resolution B3.}.
The solar radius is given in \tab{tab:Sun} and described in \sect{sect:sun}, and its uncertainty by far dominates the uncertainty in the solar surface gravity.

\subsection{Mass determination from evolutionary tracks}
\label{sect:mass}

\begin{table*}
   \caption{Masses $M$, parallaxes $\pi$, and luminosities $L$, and their uncertainties (absolute, $u$, and in percent, $\%u$), for \gbs. See text for description and references.}
   \label{tab:Mass}
   \centering
   \begin{tabular}{lrrrrrrrrr}
      \hline\hline
      Name & $M$ [$M_\odot$] & $u(M)$ & $\%u(M)$ & $\pi$ [mas] & $u(\pi)$ & $\%u(\pi)$ & $\log L/L_\odot$ & $u(\log L/L_\odot)$ & $L_X/L$ $[10^{-6}]$ \\
      \hline
      {\bf F dwarfs} & & & & & & & & & \\
Procyon&1.50&0.07&5&284.52&1.27&0.45&0.839&0.022&0.72\\
HD 84937&0.75&0.04&5&12.24&0.20&1.63&0.424&0.005&\\
HD 49933&1.17&0.06&5&33.68&0.42&1.25&0.547&0.005&23\\
\hline
{\bf FGK subgiants} & & & & & & & & & \\
\delEri &1.13&0.05&5&110.62&0.22&0.20&0.468&0.000&0.08\\
HD 140283&0.68&0.17&24&17.18&0.26&1.51&0.612&0.009&\\
\epsFor&0.91&0.16&17&31.05&0.36&1.16&0.665&0.020&\\
\etaBoo&1.64&0.07&5&87.77&1.24&1.41&0.953&0.006&0.31\\
\betHyi&1.15&0.04&3&134.07&0.11&0.08&0.546&0.011&0.48\\
\hline
{\bf G dwarfs} & & & & & & & & & \\
\alfCen A&1.11&0.04&3&747.10&1.20&0.16&0.182&0.004&0.38\\
HD 22879&0.75&0.06&7&39.13&0.57&1.46&0.071&0.005&\\
\muCas&0.58&0.08&15&132.40&0.82&0.62&-0.350&0.005&\\
\tauCet&0.71&0.03&4&273.96&0.17&0.06&-0.315&0.005&0.59\\
\alfCen B&0.93&0.06&6&747.10&1.20&0.16&-0.298&0.006&\\
18 Sco&1.02&0.06&6&71.93&0.37&0.51&0.020&0.023&\\
\muAra&1.19&0.06&5&64.48&0.31&0.48&0.248&0.009&\\
\betVir&1.34&0.04&3&91.50&0.22&0.24&0.554&0.005&1.8\\
\hline
{\bf FGK giants} & & & & & & & & & \\
Arcturus&1.03&0.21&20&88.83&0.53&0.60&2.295&0.011&\\
HD 122563&0.86&0.03&3&4.22&0.35&8.29&2.360&0.020&\\
\muLeo&1.69&0.42&25&26.27&0.16&0.61&1.715&0.020&\\
\betGem&2.30&0.40&18&96.52&0.24&0.25&1.598&0.020&\\
\epsVir&3.02&0.11&4&29.75&0.14&0.47&1.892&0.020&\\
\ksiHya&2.84&0.12&4&25.14&0.16&0.64&1.784&0.011&\\
HD 107328&1.41&0.41&29&10.60&0.25&2.36&2.060&0.021&\\
HD 220009&0.83&0.21&26&7.55&0.40&5.30&2.383&0.020&\\
\hline
{\bf M giants} & & & & & & & & & \\
\alfTau&0.96&0.41&43&48.92&0.77&1.57&2.642&0.017&\\
\alfCet&1.76&0.91&52&13.10&0.44&3.36&3.273&0.030&\\
\betAra&8.21&1.88&23&4.54&0.61&13.44&3.751&0.020&\\
\gamSge&1.11&0.82&75&12.61&0.18&1.43&2.704&0.022&\\
\psiPhe&1.00&0.40&40&9.54&0.20&2.10&3.041&0.041&\\
\hline
{\bf K dwarfs} & & & & & & & & & \\
\epsEri&0.80&0.06&8&310.95&0.16&0.05&-0.490&0.009&17\\
Gmb 1830&0.64&0.03&4&109.98&0.41&0.37&-0.666&0.005&\\
61 Cyg A&0.69&0.05&7&286.83&6.77&2.36&-0.835&0.006&5\\
61 Cyg B&0.61&0.05&8&285.89&0.55&0.19&-1.069&0.006&\\

      \hline\hline
   \end{tabular}
   \tablefoot{Column $L_X/L$ gives the ratio of X-ray luminosity measured by ROSAT \citep{1998A&AS..132..155H} to bolometric luminosity.}
\end{table*}

\begin{table}
   \caption{Mean $\alpha$-element abundances and mass corrections determined from \emph{Yonsei-Yale} grids: $\Delta M = M(+0.3)-M(0.0)$.}
   \label{tab:alpha}
   \centering
   \begin{tabular}{lllcr}
      \hline\hline
Name & \aFe & Elements & Ref. & $\Delta M [M_\odot]$ \\
      \hline
HD~84937   & +0.4  &  Mg, Ca         &  1    &    0.00 \\
HD~140283  & +0.3  &  Mg, Ca         &  1    &   +0.02 \\
\epsFor    & +0.25 &  Si, Ti         &  2    & $-$0.02 \\
HD~22879   & +0.3  &  Si, Ti         &  2    &    0.00 \\
\muCas     & +0.3  &  Mg, Si, Ca, Ti &  3    &    0.00\tablefootmark{$\dagger$} \\
\tauCet    & +0.2  &  Si, Ca, Ti     &  2, 3 &   +0.01 \\
HD~220009  & +0.2  &  Si, Ca, Ti     &  4    & $-$0.03 \\
      \hline\hline
   \end{tabular}
   \tablefoot{References: (1) \citet{2007A&A...461..261M}, (2) \citet{2005ApJS..159..141V}, (3) \citet{2005AJ....129.1063L}, (4) \citet{2007AJ....133.2464L}.
   \tablefoottext{$\dagger$}{For \muCas\ only the \aFe=+0.3 \emph{Yonsei-Yale} grid could be used, and the \aFe\ correction was assumed to be zero, as found for HD~22879.}}
\end{table}

For all stars apart from the Sun, we aimed to determine the mass (in solar units) in a homogeneous way by visual interpolation in two different grids of evolutionary tracks. We used the fundamental \Teff\ value (\sect{sect:tefflogg}), the stellar luminosity, and the spectroscopic metallicity as constraints.
The luminosity was calculated from the bolometric flux (see \sect{sect:fbol}) and the parallax (see \tab{tab:Mass}).
For the metallicity, we used the [Fe/H] value from Table~3 in Paper~III, with an uncertainty obtained by quadratically summing all $\sigma$ and $\Delta$ columns in that table (see \tab{tab:general}).
For the two stars HD~122563 and Gmb~1830, and the two components of \Cyg, the \Teff\ and $L$ values lie outside the range of the selected grids. Thus, we adopted the masses determined by \citet{2012A&A...545A..17C} 
and \citet{2008A&A...488..667K}, 
respectively, which are based on CESAM2k stellar evolution models \citep{1997A&AS..124..597M,2008Ap&SS.316...61M}.

Throughout the discussion in this section, one should keep in mind that the possible uncertainties in the mass are less important for determining surface gravity than the uncertainty of the radius.
We selected the \emph{Padova} stellar evolution models by \citet{2008A&A...484..815B,2009A&A...508..355B}\footnote{\url{http://stev.oapd.inaf.it/YZVAR/}} 
and the \emph{Yonsei-Yale} models by \citet{2003ApJS..144..259Y}, \citet{2004ApJS..155..667D}\footnote{\url{http://www.astro.yale.edu/demarque/yystar.html}}. 
These groups have published grids of stellar evolution tracks for a wide range of masses and metallicities.
The grid metallicities are given by [Fe/H] = $\log_{10}(Z_0/X_0) - \log_{10}(Z_{0,\odot}/X_{0,\odot})$, where $Z_0$ and $X_0$ are the initial mass fractions of metals and hydrogen, respectively.
For the \emph{Padova} models, $Z_{0,\odot}=0.017$ and $X_{0,\odot}=0.723$, and the metallicities range from +0.75 to $-2.26$~dex.
For the \emph{Yonsei-Yale} models, $Z_{0,\odot}=0.0181$ and $X_{0,\odot}=0.7149$ \citep{2001ApJS..136..417Y}, and the metallicities range from +0.78 to $-3.29$~dex.
The initial mass fraction of helium $Y_0$ was set to $Y_0 = 0.23+2Z_0$.
For the \emph{Yonsei-Yale} models, grids for different $\alpha$-element abundances are available.
For stars with enhanced \aFe\ ratios, \emph{Yonsei-Yale} tracks with \aFe=+0.3 were used, based on $\alpha$-element abundances from the literature (see \tab{tab:alpha}). In these cases, the mass was also determined from solar \aFe\ tracks, and a correction determined in this way was applied to the masses obtained from the \emph{Padova} tracks. The correction values are given in \tab{tab:alpha} and are found to be insignificant.

\begin{figure}[ht]
   \begin{center}
      \resizebox{\hsize}{!}{\includegraphics[trim=50 50 50 50,clip]{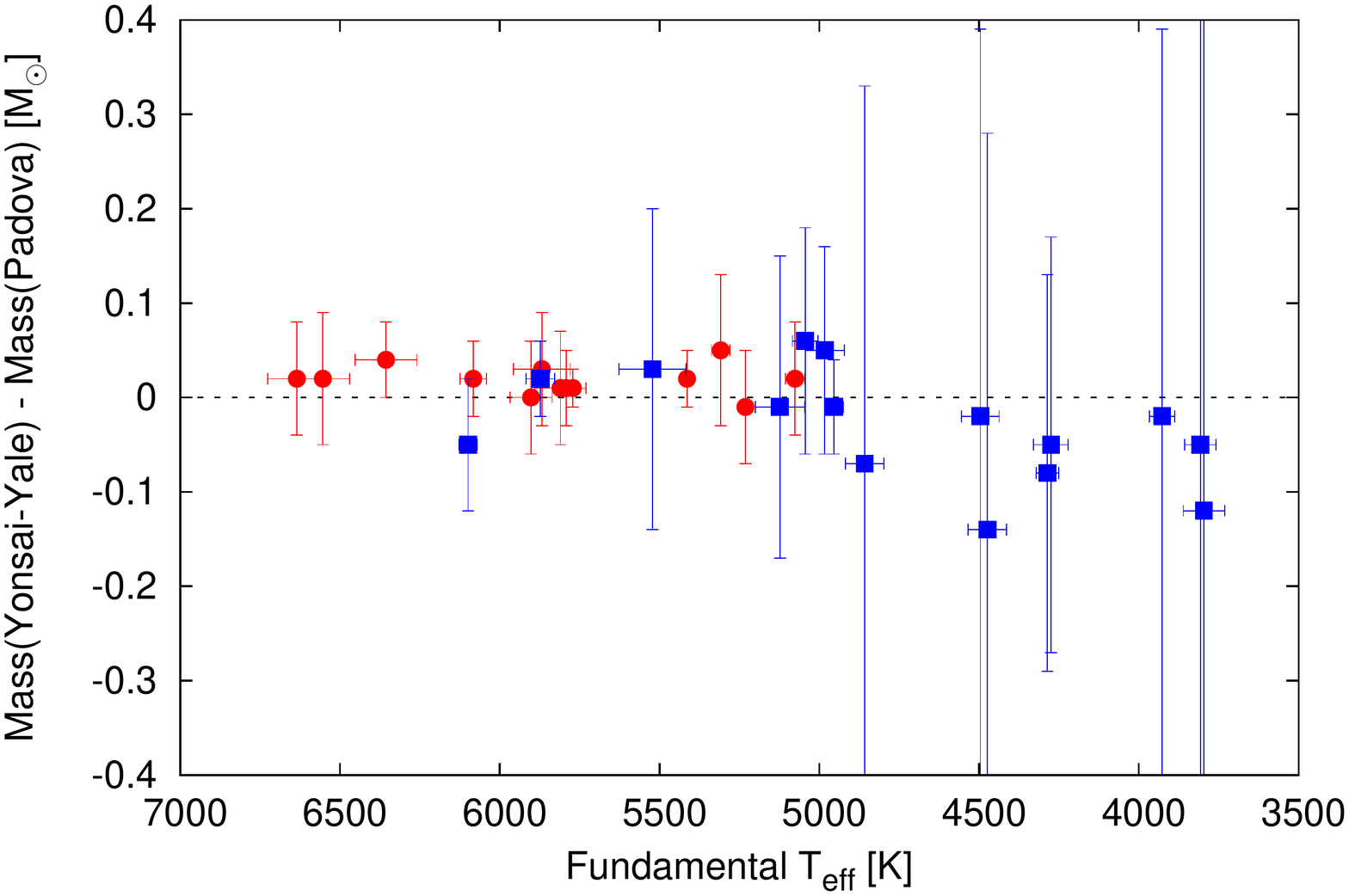}}
   \end{center}
   \caption{Difference between masses determined from \emph{Yonsei-Yale} and \emph{Padova} grids as a function of fundamental \Teff. Red circles are dwarfs (\logg $\ge$ 4), blue squares are giants and subgiants.}
   \label{fig:massPYY}
\end{figure}

\begin{figure}[t]
   \begin{center}
      \resizebox{\hsize}{!}{\includegraphics[trim=50 50 50 50,clip]{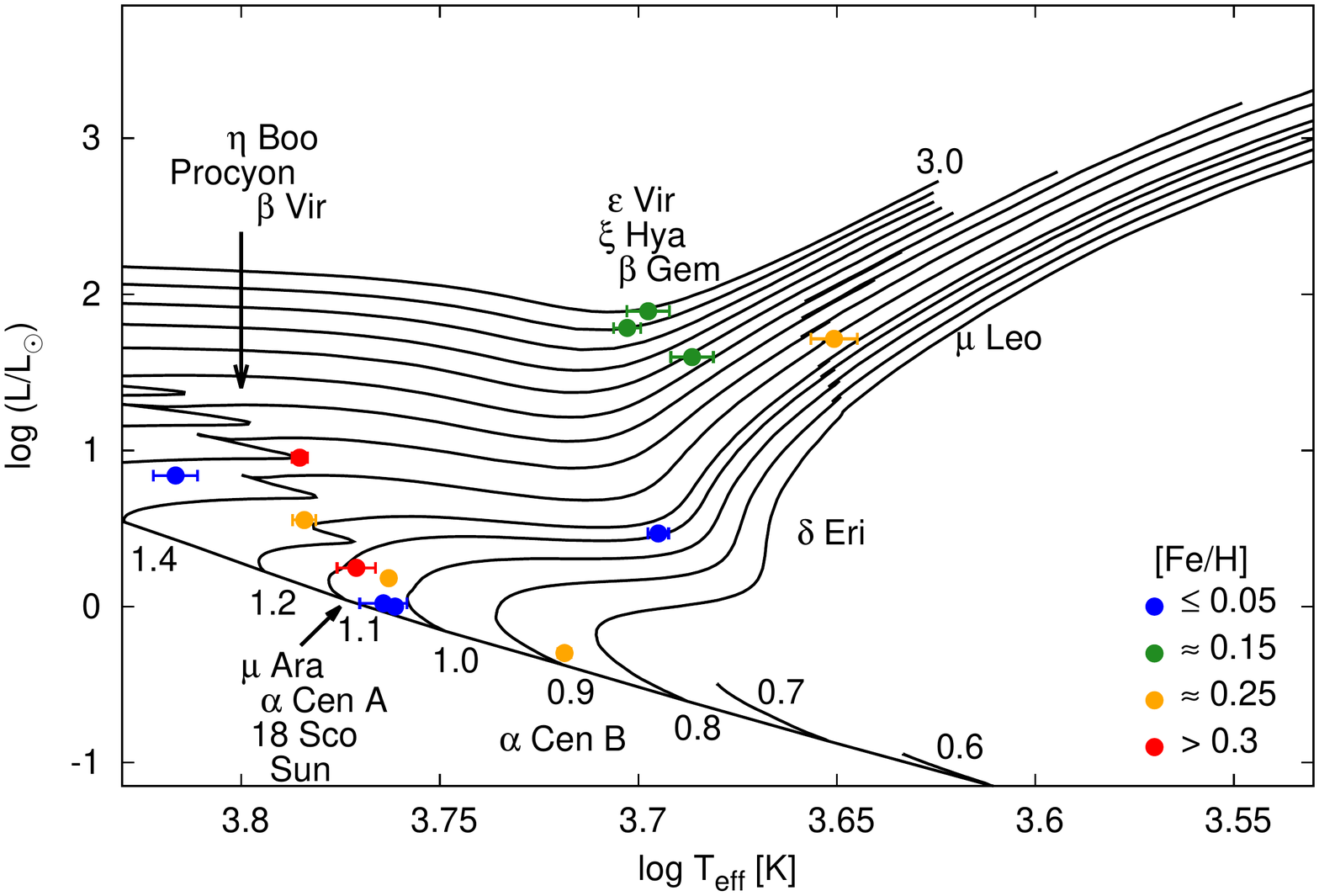}}
   \end{center}
   \caption{HR-diagram with fundamental \Teff, and $L$ from \Fbol\ and parallax for \gbs\ with metallicities between +0.4 and $0.0$~dex. \emph{Yonsei-Yale} evolutionary tracks for \FeH=+0.05 and \aFe=+0.0, labelled with mass in $M_\odot$. Between 1.4 and 3.0~$M_\odot$ the step is 0.2~$M_\odot$. The mass can be read off the figure only for \delEri, which has the same metallicity as the tracks. For lower metallicities (blue points) the tracks shift towards higher \Teff\ and $L$, and the points will be located on tracks with lower masses (cf. \fig{fig:HRD3}). The opposite applies to stars with higher metallicities (green, orange, red points), which will fall on tracks with higher masses than those shown (e.g. $\approx1.2~M_\odot$ for \muAra).}
   \label{fig:HRD1a}
\end{figure}

\begin{figure}[ht]
   \begin{center}
      \resizebox{\hsize}{!}{\includegraphics[trim=50 50 50 50,clip]{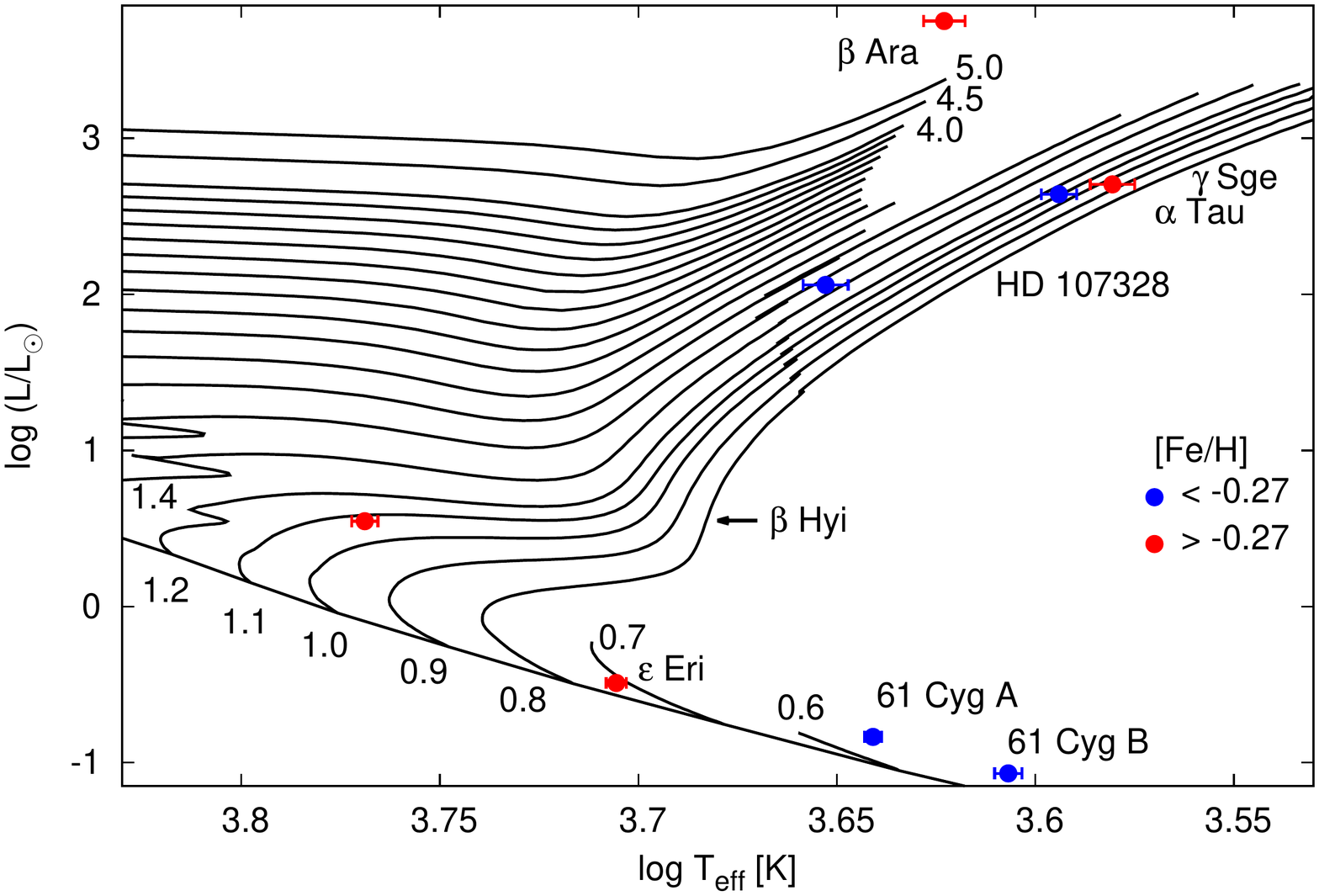}}
   \end{center}
   \caption{Same as \fig{fig:HRD1a} for stars with metallicities between $0.0$ and $-0.4$~dex. \emph{Yonsei-Yale} evolutionary tracks for \FeH=$-0.27$ and \aFe=+0.0, labelled with mass in $M_\odot$. Between 1.4 and 4.0~$M_\odot$ the step is 0.2~$M_\odot$. Note that none of the stars has exactly the same metallicity as the shown tracks.}
   \label{fig:HRD1b}
\end{figure}

\begin{figure}[ht]
   \begin{center}
      \resizebox{\hsize}{!}{\includegraphics[trim=50 50 50 50,clip]{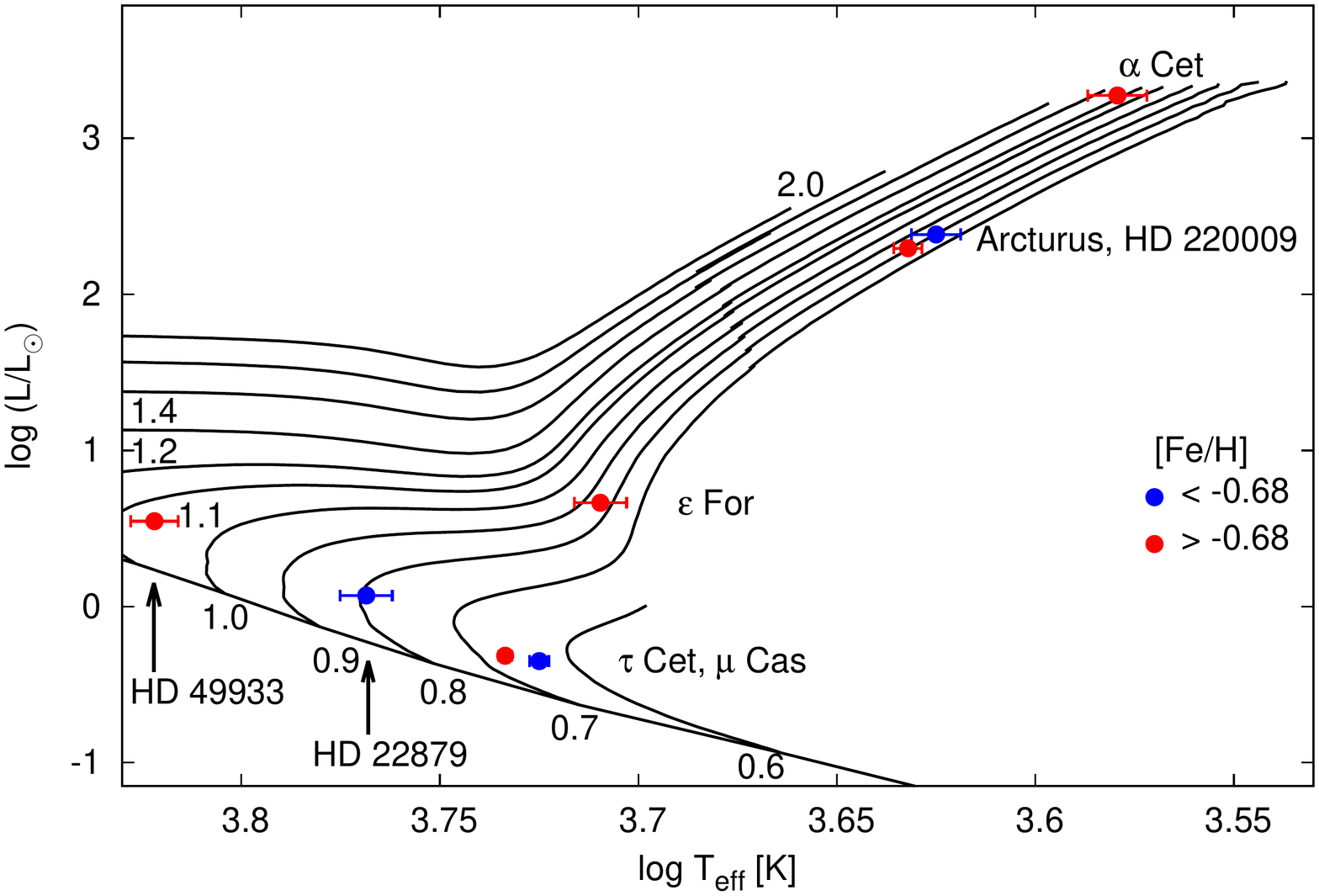}}
   \end{center}
   \caption{Same as \fig{fig:HRD1a} for stars with metallicities between $-0.4$ and $-0.9$~dex. \emph{Yonsei-Yale} evolutionary tracks for \FeH=$-0.68$ and \aFe=+0.3, labelled with mass in $M_\odot$. The step is 0.2~$M_\odot$ between 1.4 and 2.0~$M_\odot$. HD~220009 with $M\approx0.8~M_\odot$ has a metallicity closest to that of the shown tracks. HD~49933 and HD~22879 lie at the maximum and the minimum of the metallicity range, thus falling on higher (1.2~$M_\odot$) and lower mass (0.7~$M_\odot$) tracks, respectively, than those shown.}
   \label{fig:HRD2}
\end{figure}

\begin{figure}[ht]
   \begin{center}
      \resizebox{\hsize}{!}{\includegraphics[trim=50 50 50 50,clip]{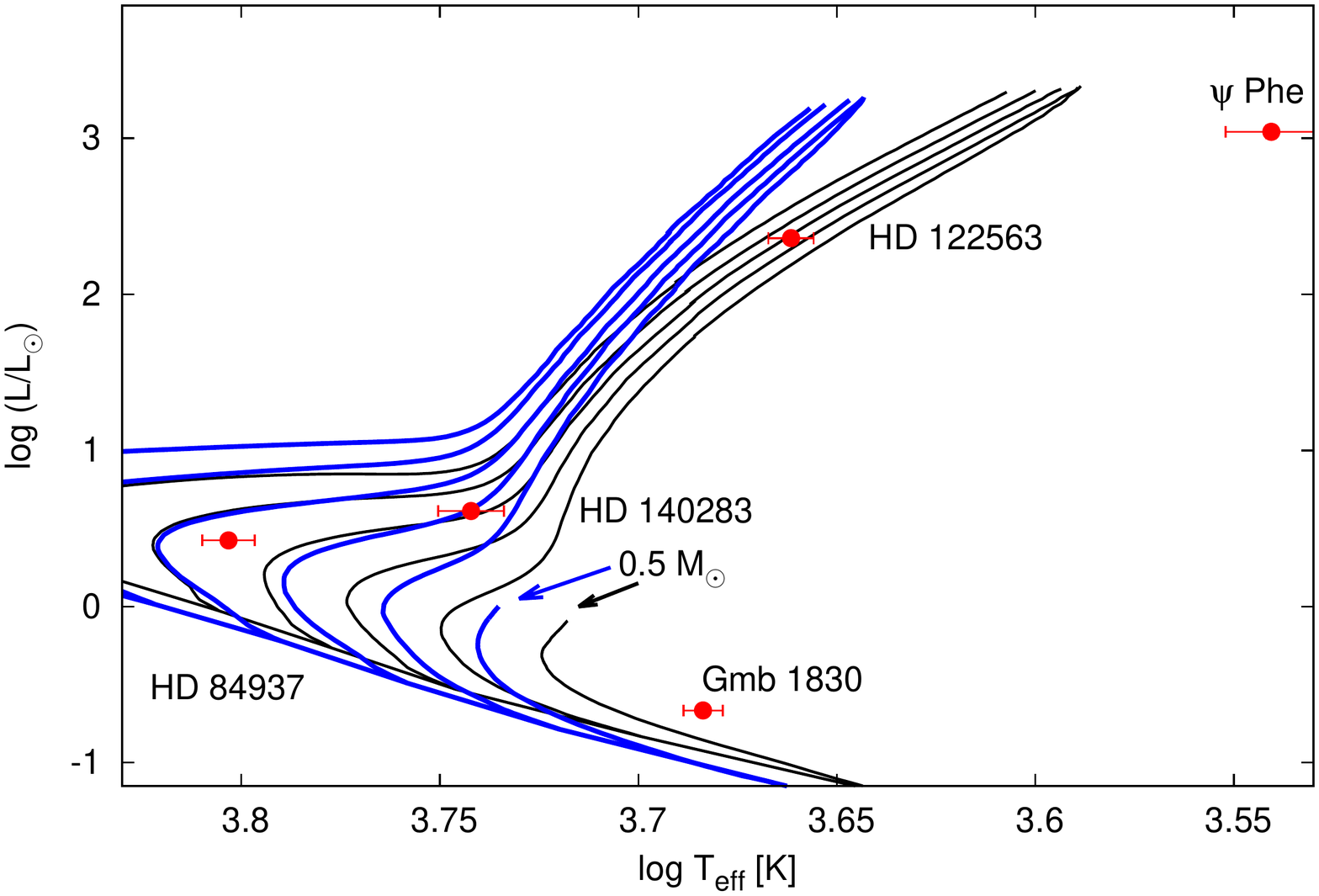}}
   \end{center}
   \caption{Same as \fig{fig:HRD1a} for stars with metallicities lower than $-1.2$~dex. \emph{Yonsei-Yale} evolutionary tracks for \FeH=$-1.29$ (black, to the right) and $-2.29$ (blue, to the left), and \aFe=+0.3, for masses between 0.5 and 1.0~$M_\odot$, in steps of 0.1~$M_\odot$. Clearly, \psiPhe\ (\FeH=$-1.2$), Gmb~1830 ($-1.5$) and HD~122563 ($-2.6$) lie outside the range of the tracks with corresponding metallicities. HD~84937 and HD~140283 have metallicities close to the $-2.29$ tracks, resulting in derived masses of about 0.75 and 0.7~$M_\odot$, respectively.}
   \label{fig:HRD3}
\end{figure}

The final adopted mass value is the average from the \emph{Padova} and \emph{Yonsei-Yale} grids (except for \betAra, see below) and is given in \tab{tab:Mass}, where we also list the parallaxes, luminosities, and relative X-ray luminosities.
The uncertainties of the masses were determined for each grid based on the uncertainties of \Teff, $L$, and [Fe/H]. See Appendix~\ref{app:masses} for more details and examples.
The uncertainties of the adopted masses were then estimated using the maximum possible deviation from the mean in any one of the two grids\footnote{If $M$ is the adopted mass, $M_P$ and $M_Y$ the masses from the \emph{Padova} and \emph{Yonsei-Yale} grids, respectively, and $\sigma_P$ and $\sigma_Y$ the corresponding uncertainties, then $\sigma_M$ = max($|M-(M_P-\sigma_P)|,|M-(M_P+\sigma_P)|,|M-(M_Y-\sigma_Y)|,|M-(M_Y+\sigma_Y)|$).}.
Applying the same method using the solar \Teff\ and $L$ results in a mass of $1.01\pm0.02~M_\odot$.

The main difference between the two sets of evolutionary tracks is that the \emph{Yonsei-Yale} models take the diffusion of helium into account, while the \emph{Padova} models do not include diffusion at all. Other differences can be found in the assumptions for microscopic physics, in the treatment of convective core overshoot, or in the boundary conditions at the surface.
However, \figu{fig:massPYY} shows that the differences between the masses determined from the two grids are much lower than the uncertainties. 
\figus{fig:HRD1a} to \ref{fig:HRD3} show the locations of all \gbs\ in the theoretical HR-diagram for several metallicity groups, together with \emph{Yonsei-Yale} evolutionary tracks.

We note that the best possible modelling of a star's location in the HR-diagram should take other parameters into account apart from metallicity and $\alpha$-element abundances.
Examples of these are the mixing length used for modelling convection, the initial helium abundance, or diffusion of chemical elements other than helium.
When using a fixed set of evolutionary tracks for all stars, we do not have the possibility of arbitrarily varying these parameters, but detailed modelling of each star is beyond the scope of this article.
Another shortcoming of the method is that we did not consider evolution beyond the red giant stage. 
Stars with luminosities greater than about 1.6~$L_\odot$ could be in a stage after the onset of core helium burning.
Tracks for the corresponding ``horizontal branch'' (HB) models are included in the \emph{Padova} grid, and we used these to evaluate the effect of ambiguous evolutionary stages on mass determination for the giants in our sample.
Most of the stars are in fact not located on the HB tracks. The only affected stars are \gamSge, for which the mass derived from HB tracks would be lower by about 0.1~$M_\odot$, as well as \muLeo\ and HD~107328, for which the HB mass would be lower by about 0.4~$M_\odot$.

The interpolation between evolutionary tracks was only made with respect to mass. We did not try to estimate an age for any of the stars. However, the model grids include models with assigned ages that are greater than the age of the Universe, and we noticed that a few of the stars lie close to or in the areas in the HR diagram corresponding to such models. This points towards problems with the model physics.
Two of the metal-poor dwarfs, HD~22879 (\fig{fig:HRD2}) and HD~84937 (\fig{fig:HRD3}), are located at the main-sequence turn-off for their masses of $\sim$0.7$~M_\odot$, and neighbouring models have presumed ages between 12 and 24~Gyr.
HD~84937 has a metallicity similar to the globular cluster NGC~6397, for which \citet{2007ApJ...671..402K} 
estimated the effects of atomic diffusion.
They find a surface iron abundance of turn-off stars on average $0.16$~dex lower than that of red-giant-branch stars. This indicates that more metal-rich evolutionary models would be more appropriate for determining the mass of HD~84937. A slightly higher mass would be derived in this case with models of more reasonable ages.
For more details on the effects of atomic diffusion see also \citet{2011A&A...533A..59J} or \citet{2015A&A...575A..26C}.
In the case of the metal-poor subgiant HD~140283 (\fig{fig:HRD3}), which has passed the turn-off of the 0.7$~M_\odot$ track, the neighbouring models have ages between 14 and 37~Gyr.
For the metal-deficient dwarfs \muCas\ and \tauCet\ (\fig{fig:HRD2}), which are located on the main sequence near 0.6 to 0.7~$M_\odot$, the models would predict ages that are too large (more than 20~Gyr).

In the following, we give further comments on the mass determination for a few individual stars.
For HD~140283 the mass determined from the \emph{Padova} models was obtained by extrapolation in metallicity, because the lowest metallicity of the grid is 0.1~dex higher than the stellar metallicity. Both
\etaBoo\ and \betVir\ (\fig{fig:HRD1a}) lie in a post-main-sequence region of the HR diagram where the evolutionary tracks form a loop. The range of masses given by the high-$L$ and the low-$L$ parts of the loop provides the main contribution to the uncertainty in mass in this case.
For \alfCet\ (\fig{fig:HRD2}), the gravity used in Paper~III was calculated with a mass determined for solar metallicity (3~$M_\odot$), while the low metallicity determined in Paper~III (\FeH=$-0.45$~dex) results in a mass of 1.8~$M_\odot$ and a gravity that is lower by 0.2~dex.
For \betAra, only the \emph{Padova} grids were used, since the \emph{Yonsei-Yale} grids are only available up to 5~$M_\odot$ (\fig{fig:HRD1b}). This star lies in a region that is crossed several times by the model tracks. This, together with the uncertain metallicity, results in a more than 20\% uncertainty on the mass.
For \psiPhe, the \Teff\ and $L$ values lie outside the range of the grid values for metallicities lower than $-0.3$~dex (\fig{fig:HRD3}). Thus, the mass cannot be determined using the metallicity determination from Paper~III ($-1.2\pm0.4$~dex). Possible mass values range from $1.3\pm0.3$~$M_\odot$ at solar metallicity to $0.8\pm0.2$~$M_\odot$ at \FeH=$-0.3$~dex, corresponding to \logg(\ulogg)=0.6 and 0.4, respectively.

For the \Cyg\ system, the \Teff\ and $L$ values lie outside the range of the grid values. That is, the \emph{Padova} models indicate a mass lower than 0.6~$M_\odot$ for both components. In the \emph{Yonsei-Yale} grids, the \Cyg~A values lie close to the 0.6~$M_\odot$ track, and the \Cyg~B values close to 0.5~$M_\odot$, but beyond the point with the highest age of 30 Gyr (\fig{fig:HRD1b}). Thus, we resorted to using the masses derived by \citet{2008A&A...488..667K} 
for this system. Since \citet{2008A&A...488..667K} did not derive uncertainties for the masses, we adopted uncertainties of 0.05~$M_\odot$ for both components (similar to \epsEri).
\footnote{For the metallicity determination in Paper~III, we used gravities based on extrapolated masses of 0.4~$M_\odot$ for both components, which are probably too low. However, for these stars, the change in metallicity caused by a change in \logg\ is negligible (see Table~3 in Paper~III).}

\subsection{Other determinations of mass}
\label{sect:othermass}

Our sample includes both components or the primary ones of several visual binary systems.
The masses of the two components of \Cyg\ were derived by \citet{2006Ap.....49..386G} 
from dynamical modelling of astrometric observations to be 0.74 and 0.46~$M_\odot$. The observations span 40 years, that is, about 6\% of the 700-year orbit. \citet{2006Ap.....49..386G} derived a somewhat higher mass ratio (1.6) than \citet[1.1, adopted here]{2008A&A...488..667K}, while the values for the total mass of the system are similar (1.2 and 1.3~$M_\odot$, respectively).
For \alfCen, precise masses were determined by \citet{2002A&A...386..280P} 
from dynamical modelling of combined astrometric and spectroscopic observations.
These masses (A: 1.105 $\pm$ 0.007~$M_\odot$, B: 0.934 $\pm$ 0.006~$M_\odot$) are in excellent agreement with the ones derived here from the stellar model grids.
\citet{1995ApJ...450..380D} used astrometric data for \muCas\ and its 5~mag fainter M-dwarf companion GJ~53~B to derive a mass of 0.74$\pm$0.06$M_\odot$ for \muCas, which is somewhat higher than the one determined here.
The mass of Procyon was determined by \citet{2000AJ....119.2428G} and \citet{2006AJ....131.1015G} from astrometric measurements of the 40-year orbit with its white-dwarf companion. The authors obtained values of 1.50 $\pm$ 0.04~$M_\odot$ and 1.43 $\pm$ 0.03~$M_\odot$ from observations spanning 83 and 18 years, respectively.
These values agree with the one derived from the stellar model grids.

\begin{table}
   \caption{Seismic data (large frequency separation \Dnu), masses ($M$) from Eq.~\ref{equ:massA}, and their uncertainties ($u$), for a subset of \gbs.}
   \label{tab:MassA}
   \centering
   \begin{tabular}{lrrccr}
      \hline\hline
Name & $M$ [$M_\odot$] & $u(M)$ & \Dnu\ [$\mu$Hz] & $u$(\Dnu) & Ref.\\
      \hline
Procyon    & 1.40 & 0.06 & 55     & 1    & 1  \\
HD~49933   & 1.14 & 0.10 & 85.2   & 0.5  & 2  \\
\delEri    & 1.32 & 0.04 & 43.8   & 0.1  & 3  \\
\etaBoo    & 1.68 & 0.08 & 39.9   & 0.1  & 4  \\
\betHyi    & 1.06 & 0.03 & 57.24  & 0.16 & 5  \\
\alfCen~A  & 1.13 & 0.02 & 106    & 1    & 6  \\
\tauCet    & 0.78 & 0.01 & 169.3  & 0.3  & 7  \\
\alfCen~B  & 0.92 & 0.01 & 161.38 & 0.06 & 8  \\
18~Sco     & 1.02 & 0.03 & 134.4  & 0.3  & 9  \\
\muAra     & 0.91 & 0.06 & 90     & 1    & 10 \\
\betVir    & 1.41 & 0.05 & 72.07  & 0.10 & 11 \\
Arcturus   & 0.62 & 0.08 & 0.83   & 0.05 & 12 \\
\betGem    & 1.96 & 0.09 & 7.14   & 0.12 & 13 \\
\ksiHya    & 2.94 & 0.15 & 7.11   & 0.14 & 14 \\
      \hline\hline
   \end{tabular}
   \tablefoot{References for \Dnu:
1  \ldots \citet{2010ApJ...713..935B}, estimated from their Fig.~11; 
2  \ldots \citet{2010A&A...509A..77K}; 
3  \ldots \citet{2003Ap&SS.284...21B}; 
4  \ldots \citet{2005A&A...434.1085C}; 
5  \ldots \citet{2007ApJ...663.1315B}; 
6  \ldots \citet{2001A&A...374L...5B}; 
7  \ldots \citet{2009A&A...494..237T}, estimated from the mean density of 2.21$\pm$0.01~gcm$^{-3}$ determined by the authors; 
8  \ldots \citet{2005ApJ...635.1281K}, oscillations discovered by \citet{2003A&A...406L..23C}; 
9  \ldots \citet{2011A&A...526L...4B}; 
10 \ldots \citet{2005A&A...440..609B}; 
11 \ldots \citet{2005NewA...10..315C}; 
12 \ldots \citet{2003ApJ...591L.151R}; 
13 \ldots \citet{2012A&A...543A..98H}; 
14 \ldots \citet{2002A&A...394L...5F}, their Eq.~1. 
   }
\end{table}

As an alternative method for determining masses of single stars we can make use of measured pulsation frequencies (asteroseismology). For solar-like oscillations (convection-powered p-modes), the power spectrum shows a series of peaks with constant frequency separation \Dnu. These correspond to pulsation modes with the same spherical harmonic degree $l$, but different radial orders $n$. It has been shown \citep[e.g.][]{1995A&A...293...87K} that this so-called large frequency separation is proportional to the square root of the mean stellar density.
Thus, we can use \Dnu\ measurements, together with the radius determined from \Ang\ and the parallax, for a \emph{seismic} mass estimation:
\begin{equation}
   \label{equ:massA}
   \frac{\Delta\nu}{\Delta\nu_\odot} \approx \left(\frac{\bar\rho}{\bar\rho_\odot}\right)^{1/2}
   \Rightarrow \frac{M}{M_\odot} \approx \left(\frac{\Delta\nu}{\Delta\nu_\odot}\right)^2 \left(\frac{R}{R_\odot}\right)^3,
\end{equation}
where $\Delta\nu_\odot = 135.229\pm0.003~\mu$Hz \citep{2011A&A...526L...4B}.
We compiled measurements of \Dnu\ for about half of our sample from the literature, starting from the references given in \citet{2010MNRAS.405.1907B} and \citet{2010A&A...509A..77K}. All of these stars except \muAra\ have interferometric measurements of \Ang. The data, references, and seismic masses are given in Table ~\ref{tab:MassA}.
For most stars, the seismic masses agree with those determined from the stellar model grids within $\pm0.1~M_\odot$.
Larger deviations are obtained for \delEri, \muAra, and Arcturus (+0.2, $-0.3$, and $-0.4~M_\odot$, respectively).
As mentioned above, the \Teff\ and radius of \muAra\ are uncertain owing to a calibrated angular diameter.
In the case of Arcturus, the \Dnu\ value (the lowest one of the sample) is rather uncertain, since it is close to the observational frequency resolution. Also, it is not yet certain that the regularities in the observed time series of Arcturus are due to p-mode oscillations \citep{2003ApJ...591L.151R}.

For several stars, detailed investigations of the evolutionary state have been
done with the CESAM2k code \citep{1997A&AS..124..597M,2008Ap&SS.316...61M},
taking seismic data into account when available.
For \alfCen~A and B, \citet{2002A&A...392L...9T} 
constrained their models by measured values of \Teff, $L$, \FeH, and both the large and small frequency separations. The age, initial helium content, initial metallicity, mixing length parameter, and mass were varied, and the best-fit masses were 1.100$\pm$0.006 and 0.907$\pm$0.006~$M_\odot$ for the A and B components, respectively.
\citet{2004AaA...413..251K} 
explored evolutionary models for Procyon with two different masses and find that models with a mass of 1.42~$M_\odot$, with or without diffusion of heavy elements, predicted oscillation frequencies consistent with the observed large frequency separation. The model with a higher mass of 1.5~$M_\odot$ predicted a \Dnu\ value that was too high by about 3\%.
For \delEri, \etaBoo, and \ksiHya,\ \citet{2005AaA...436..253T} 
computed best-fit \Teff, $L$, and \FeH\ values by adjusting the model parameters of mass, age, and initial metallicity. The derived masses (1.215, 1.70, and 2.65~$M_\odot$, respectively) were not affected by including diffusion. The models for \delEri\ and \ksiHya\ predicted \Dnu\ values similar to the observed ones within 2--3\%, while the values predicted for \etaBoo\ were too large by about 5\%.

The CoRoT target HD~49933 was modelled by \citet{2011A&A...534L...3B}, 
who reproduced the interferometric radius and the observed large and small frequency separations by adjusting the mass, initial helium content, initial metallicity, and the core-overshoot and mixing-length parameters. They arrived at 1.20$\pm$0.08~$M_\odot$.
\citet{2015A&A...574A..45R} 
used a different stellar evolution code \citep{2008Ap&SS.316...75R} and applied a model-fitting technique to the observed frequencies of HD~49933, which uses constraints that do not depend on the structure of the stellar surface layers. 
Best-fit models with masses ranging from 1.1 to 1.3~$M_\odot$ were found.
Finally, the metal-poor subgiant HD~140283 was investigated in detail by \citet{2015A&A...575A..26C}. 
Interferometric, spectroscopic, and photometric data were used to fine-tune stellar evolution models including diffusion.
The correlated effects of varying mass, initial helium abundance, and mixing-length parameter were evaluated.
Best-fit models with masses ranging from 0.77 to 0.81~$M_\odot$ were obtained.
In most of these cases, the masses obtained by detailed modelling agree with the masses estimated from the \emph{Padova} and \emph{Yonsei-Yale} grids within the uncertainties (\tab{tab:Mass}). The masses for \delEri\ and \ksiHya\ show marginally significant discrepancies of 7--8\%. The largest difference of 13--19\% is seen for HD~140283, which is, however, expected considering the estimated uncertainty of 24\%. 

\section{Results and comparisons}
\label{sect:results}

\subsection{Fundamental effective temperature and surface gravity}
\label{sect:tefflogg}

\begin{table}
   \caption{Fundamental \Teff\ and \logg\ values and their uncertainties (absolute, $u$, and in percent, $\%u$) for \gbs. Values in square brackets are uncertain and should not be used as a reference for calibration or validation purposes (see \sect{sect:Teffloggsummary}).}
   \label{tab:Teff_logg}
   \centering
   \begin{tabular}{lp{5mm}lllp{5mm}}
      \hline\hline
Name & \Teff & $u$(\Teff)      & $\%u$(\Teff) & \logg & $u$(\logg) \\
     & \multicolumn{2}{c}{[K]} &              & \multicolumn{2}{c}{[\ulogg]} \\
\hline
\multicolumn{6}{p{0.95\hsize}}{\bf F dwarfs}\\
      Procyon &
6554 & 84 & 1.28 &
4.00 & 0.02 \\
HD 84937 &
6356 & 97 & 1.52 &
4.06 & 0.04 \\
HD 49933 &
6635 & 91 & 1.38 &
4.20 & 0.03 \\
\hline
\multicolumn{6}{p{0.95\hsize}}{\bf FGK subgiants}\\
\delEri  &
4954 & 30 & 0.61 &
3.76 & 0.02 \\
HD 140283 &
[5522] & [105] & [1.91] &
3.58 & 0.11 \\
\epsFor &
5123 & 78 & 1.53 &
[3.52] & [0.08] \\
\etaBoo &
6099 & 28 & 0.45 &
3.79 & 0.02 \\
\betHyi &
5873 & 45 & 0.77 &
3.98 & 0.02 \\
\hline
\multicolumn{6}{p{0.95\hsize}}{\bf G dwarfs}\\
\alfCen A &
5792 & 16 & 0.27 &
4.31 & 0.01 \\
HD 22879 &
5868 & 89 & 1.52 &
4.27 & 0.04 \\
Sun &
5771 & 1 & 0.01 &
4.4380 & 0.0002 \\
\muCas &
5308 & 29 & 0.54 &
[4.41] & [0.06] \\
\tauCet &
5414 & 21 & 0.39 &
[4.49] & [0.02] \\
\alfCen B &
5231 & 20 & 0.38 &
4.53 & 0.03 \\
18 Sco &
5810 & 80 & 1.38 &
4.44 & 0.03 \\
\muAra &
[5902] & [66] & [1.12] &
4.30 & 0.03 \\
\betVir &
6083 & 41 & 0.68 &
4.10 & 0.02 \\
\hline
\multicolumn{6}{p{0.95\hsize}}{\bf FGK giants}\\
Arcturus &
4286 & 35 & 0.82 &
[1.64] & [0.09] \\
HD 122563 &
4587 & 60 & 1.31 &
1.61 & 0.07 \\
\muLeo &
4474 & 60 & 1.34 &
2.51 & 0.11 \\
\betGem &
4858 & 60 & 1.23 &
2.90 & 0.08 \\
\epsVir &
4983 & 61 & 1.21 &
2.77 & 0.02 \\
\ksiHya &
5044 & 40 & 0.78 &
2.87 & 0.02 \\
HD 107328 &
4496 & 59 & 1.32 &
2.09 & 0.13 \\
HD 220009 &
[4217] & [60] & [1.43] &
[1.43] & [0.12] \\
\hline
\multicolumn{6}{p{0.95\hsize}}{\bf M giants}\\
\alfTau &
3927 & 40 & 1.01 &
1.11 & 0.19 \\
\alfCet &
3796 & 65 & 1.71 &
0.68 & 0.23 \\
\betAra &
[4197] & [50] & [1.20] &
[1.05] & [0.15] \\
\gamSge &
3807 & 49 & 1.28 &
1.05 & 0.32 \\
\psiPhe &
[3472] & [92] & [2.65] &
[0.51] & [0.18] \\
\hline
\multicolumn{6}{p{0.95\hsize}}{\bf K dwarfs}\\
\epsEri &
5076 & 30 & 0.60 &
4.61 & 0.03 \\
Gmb 1830 &
[4827] & [55] & [1.14] &
4.60 & 0.03 \\
61 Cyg A &
4374 & 22 & 0.49 &
4.63 & 0.04 \\
61 Cyg B &
4044 & 32 & 0.78 &
4.67 & 0.04 \\

      \hline\hline
   \end{tabular}
\end{table}

\begin{figure}[ht]
   \begin{center}
      \resizebox{\hsize}{!}{\includegraphics[trim=0 0 0 0,clip]{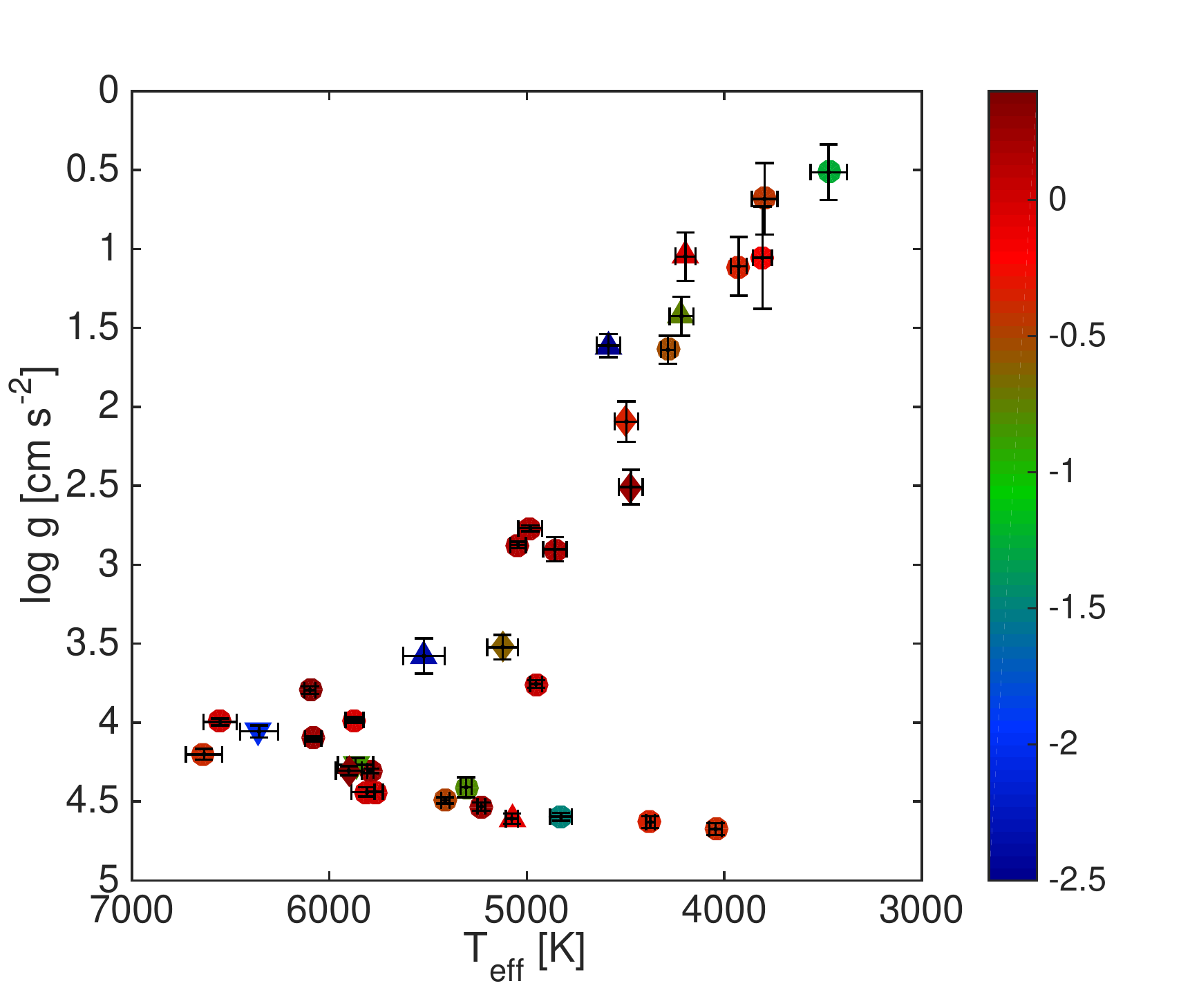}}
   \end{center}
   \caption{Fundamental \Teff\ and \logg\ values for \gbs. Colour indicates \FeH\ (NLTE) as determined in Paper~III. Circles: stars for which both \Ang\ and \Fbol\ have been measured; triangles up: stars with only \Ang\ measured; triangles down: stars with only \Fbol\ measured; diamonds: both calibrated.}
   \label{fig:TGMdirect}
\end{figure}

The input data discussed in \sects{sect:teff} and \ref{sect:logg} and listed in \tabs{tab:parameters} and \ref{tab:Mass} were used to calculate the adopted fundamental values of \Teff\ and \logg\ for the current sample of \gbs. These values and their uncertainties are given in \tab{tab:Teff_logg}.
\figu{fig:TGMdirect} shows the distribution of the stars in the \Teff--\logg\ plane, with metallicity indicated by the symbol's colour (cf. \tab{tab:general}). 
The sample covers the expected locations of FGK-type dwarfs, subgiants, and giants fairly well.
It is obvious that stars with metallicities around the solar value dominate.
However, the metal-poor stars are distributed regularly over the parameter space.
The uncertainties in \Ang\ and \Fbol\ listed in \tabs{tab:parameters} are below 5\% for all stars (except \Fbol\ for two M giants), propagating to \Teff\ uncertainties below 1\% for half of the stars and below 2\% otherwise (except for \psiPhe).
Uncertainties in \logg\ are below 0.1~dex except for the coolest giants (up to 0.3~dex).

In \fig{fig:TGMdirect}, the shape of the symbol indicates the quality of the input angular diameters and bolometric fluxes.
Twenty-two stars have both measured angular diameters and integrated bolometric flux values, which is two thirds of the current sample (disregarding the Sun, see \tab{tab:parameters}, rows without asterisks).
Five stars have measured \Ang\ values, but calibrated bolometric fluxes: the K~dwarf \epsEri, the metal-poor (sub)giants HD~140283, HD~122563, and HD~220009, and the M~giant \betAra.
Two metal-poor dwarfs have integrated bolometric fluxes, but indirect \Ang\ values (HD~22879, HD~84937).
Lastly, for four stars the angular diameter is currently not directly measured, and the bolometric flux is determined from a calibration (the metal-rich dwarf \muAra, the subgiant \epsFor, and the giants \muLeo\ and HD~107328).

\begin{figure}[ht]
   \begin{center}
      \resizebox{\hsize}{!}{\includegraphics[trim=0 0 0 0,clip]{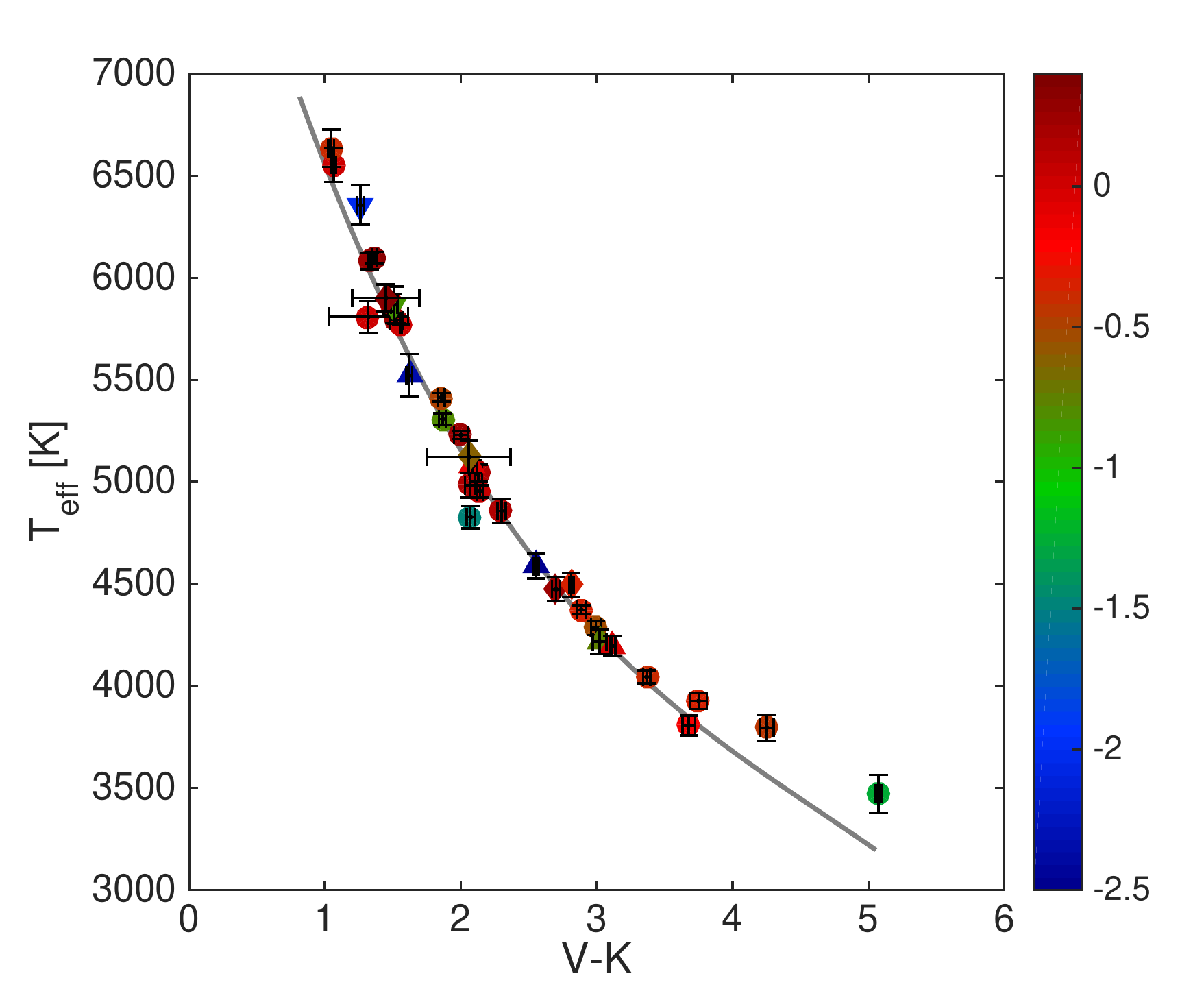}}
   \end{center}
   \caption{Fundamental \Teff\ as a function of $V-K$ colour index. Symbols and colours are as in \fig{fig:TGMdirect}. $V$ magnitudes are mean values extracted from the GCPD \citep{1997A&AS..124..349M}. 
   $K$ magnitudes were taken from the 2MASS catalogue \citep{2003yCat.2246....0C} if the entry had quality flag (Qflg) ``A''. Otherwise they are mean values taken from the Catalog of Infrared Observations \citep{CIO}, if available, and transformed to the 2MASS system \citep[][Eq.~A1]{2001AJ....121.2851C}. Stars with large error bars in $V-K$ have $K$ magnitudes from 2MASS with Qflg ``D''.
   $V-K$ for the Sun was taken from \citet{2012ApJ...761...16C}. 
   Grey line: empirical relation derived by \citet[][their Eq.~2 and Table~8, row $(V-K)^c$]{2013ApJ...771...40B}.}
   \label{fig:VKT}
\end{figure}

The colour index $V-K$  has high sensitivity to effective temperature and low sensitivity to metallicity \citep[see e.g.][]{2013ApJ...771...40B}. 
As can be seen in \fig{fig:VKT}, the stars in our sample follow a tight relation in the $V-K$ versus fundamental \Teff\ diagram.
\fig{fig:VKT} shows the empirical relation derived by \citet{2013ApJ...771...40B} based on 111 FGK dwarfs with measured angular diameters and represented by a third-order polynomial (their Eq.~2) using the coefficients given in their Table~8, row $(V-K)^c$. Excellent agreement is evident, except for the warmest and coolest stars.
We note the deviating point at $V-K\approx2$ corresponding to Gmb~1830, which is discussed in \sect{sect:Gmb1830}.

In \sects{sect:Teff_other} to \ref{sect:logg_other}, we present comparisons of the fundamental \Teff\ and \logg\ values with spectroscopic and photometric determinations, and with estimates based on parallaxes and asteroseismic data, and we discuss several cases in detail.
The impatient reader may at this point skip to \sect{sect:Teffloggsummary}, where we give a brief summary of the status and conclusions for each star, and refer to the detailed discussions, as appropriate.

\subsection{Comparison of fundamental \Teff\ to other methods}
\label{sect:Teff_other}

\begin{table*}
   \caption{Mean \Teff\ and \logg\ values from a compilation of spectroscopic, photometric, and parallax-based determinations published between 2000 and 2012.
The stars are ordered by increasing fundamental temperature.}
   \label{tab:lit_teff_logg}
   \centering
   \begin{tabular}{lrrrrrrrrrrrr}
\hline\hline
Name & \Teff & $\sigma$ & $N$ & \Teff & $\sigma$ & $N$ & \logg & $\sigma$ & $N$ & \logg & $\sigma$ & $N$ \\
 & \multicolumn{3}{l}{[K]}           & \multicolumn{3}{l}{[K]} 
 & \multicolumn{3}{l}{[\ulogg]}      & \multicolumn{3}{l}{[\ulogg]} \\
 & \multicolumn{3}{l}{spectroscopic} & \multicolumn{3}{l}{photometric} 
 & \multicolumn{3}{l}{spectroscopic} & \multicolumn{3}{l}{parallax} \\
\hline
          \alfCet & 3834 & 220 &  7 & 3724 &  87 &  2 &  1.29 &  0.43 &  7 &  0.73 &  0.30 &  1\\
       \gamSge & 4064 & 106 &  2 & 3877 &  41 &  1 &  1.52 &  0.32 &  2 &       &       &   \\
       \alfTau & 3987 & 246 & 14 & 3887 &  86 &  2 &  1.42 &  0.31 & 14 &  1.20 &  0.30 &  1\\
     61 Cyg ~B & 4400 & 100 &  1 &      &     &    &  4.20 &  0.10 &  1 &       &       &   \\
     HD 220009 & 4402 & 111 &  2 & 4484 & 100 &  1 &  1.95 &  0.34 &  2 &  2.03 &  0.10 &  1\\
      Arcturus & 4326 &  68 &  6 & 4274 &  83 &  5 &  1.70 &  0.20 &  6 &  1.82 &  0.15 &  5\\
     61 Cyg ~A & 4655 & 138 &  3 & 4300 & 170 &  2 &  4.49 &  0.17 &  3 &  4.50 &  0.20 &  1\\
        \muLeo & 4590 &  65 &  5 & 4425 & 100 &  1 &  2.65 &  0.22 &  5 &  2.24 &  0.10 &  1\\
     HD 107328 & 4514 & 100 &  1 & 4436 & 125 &  2 &  1.94 &  0.10 &  1 &  2.09 &  0.27 &  2\\
     HD 122563 & 4509 &  75 &  5 & 4795 & 107 &  1 &  1.23 &  0.30 &  4 &  1.44 &  0.24 &  2\\
      Gmb 1830 & 5087 & 100 & 11 & 5133 &  36 &  4 &  4.75 &  0.21 &  9 &  4.69 &  0.05 &  6\\
       \betGem & 4952 &  33 &  4 & 4795 &  87 &  4 &  3.10 &  0.04 &  4 &  2.76 &  0.06 &  3\\
       \delEri & 5085 &  51 &  9 & 5004 &  20 &  3 &  3.89 &  0.12 & 10 &  3.92 &  0.24 &  3\\
       \epsVir & 5115 &  29 &  3 & 5055 &  24 &  3 &  3.05 &  0.11 &  3 &  2.78 &  0.17 &  2\\
       \ksiHya & 5045 &  80 &  1 &      &     &    &  2.81 &  0.08 &  1 &       &       &   \\
       \epsEri & 5114 &  68 & 11 & 5063 &  80 &  6 &  4.51 &  0.14 & 12 &  4.61 &  0.02 &  4\\
       \epsFor & 5093 &  14 &  4 & 5135 & 155 &  4 &  4.07 &  0.30 &  5 &  3.61 &  0.10 &  5\\
     \alfCen~B & 5194 &  33 &  5 & 5131 & 110 &  4 &  4.49 &  0.07 &  5 &  4.58 &  0.06 &  2\\
        \muCas & 5341 &  92 &  7 & 5338 &  82 &  4 &  4.51 &  0.20 &  8 &  4.62 &  0.09 &  4\\
       \tauCet & 5326 &  45 & 12 & 5395 &  92 &  5 &  4.56 &  0.18 & 14 &  4.62 &  0.07 &  2\\
     HD 140283 & 5692 & 102 &  7 & 5774 &  77 &  7 &  3.54 &  0.15 &  3 &  3.69 &  0.03 &  8\\
     \alfCen~A & 5816 &  39 &  6 & 5681 & 117 &  6 &  4.30 &  0.13 &  7 &  4.32 &  0.10 &  2\\
        18 Sco & 5789 &  30 &  9 & 5759 &  72 &  5 &  4.39 &  0.07 & 10 &  4.45 &  0.05 &  4\\
      HD 22879 & 5840 &  73 &  9 & 5844 &  87 & 10 &  4.37 &  0.15 &  8 &  4.33 &  0.10 &  9\\
       \betHyi & 5875 &  87 &  3 & 5815 &  95 &  3 &  4.07 &  0.21 &  4 &  3.98 &  0.03 &  3\\
        \muAra & 5783 &  46 &  9 & 5703 &  93 &  3 &  4.29 &  0.13 &  9 &  4.26 &  0.12 &  2\\
       \betVir & 6138 &  54 &  6 & 6154 &  89 &  4 &  4.13 &  0.14 &  7 &  4.14 &  0.04 &  3\\
       \etaBoo & 6085 & 106 &  8 & 6041 & 140 &  2 &  3.94 &  0.18 &  8 &  3.77 &  0.10 &  2\\
      HD 84937 & 6340 &  41 &  7 & 6352 &  63 &  6 &  4.02 &  0.09 &  6 &  4.05 &  0.05 &  6\\
       Procyon & 6601 & 148 &  9 & 6610 &  82 &  6 &  4.16 &  0.30 &  9 &  4.01 &  0.05 &  4\\
      HD 49933 & 6661 & 110 &  5 & 6741 & 106 &  2 &  4.23 &  0.10 &  5 &  4.21 &  0.50 &  1\\

\hline\hline
   \end{tabular}
   \tablefoot{Columns headed $\sigma$ give standard deviations if number of determinations $N\ge3$, or combined linear uncertainties ($N=2$) or individual uncertainties ($N=1$) quoted in the publications.
   \Teff\ values from \citet{2009A&A...497..497G} with uncertainties larger than 200~K due to large uncertainties in 2MASS magnitudes were not included.\\
\emph{References for spectroscopic \Teff\ and \logg}:
\citet{2000A&A...356..570S},
\citet{2000A&A...364..249M},
\citet{2000AJ....120.1841F},
\citet{2000ApJ...530..783W},
\citet{2001A&A...367..253F},
\citet{2001A&A...370..951M},
\citet{2001A&A...373.1019S},
\citet{2002AJ....123.1647S},
\citet{2003A&A...398..363S},
\citet{2003AJ....125.2664L},
\citet{2003AJ....126.2015H},
\citet{2003ARep...47..422M},
\citet{2003MNRAS.341..199R},
\citet{2004A&A...415.1153S},
\citet{2004A&A...425..683B},
\citet{2004AN....325....3F},
\citet{2005A&A...437.1127S},
\citet{2005AJ....129.1063L},
\citet{2005ApJS..159..141V},
\citet{2005PASJ...57...27T},
\citet{2006A&A...448..341G},
\citet{2006A&A...458..609D},
\citet{2007A&A...468..663T},
\citet{2007A&A...475.1003H},
\citet{2008A&A...478..529M},
\citet{2008A&A...487..373S},
\citet{2008A&A...488..653P},
\citet{2008MNRAS.384..173F},
\citet{2009A&A...506..235B},
\citet{2010MNRAS.403.1368G},
\citet{2010MNRAS.405.1907B},
\citet{2011A&A...526A..71D},
\citet{2011A&A...526A..99S},
\citet{2011A&A...528A.121B},
\citet{2012A&A...541A..40M},
\citet{2012A&A...542A..84D},
\citet{2012A&A...542A.104B},
\citet{2012A&A...543A.160T},
\citet{2012A&A...547A.108L},
\citet{2012MNRAS.422..542P}.\\
\emph{References for photometric \Teff\ and parallax-based \logg}:
\citet{2000A&A...353..722N},
\citet{2000A&A...354..169G},
\citet{2000A&A...362.1077Z},
\citet{2002A&A...390..235N},
\citet{2002AJ....124.2224Z},
\citet{2004A&A...420..183A},
\citet{2004ARep...48..492G},
\citet{2005A&A...440..321J},
\citet{2005MNRAS.364..712Z},
\citet{2006A&A...449..127Z},
\citet{2006A&A...454..833K},
\citet{2006ApJ...652.1604B},
\citet{2006MNRAS.367.1329R},
\citet{2007A&A...465..271R},
\citet{2008A&A...484L..21M},
\citet{2009A&A...508L..17R},
\citet{2010A&A...511L..10N},
\citet{2011A&A...530A.138C}.\\
\emph{References for spectroscopic \Teff\ and parallax-based \logg}:
\citet{2003A&A...404..187G},
\citet{2003A&A...407..691K},
\citet{2003A&A...410..527B},
\citet{2006A&A...451.1065G},
\citet{2012ApJ...756...36R},
\citet{2012ApJS..203...27R}.\\
\emph{References for photometric \Teff\ and spectroscopic \logg}:
\citet{2006MNRAS.370..163B},
\citet{2012ApJ...753...64I}.\\
\emph{References for photometric \Teff\ only}:
\citet{2005ApJ...626..446R},
\citet{2006A&A...450..735M},
\citet{2009A&A...497..497G},
\citet{2010A&A...512A..54C},
\citet{2010A&A...523A..71G}.\\
\emph{References for spectroscopic \logg\ only}:
\citet{2004A&A...418..551M},
\citet{2004A&A...425..187T},
\citet{2004A&A...427..933K}.\\
\emph{References for parallax-based \logg\ only}:
\citet{2008A&A...489..923M},
\citet{2008A&A...492..823B},
\citet{2011ApJ...743..135R}.\\
\emph{References containing values in three or four categories}:
\citet{2005A&A...433..647A},
\citet{2007A&A...468..679S},
\citet{2007AJ....133.2464L}.
   }
\end{table*}

\begin{figure*}[ht]
   \begin{center}
      \resizebox{0.8\hsize}{!}{\includegraphics[]{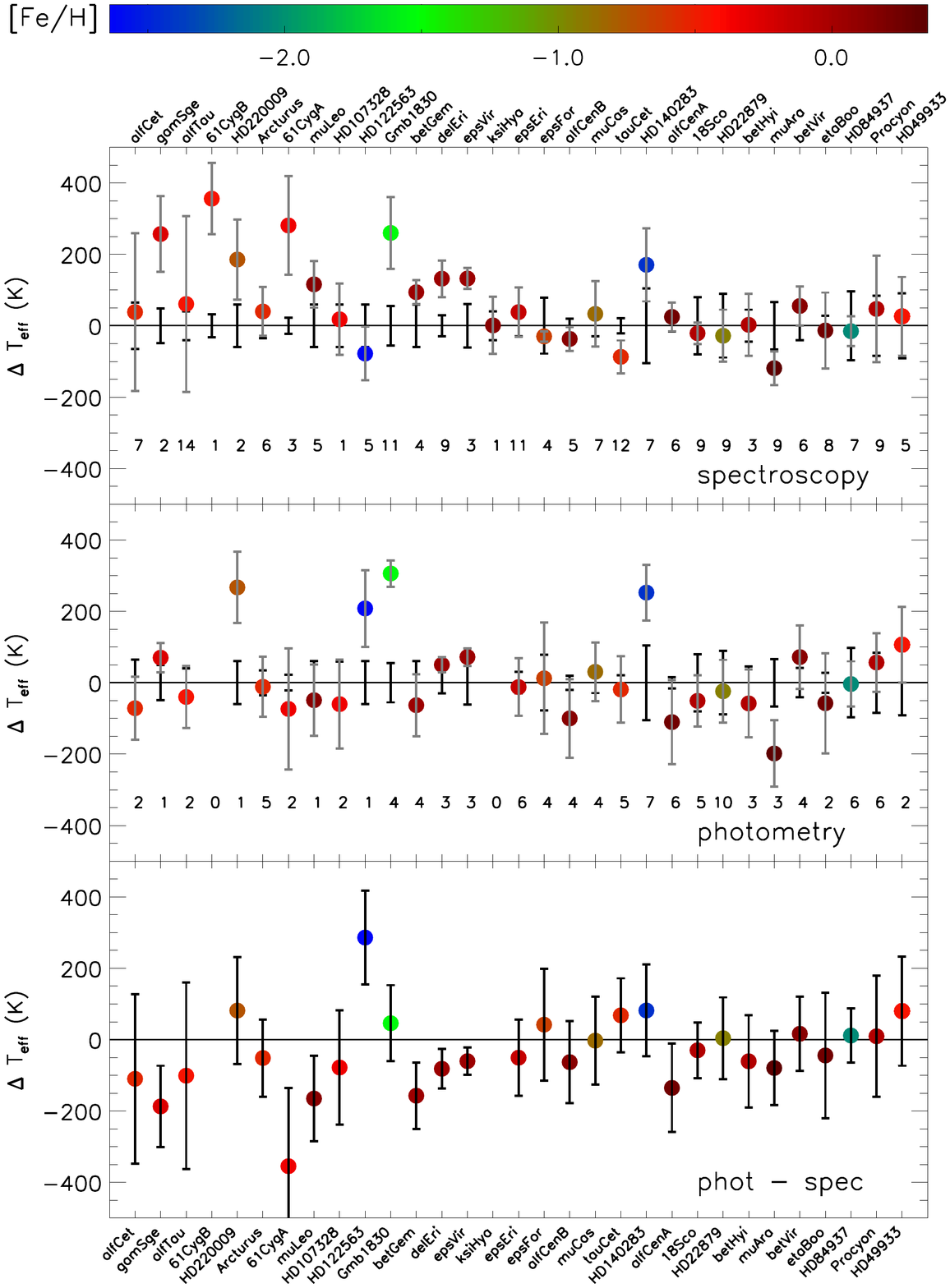}}
   \end{center}
   \caption{Comparison of the fundamental temperature values with mean values of spectroscopic and photometric determinations compiled from the literature. The stars are ordered by increasing fundamental temperature from left to right. The two upper panels display the difference in the literature \Teff\ from the fundamental \Teff. Black error bars centred on zero represent the uncertainty in fundamental \Teff\ for each star, while grey error bars represent the standard deviation of the mean for $N\ge3$, and combined and individual uncertainties in the case of two and one determinations, respectively. The number of determinations for each star are indicated in the two panels. The bottom panel compares the photometric and spectroscopic temperature measurements with error bars representing combined standard deviations. Symbol colour indicates metallicity from \tab{tab:general}. See text and \tab{tab:lit_teff_logg} for data, references, and discussion.}
   \label{fig:teff_comparison}
\end{figure*}

The sample of \gbs\ was selected to include bright and well-known stars. Thus, many studies reporting temperatures can be found in the literature. In addition to our fundamental method the two main approaches to determine effective temperatures are through spectroscopic analysis or relations with photometric colour indices, the latter mostly based on the infrared flux method (IRFM). Spectroscopic temperature determinations are usually based on the requirement of excitation equilibrium of neutral iron lines or on fitting the profiles of Balmer lines.
We queried the PASTEL catalogue\footnote{\url{http://vizier.u-strasbg.fr/viz-bin/Cat?B/pastel}, Version 17-May-2013} \citep{2010A&A...515A.111S} for temperatures of \gbs\ published between 2000 and 2012. We supplemented the results with some additional data and classified the \Teff\ determinations by method. Duplicate values and those outside the two categories were removed. 
We compiled 191 \Teff\ determinations using spectroscopic methods and 108 values using photometric calibrations.
Ten or more \gbs\ were analysed spectroscopically by
\citet{2005ApJS..159..141V}, 
\citet{2005AJ....129.1063L}, 
and \citet{2010MNRAS.405.1907B}, 
while photometric temperature determinations have been published for more than ten stars by
\citet{2004A&A...420..183A}, 
\citet{2005ApJ...626..446R}, 
\citet{2007A&A...465..271R}, 
and \citet{2011A&A...530A.138C}. 

In \tab{tab:lit_teff_logg} we list the mean and standard deviation or quoted uncertainties of the compiled \Teff\ values, as well as the number of determinations used for each star and method. References are given in the table notes.
These values are compared in \fig{fig:teff_comparison}, where we plot the difference between the spectroscopic and the fundamental \Teff\ values (cf. \tab{tab:Teff_logg}) in the top panel, and the difference between the photometric and fundamental values in the middle panel.
The uncertainties in the fundamental values and the $\sigma$ values from \tab{tab:lit_teff_logg} are represented as separate error bars.
The bottom panel shows the difference between the mean spectroscopic and photometric values with combined uncertainties.
The stars are ordered by increasing fundamental temperature from left to right and the colour of the symbol indicates the metallicity.
Two stars, \psiPhe\ and \betAra, are not included in this comparison. The reason is that \psiPhe\ has no entry in PASTEL, while for \betAra\ there is only one entry: \citet{1979ApJ...232..797L} 
quotes a temperature about 400~K higher than the fundamental one.

There is generally good agreement between the \Teff\ values obtained from the different approaches, when taking the uncertainties into account.
For stars with \Teff\ $<5000$~K, the mean spectroscopic \Teff\ values are almost always greater than the fundamental values, although the difference is not significant in most cases.
Also, the error bars for spectroscopic methods become larger towards cooler stars.
At the hot end, the spectroscopic \Teff\ values of Procyon and HD~49933 show somewhat larger than typical scatter.
For the FG dwarfs 18~Sco, HD~22879, and HD~84937, all three values agree very well, and the standard deviations of the spectroscopic and photometric values (each based on five to ten determinations) are smaller than the uncertainties in the fundamental values.
Also for Arcturus, the most widely used giant among the \gbs, the three \Teff\ values agree well. Furthermore, \citet{2011ApJ...743..135R} 
used an additional method to derive the effective temperature of Arcturus from a fit of model SEDs scaled by the square of the angular diameter (21.06$\pm$0.17~mas) to the observed SED.
They obtained \Teff=4286$\pm$30~K, which is in excellent agreement with the value given in \tab{tab:Teff_logg}.
In the following, we discuss those stars with discrepant values (non-overlapping error bars).
For the spectroscopic methods, all of these have fundamental \Teff\ lower than 5000~K.
An extensive discussion of the parameters of \alfTau\ and \alfCet\ can be found in \citet{2012A&A...547A.108L}.

\subsubsection{The M giant \gamSge}
\label{sect:gamSgeTeff}
%
%
%
For \gamSge, \citet{2006A&A...458..609D} 
and \citet{2007A&A...475.1003H} 
obtained a higher spectroscopic \Teff\ than the fundamental one: by 170~K or 2$\sigma$\footnote{The symbol $\sigma$ refers to the combined uncertainty (the uncertainties added in quadrature).}, and by 340~K or more than 3$\sigma$, respectively.
Both used excitation equilibrium of Fe~I lines as a constraint for \Teff, based on equivalent widths determined by two different procedures from spectra with similar quality ($R=\lambda/\Delta\lambda=50\,000-60\,000$), and different line lists, atmospheric models, and radiative transfer codes.
Using different selections of Fe lines and atomic data can have a large impact on the result of a spectroscopic analysis, in particular for cool stars \citep[e.g.][]{2012A&A...547A.108L}. This and other differences in analysis could explain the difference between the two spectroscopic values.
Furthermore, \citet{2007A&A...475.1003H} compared in their Fig.~2 their \Teff\ values for 250 stars from their sample with literature determinations, and found differences of up to 300~K at temperatures around 4000~K.
On the other hand, the photometric \Teff\ by \citet{2005ApJ...626..446R} agrees with the fundamental value.
We conclude that spectroscopic \Teff\ determinations for \gamSge\ are probably overestimated.

\subsubsection{The \Cyg\ system}
\label{sect:61CygTeff}
%
%
%
For \Cyg~A, the spectroscopic \Teff\ determined in three publications is on average 280~K (2$\sigma$) higher than the fundamental value (4370~K),
which is based on direct measurements of both angular diameter and bolometric flux and has a small formal uncertainty. 
The three individual values determined by \citet{2003AJ....126.2015H}, \citet{2005AJ....129.1063L}, and \citet{2005A&A...433..647A} are 4800~K, 4640~K, and 4525~K, respectively.
All studies are based on an equivalent width analysis, and in all cases, the effective temperature was determined by forcing the excitation equilibrium of Fe~I lines (about 240 lines in the first two studies and 26 in the third one). \citet{2005A&A...433..647A} did not measure any Fe~II lines and could not simultaneously solve for ionization balance, as was done in the other two studies (based on 3--4 Fe~II lines).

Equivalent width analyses of stars with such low temperatures are questionable, because most lines will be blended, not the least by molecular bands.
\citet{2003AJ....126.2015H} compare their temperatures with previous studies in their Fig.~10, which shows that their temperatures might be overestimated by up to 350~K for stars around 5000~K.
Also the comparison with \citet{2004AN....325....3F} in Fig.~9 of \citet{2005AJ....129.1063L} shows a systematic offset of 80~K towards higher \Teff\ values over the whole \Teff\ range. 
On the other hand, the \Teff\ value published by \citet{2008A&A...489..923M} 
using a different method (\Teff=4236$\pm$11~K) is lower than the fundamental value.
It is based on a calibration of line-depth ratios with a very low formal uncertainty. The absolute scale is given by the \Teff\ values of their calibrator stars, which are determined by the IRFM \citep{2004A&A...427..933K}. 
The better agreement of this latter method with the fundamental value is in line with the good agreement of the photometric \Teff\ determination by \citet{2011A&A...530A.138C}.

%
%
For the secondary component in the \Cyg\ system, the spectroscopic \Teff\ included in \tab{tab:lit_teff_logg} \citep{2005AJ....129.1063L} is 360~K (3$\sigma$) higher than the fundamental value. An additional spectroscopic determination of 4120$\pm$100~K was published by \citet{1999ApJ...523..234T}, which agrees with the fundamental value of 4040$\pm$30~K.
No photometric \Teff\ value published after 2000 is available, but two earlier determinations by \citet{1996AaAS..117..227A} and \citet{1998A&A...339..858D} resulted in 3800~K and 4000~K, respectively, close to the fundamental value.
Furthermore, the line-depth ratio method gives \Teff=3808$\pm$26~K for \Cyg~B \citep{2003A&A...411..559K}.
This is the coolest dwarf in our sample, and the same caveats with regard to equivalent width analyses as for \Cyg~A hold.

\subsubsection{\delEri\ and \epsVir}
\label{sect:delEriepsVirTeff}
%
%
%
%
For \delEri\ and \epsVir, solar-metallicity (sub)giants with \Teff\ about 5000~K, the means of the spectroscopic \Teff\ determinations are higher than the fundamental values by 130~K, or about 2$\sigma$. In both cases, the photometric \Teff\ values show better agreement.
%
The star \delEri\ is the coolest subgiant in our sample\footnote{For another subgiant, \epsFor, with a slightly higher \Teff\ and almost five times lower metallicity than \delEri, we do not see any discrepancy between the different \Teff\ determinations. However, in this case, the input data for the fundamental value are based on calibrations.}.
Three of the nine spectroscopic \Teff\ values are within 1--2\% of the fundamental value, and they agree with the latter within the uncertainties.
These studies used three different methods. The first two were based on excitation equilibrium for 147 and 116 Fe~I lines with abundances derived from equivalent-width modelling and line-profile fitting (\citealt{2003A&A...410..527B} and \citealt{2010MNRAS.405.1907B}, respectively). The third study used model fits for the wings of the Balmer lines \citep{2008MNRAS.384..173F}.
The discrepant studies are based on either excitation equilibrium for about 40 Fe~I lines \citep{2001A&A...373.1019S,2004A&A...415.1153S,2005A&A...433..647A}, excitation equilibrium for about 400 or 250 Fe~I lines \citep{2005AJ....129.1063L,2008A&A...487..373S}, or on spectrum synthesis for a mixture of weak and strong lines \citep{2005ApJS..159..141V}.
In conclusion, for stars like \delEri,\ it seems that the preferred method is Balmer-line fitting for spectroscopic \Teff\ determination. If the excitation equilibrium method is chosen, the Fe lines and atomic data should be carefully selected, possibly guided by the two studies above, which reproduced the fundamental \Teff\ value of \delEri. 

In the case of \epsVir,\ there are three discrepant studies, two of which are based on excitation equilibrium for either a few Fe~I lines (20 lines by \citealt{2007A&A...475.1003H}) or numerous lines ($>$350 by \citealt{2007AJ....133.2464L})\footnote{The third study \citep{2006A&A...458..609D} with \Teff\ closest to the fundamental value used the same method with an unspecified number of Fe~I lines.}.
Our sample includes two stars with similar parameters to \epsVir.
For \betGem\ (included in the same three studies), we see a similar tendency, but with lower significance.
For \ksiHya, there is no discrepancy with the single spectroscopic \Teff\ value by \citet[][line profile fitting for 99 Fe~I lines]{2010MNRAS.405.1907B}.
\citet{1980A&A....92...70R} analysed high-resolution spectra of \betGem\ and derived systematically lower abundances for low-excitation than for high-excitation Fe~I lines (with differences up to 0.4~dex). On the basis of kinetic-equilibrium studies available at the time, they argued that this abundance separation could be caused by a combination of two non-LTE effects (non-thermal ionization and excitation).
Recent studies of non-LTE line formation of Fe \citep{2012MNRAS.427...27B,2012MNRAS.427...50L} 
indeed predict differential non-LTE corrections for stars similar to \betGem\ and \epsVir\ in the sense described above; however, the magnitude of the effect is much smaller.
Interpolation in these calculations, available at the \emph{INSPECT database}\footnote{\url{http://www.inspect-stars.com/}}, at \Teff=5000~K, \logg=2.8, and \FeH=0.1 results in mean non-LTE corrections of 0.01~dex and 0.04~dex for lines with lower level energies above and below 3.5~eV, respectively.

\subsubsection{Metal-poor stars with discrepant photometric \Teff}
\label{sect:Teffphot}
%
%
For HD~220009 the photometric \Teff\ is higher (4500~K) than the fundamental one (4200~K), 
while the two spectroscopic \Teff\ values (4400~K) almost agree with the fundamental one within the uncertainties.
The photometric \Teff\ by \citet{2007AJ....133.2464L} 
was derived from theoretical calibrations of a variety of narrow- and broad-band colour indices.
It might be affected by systematic errors, because earlier photometric \Teff\ determinations based on the IRFM resulted in lower values around 4200~K \citep{1998A&A...339..858D,1999A&AS..139..335A}. 
The fundamental value is based on a preliminary direct angular diameter and a calibrated bolometric flux. 

%
%
For the metal-poor giant HD~122563, a similar situation to that of HD~220009 is observed. Here, the photometric \Teff\ is from \citet{2009A&A...497..497G} 
and is based on the IRFM and 2MASS colour indices.
In this case, the 2MASS data have poor quality, and several other photometric \Teff\ determinations giving lower values around 4600~K can be found (e.g.
\citealt{1996ApJ...471..254R}, 
from theoretical $B-V$ and $R-I$ calibrations,
\citealt{1999A&AS..139..335A}, 
from the IRFM and colour indices in the TCS\footnote{Telescopio Carlos S\'anchez} system,
and \citealt{2014MNRAS.439.2060C} 
from the IRFM using Johnson $JHK$ photometry).
The fundamental value is based on a direct angular diameter but a calibrated bolometric flux. 

%
%
For the metal-poor subgiant HD~140283, the photometric \Teff\ value determined by seven authors is 250~K (2$\sigma$) higher than the fundamental one (5500~K).
Three publications using the IRFM calibration by \citet{1996A&A...313..873A} report \Teff\ values around 5700~K.
Three other publications that apply more recent, 2MASS-based IRFM implementations \citep{2009A&A...497..497G,2010A&A...512A..54C} arrived at \Teff\ values around 5800~K.
The highest value of $\sim$5900~K is given by \citet{2006A&A...450..735M}, who used a spectral energy distribution fit method with $V$ and 2MASS photometry. Discrepant \Teff\ values at low metallicities for the latter method were noticed by \citet{2010A&A...512A..54C}.
The spectroscopic \Teff\ value is higher than the fundamental one by 170~K, which is not significant given the uncertainties. The published spectroscopic determinations span a wide range of values, even when using the same diagnostics (e.g. 5560~K and 5810~K by \citealt{2003A&A...404..187G} and \citealt{2003A&A...407..691K}, respectively, from Balmer line fitting).
\citet{2010A&A...512A..54C} observe that for metal-poor stars, their \Teff\ scale is 100--200~K hotter than the spectroscopic one. They state that the IRFM is less model dependent than the spectroscopic \Teff\ determinations, but can be affected by reddening.

This star has the smallest measured angular diameter in our sample with an associated uncertainty larger than for all other stars, but still below 4\%. The interferometric observations were obtained at optical wavelengths, and they sample the visibility curve well.
The bolometric flux determination is somewhat problematic due to the possible effect of an unknown amount of interstellar extinction (see discussion in \citealt{2015A&A...575A..26C}). We used the calibrated value from \citet{1996AaAS..117..227A}, 
who estimated zero reddening for HD~140283 using two different photometric methods. 
Also, according to the maps of the local ISM by \citet{2014A&A...561A..91L}, 
HD~140283 should be located within a local cavity with negligible reddening.
Furthermore, we did not detect any interstellar absorption in the high-resolution spectra of this star  (Paper~II), 
for neither the Na~D lines nor the diffuse interstellar bands.
The bolometric flux adopted here ($38.6\pm0.8\cdot10^{-9}$mWm$^{-2}$) is similar to the flux adopted by \citet{2015A&A...575A..26C} for zero extinction ($38.9\pm6.6\cdot10^{-9}$mWm$^{-2}$), resulting in \Teff=5500$\pm$100~K. For the extreme case with an extinction $A_V=0.1$~mag, \citet{2015A&A...575A..26C} derived \Fbol=$42.2\pm6.7\cdot10^{-9}$mWm$^{-2}$ and \Teff=5600$\pm$100~K.

\subsubsection{The G dwarfs \tauCet\ and \muAra}
\label{sect:tauCetmuAraTeff}
%
%
For \tauCet\ (\FeH=$-0.5$), the mean of twelve spectroscopic \Teff\ determinations is about 90~K (1.7$\sigma$) lower than the fundamental one (5410~K), which is based on direct measurements of angular diameter and bolometric flux.
However, this includes an individual value of 5420$\pm$25~K by \citet{2005PASJ...57...27T} 
with very good agreement.
In fact, half of the spectroscopic measurements are consistent with the fundamental value within the uncertainties.
The mean of the five photometric \Teff\ determinations agrees with the fundamental one, but includes individual values that are 100~K higher and lower.
We cannot identify any systematic uncertainties affecting the fundamental determination.

%
%
For the metal-rich G dwarf \muAra, both the mean spectroscopic and the mean photometric \Teff\ are lower than the fundamental one (5900~K) by 120~K (1.5$\sigma$) and 200~K (1.7$\sigma$), respectively.
Eight of the nine spectroscopic determinations lie within the interval 5780 to 5820~K. Two of the photometric determinations (5700 and 5800~K) are by the same first author \citep{2005ApJ...626..446R,2007A&A...465..271R}.
The third one, 5600~K by \citet{2006MNRAS.370..163B}, 
is based on a very rough relation between $B-V$ colour index and temperature 
derived from the assumption that a star radiates as a black body and used the absolute magnitudes of the Sun for calibration.
However, the fundamental value is based on indirect determinations of both angular diameter and bolometric flux. 
Thus, we cannot currently conclude about the reliability of the non-fundamental values. 

\subsubsection{The metal-poor K dwarf Gmb~1830}
\label{sect:Gmb1830}
%
%
The metal-poor K dwarf Gmb~1830 represents an extreme case where the spectroscopic and photometric temperatures agree well (5100~K), but both are significantly higher than the fundamental one (4800~K).
%
%
%
%
%
The 11 spectroscopic \Teff\ determinations can be divided into two groups. Five authors fit one or more synthetic Balmer line profiles to the observed spectra \citep{2000A&A...364..249M, 2000A&A...362.1077Z, 2001A&A...370..951M, 2003A&A...407..691K, 2006A&A...451.1065G}, and five others performed an equivalent width analysis of Fe~I lines, with excitation equilibrium as a constraint for \Teff\ \citep{2000AJ....120.1841F, 2002AJ....123.1647S, 2003AJ....126.2015H, 2005AJ....129.1063L, 2005PASJ...57...27T}.
The analyses of the first group are based on similar spectra and theoretical models, while those of the second group differ in spectroscopic data, atmospheric models, and line data. Within the second group, the studies by \citet{2000AJ....120.1841F} and \citet{2002AJ....123.1647S} are the most alike regarding data and models; however, their results differ by 150~K.
An additional method is applied by \citet{2005ApJS..159..141V}, who performed a global fit of synthetic spectra to observed spectra in several wavelength intervals containing weak lines of various elements and the strong Mg~Ib triplet lines.
All but two of these works obtain \Teff\ values in the interval of about 5000~K to 5100~K, and no systematic difference between methods can be discerned.
%
%
%
The photometric determinations include three different approaches. 
\citet{2000A&A...354..169G} used theoretical calibrations for $B-V$ and $b-y$ colour indices, while \citet{2006A&A...454..833K} applied an empirical calibration of $R-I$ colour indices based on spectroscopic temperatures for GK dwarfs. The determinations by \citet{2009A&A...497..497G} and \citet{2011A&A...530A.138C} are both based on the IRFM with 2MASS colour indices.
Again, all values lie in a narrow interval of 5080~K to 5170~K, which is slightly higher than the spectroscopic values.

According to the discussion above, it seems that determinations using a wide variety of stellar atmosphere modelling and input data result in a consistent range of \Teff\ values.
%
%
The quoted spectroscopic analyses could be affected by systematic uncertainties owing to unrealistic atmospheric models and model spectra, assuming hydrostatic and local thermodynamic equilibrium.
However, \citet{2013MNRAS.429..126R} 
show, for somewhat warmer dwarfs and subgiants (\Teff$>$5500~K) at a metallicity of $-1.5$~dex, that including non-LTE corrections for Fe lines in the analysis results in about 100~K higher \Teff\ values than an LTE analysis. This points in a direction away from the fundamental value.
Furthermore, effective temperatures derived from Balmer-line fitting by the same authors for two dwarf stars bracketing the metallicity of Gmb~1830 agree well with the fundamental values (HD~22879 -- 5800$\pm$100~K, HD~84937 -- 6315$\pm$100~K).
See also the discussion in Paper~III, where the line-by-line abundance determinations for Fe lines using the fundamental \Teff\ and \logg\ values resulted in a considerable deviation from excitation and ionization equilibrium.
It is therefore necessary to question the correctness of the fundamental value.


\begin{figure}[t]
   \begin{center}
      \resizebox{\hsize}{!}{\includegraphics{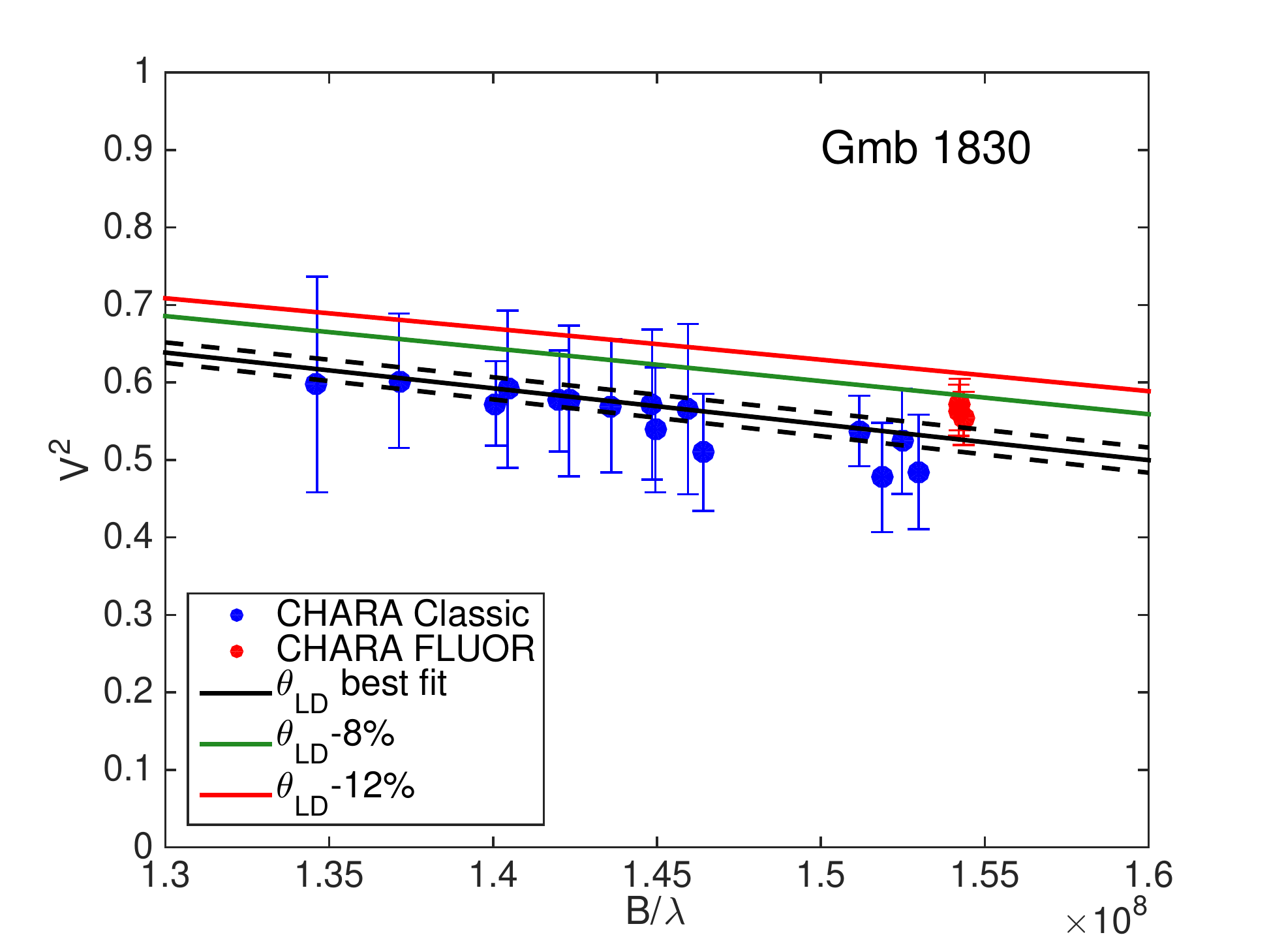}}
   \end{center}
   \caption{Squared visibility measurements for Gmb~1830 from two instruments at the CHARA array \citep[their Table~2]{2012A&A...545A..17C} compared to model visibility curves including limb darkening and assuming the best-fit angular diameter (black solid line) and its formal uncertainty (black dashed lines), as well as angular diameters that were decreased by 8 and 12\% (green and red solid lines, respectively).}
   \label{fig:ang_Gmb1830}
\end{figure}

The fundamental \Teff=4830$\pm$60~K is based on direct measurements of both angular diameter (with a 2\% uncertainty) and bolometric flux, and for each quantity there is an additional direct measurement that agrees within 2.5 and 1\%, respectively (see \fig{fig:ang_fbol_comparison}).
However, the angular diameter measurements are mostly based on the same interferometric data.
An increase in \Teff\ of 200~K (4\%) would require a decrease in \Ang\ by 8\%, while an increase of 300~K (6\%) would require a decrease in \Ang\ by 12\%.
Considering that the data for Gmb~1830 span only 6\% of the first lobe of the visibility curve, it is conceivable that they could represent an angular diameter smaller than accounted for by the formal uncertainties. 
\figu{fig:ang_Gmb1830} shows the data from \citet[their Table~2]{2012A&A...545A..17C} 
compared to model visibility curves including limb darkening and assuming the best-fit or smaller angular diameters.
A decrease of 12\% is rather unlikely, as the model curve would only pass through one-third of the data points including error bars.
For a decrease of 8\%, the model curve would pass through two-thirds of the measurements.
Furthermore, fitting the data from the two instruments separately results in angular diameters differing by about 7\%.
It is possible that the data are affected by additional uncertainties not included in the formal error, because of either the reduction procedure or the calibrator stars used.
Further interferometric observations at longer baselines and/or shorter wavelengths are clearly needed to resolve or confirm the \Teff\ discrepancy for Gmb~1830.

\subsection{Estimation of surface gravity from seismic data}
\label{sect:logg_seismic}

\begin{table}
   \caption{Seismic data (frequency of maximum power \numax), \logg\ estimates from Eq.~\ref{equ:loggA}, and their uncertainties ($u$), for a subset of \gbs.}
   \label{tab:loggA}
   \centering
   \begin{tabular}{lrrrrr}
      \hline\hline
Name & \logg & $u$(\logg) & \numax & $u$(\numax) & Ref.\\
& \multicolumn{2}{c}{[\ulogg]} & \multicolumn{2}{c}{[$\mu$Hz]} & \\
      \hline
Procyon    & 3.92  & 0.24 &  900  & 500 & 1  \\
HD~49933   & 4.19  & 0.01 &  1657 & 28  & 2  \\
\delEri    & 3.75  & 0.02 &  700  & 35  & 3  \\
\etaBoo    & 3.83  & 0.02 &  750  & 38  & 4  \\
\betHyi    & 3.94  & 0.02 &  1000 & 50  & 5  \\
\alfCen~A  & 4.32  & 0.02 &  2400 & 120 & 5  \\
\tauCet    & 4.58  & 0.02 &  4490 & 225 & 6  \\
\alfCen~B  & 4.53  & 0.02 &  4100 & 205 & 5  \\
18~Sco     & 4.44  & 0.02 &  3170 & 159 & 7  \\
\muAra     & 4.24  & 0.02 &  2000 & 100 & 8  \\
\betVir    & 4.10  & 0.02 &  1400 & 70  & 9  \\
Arcturus   & 1.44  & 0.01 &  3.7  & 0.1 & 10 \\
\betGem    & 2.84  & 0.02 & 87.0  & 4.4 & 11 \\
\ksiHya    & 2.87  & 0.01 & 92.3  & 3.0 & 12 \\
      \hline\hline
   \end{tabular}
   \tablefoot{References for \numax:
1  \ldots \citet{2008ApJ...687.1180A}, Fig.~10, and \citet{2010ApJ...713..935B}, Fig.~8; 
2  \ldots \citet{2009A&A...506.1043G}; 
3  \ldots \citet{2003Ap&SS.284...21B}; 
4  \ldots \citet{2005A&A...434.1085C}, Figs.~2, 4, 5; 
5  \ldots \citet{2008ApJ...682.1370K}; 
6  \ldots \citet{2009A&A...494..237T}; 
7  \ldots \citet{2012MNRAS.419L..34M}; 
8  \ldots \citet{2005A&A...440..609B}, Fig.~7; 
9  \ldots \citet{2005NewA...10..315C}; 
10 \ldots \citet{2003ApJ...591L.151R}, Fig.~4; 
11 \ldots \citet{2012A&A...543A..98H}; 
12 \ldots \citet{2010A&A...509A..77K}. 
   }
\end{table}

Asteroseismic data provide an alternative way to estimate the surface gravity for stars with measurable p-mode oscillations. Here, we use a second global parameter derived from the power spectrum of oscillation frequencies (in addition to the large regular separation of frequencies \Dnu, see \sect{sect:othermass}), namely the frequency where the power spectrum exhibits a global maximum, \numax. It has been shown \citep[e.g.][]{2011A&A...530A.142B} 
that this parameter is proportional to the surface gravity and inversely proportional to the square root of the effective temperature.
Thus, \numax\ measurements, together with the fundamental \Teff\ value, can be used for a \emph{seismic} \logg\ estimation:
\begin{equation}
   \label{equ:loggA}
   \log g \approx \log\nu_{\rm max} + 0.5\log T_{\rm eff} - \log\nu_{\rm max, \odot}, - 0.5\log T_{\rm eff, \odot} + \log g_\odot,
\end{equation}
where $\nu_{\rm max, \odot} = 3160\pm40~\mu$Hz \citep[weighted mean of all data shown in their Fig.~3]{2013JPhCS.440a2031B}. 
Following \citet{2012MNRAS.419L..34M}, 
we compiled measurements of \numax\ for about half of our sample from the literature. (The stars are the same as those listed in \tab{tab:MassA}.) The data, references, and seismic \logg\ values are given in \tab{tab:loggA}.
The precision of the \numax\ measurement varies from star to star, because the width of the peak in the power spectrum can be more or less narrow (see e.g. Fig.~11 in \citealt{2008ApJ...687.1180A} for a comparison of several stars).
However, an uncertainty of 5\% can be assumed in most cases \citep{2012MNRAS.419L..34M}. 

For most stars, the seismic gravity value agrees with the one determined from the stellar mass and radius within $\pm0.06$~dex or 15\%.
Larger deviations are obtained for Procyon, \tauCet, and Arcturus ($-0.08$, +0.08, and $-0.2$~dex, respectively).
For Procyon, the \numax\ determination has a large uncertainty (much larger than the deviation), because the peak in the power spectrum is very broad (referred to as a ``plateau'' in \citealt{2008ApJ...687.1180A}).  
For \tauCet, our \logg\ determination might be affected by an uncertain evolutionary mass (not consistent with a reasonable age), although it agrees with the seismic mass from \Dnu\ within 10\%.
In the case of Arcturus, we have already found a deviation in its evolutionary mass from the seismic mass (\sect{sect:othermass}). The nature of the variability of Arcturus seems to be unclear, and thus the scaling relations valid for solar-like oscillations might not be applicable.
However, \citet{2007MNRAS.382L..48T} 
analysed seismic data for Arcturus obtained on a longer time scale than the data of \citet{2003ApJ...591L.151R} and detected a peak in the power spectrum at $\sim$3.5~$\mu$Hz, which they ascribe to p-mode oscillations.

It should be noted that for seven dwarfs and subgiants in our sample \citet[their Sect.~3]{2013MNRAS.431.2419C} 
derived seismic \logg\ values using a grid-based method with \Dnu, \numax, \Teff, and \FeH\ as constraints.
Their seismic \logg\ values agree with our fundamental \logg\ values within 0.02~dex for all stars (Procyon, HD~49933, Sun, 18~Sco, \alfCen~A, \alfCen~B, \betHyi).

\subsection{Comparison of fundamental \logg\ to other determinations}
\label{sect:logg_other}

\begin{figure*}[ht]
   \begin{center}
      \resizebox{0.8\hsize}{!}{\includegraphics[]{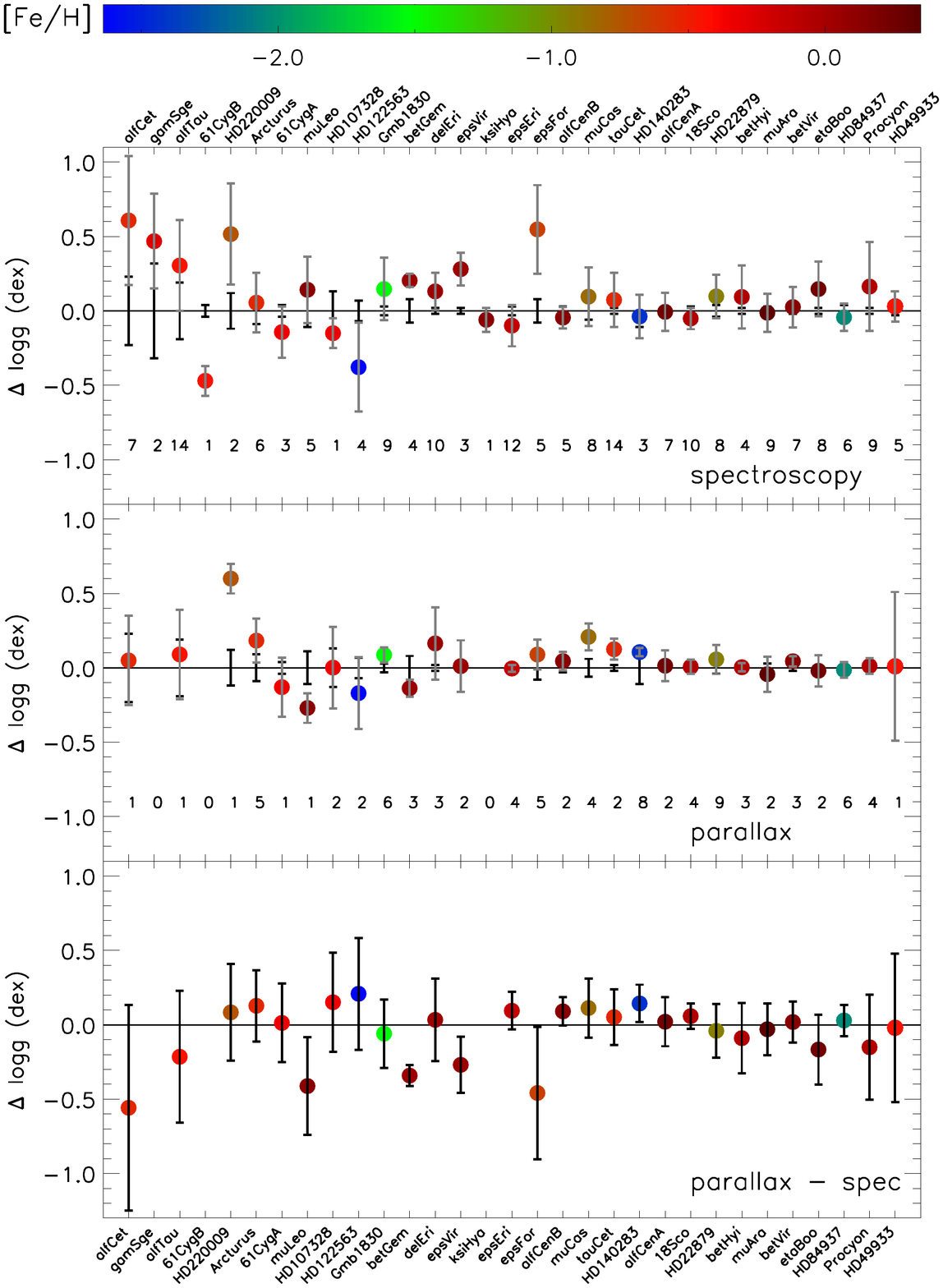}}
   \end{center}
   \caption{Comparison of the fundamental surface gravity values with mean values of spectroscopic and parallax-based determinations compiled from the literature. The stars are ordered by increasing fundamental temperature from left to right. The two upper panels display the difference in the literature \logg\ from the fundamental \logg. Black error bars centred on zero represent the uncertainty in fundamental \logg\ for each star, while grey error bars represent the standard deviation of the mean for $N\ge3$, as well as combined linear uncertainties and individual uncertainties in the case of two and one determinations, respectively. The number of determinations for each star are indicated in the two panels. The bottom panel compares the parallax-based and spectroscopic gravity measurements with error bars representing combined standard deviations. Symbol colour indicates metallicity from \tab{tab:general}. See text and \tab{tab:lit_teff_logg} for data, references, and discussion.}
   \label{fig:logg_comparison}
\end{figure*}

Two main approaches to determining surface gravity are encountered in the literature -- a spectroscopic analysis or a combination of $V$ magnitude, parallax, bolometric correction, \Teff, \FeH, and stellar model isochrones (hereafter referred to as the parallax method). Spectroscopic gravity determinations are usually based on the requirement of ionization equilibrium of lines from neutral and singly ionized iron or on fitting the pressure-broadened wings of strong metal lines\footnote{Examples for strong pressure-broadened lines are
the Mg~I~b, 
the Fe~I~5269.6\AA,
the Na~I~D, 
and the Ca~I~6162.2\AA\ lines in the optical, and
the Ca~II triplet, 
the Fe~I~8688.6\AA, and
the Mg~I~8806.8\AA\ lines in the near-IR.}.
We queried the PASTEL catalogue for gravities of \gbs\ published between 2000 and 2012. We supplemented the results with some additional data and classified the \logg\ determinations by method. Duplicate values and those outside the two categories were removed. 
We compiled 192 \logg\ determinations using spectroscopic methods and 89 values using the parallax method.
Ten or more \gbs\ were analysed spectroscopically by
\citet{2005ApJS..159..141V}, 
\citet{2005AJ....129.1063L}, 
and \citet{2010MNRAS.405.1907B}, 
while the parallax method was applied for more than ten stars by
\citet{2004A&A...420..183A} 
and \citet{2007A&A...465..271R}. 

In \tab{tab:lit_teff_logg} we list the mean and standard deviation or quoted uncertainties of the compiled \logg\ values, as well as the number of determinations used for each star and method. References are given in the table notes.
These values are compared in \fig{fig:logg_comparison}, where we plot the difference between the spectroscopic and the fundamental \logg\ values (cf. \tab{tab:Teff_logg}) in the top panel, and the difference between the parallax method and fundamental values in the middle panel.
The uncertainties in the fundamental values and the $\sigma$ values from \tab{tab:lit_teff_logg} are represented as separate error bars.
The bottom panel shows the difference between the mean spectroscopic and parallax method values with combined uncertainties.
The stars are ordered by increasing fundamental temperature from left to right, and the colour of the symbol indicates the metallicity.
As for \Teff, the stars \psiPhe\ and \betAra\ lack comparison data and are not included in this discussion.

There is in general good agreement between the gravity values obtained from the different approaches, when taking the uncertainties into account. 
For the coolest giants (\Teff$\lesssim$4300~K), the spectroscopic gravities are systematically higher than the fundamental ones, although the differences are not significant.
In the following, we discuss those stars with discrepant values (non-overlapping error bars), starting with the three dwarf stars \Cyg~B, \tauCet, and \muCas, followed by a subgiant and four giants.

\subsubsection{The dwarf stars \Cyg~B, \tauCet, \muCas}
\label{sect:dwarfslogg}
%
%
For the secondary component in the \Cyg\ system, the spectroscopic \logg\ included in \tab{tab:lit_teff_logg} \citep{2005AJ....129.1063L} is 0.5~dex (5$\sigma$) lower than the fundamental value. An additional determination of 4.4$\pm$0.2 resulting from the parallax method was published by \citet{1999ApJ...523..234T}, which is closer to the fundamental value of 4.67$\pm$0.04.
A mass of 0.46~$M_\odot$ inferred from astrometric observations (cf. \sect{sect:othermass}) would give \logg=4.55, decreasing the discrepancy even further.
No other \logg\ value published after 1980 is available.
This is the coolest dwarf in our sample.
The equivalent widths determined for such stars are uncertain due to blending.
Fe~II lines are weak and difficult to measure (see Paper~III for an extensive discussion).
The analysis of \citet{2005AJ....129.1063L} was based on Fe ionization balance, but included only one Fe~II line (and 229 Fe~I lines).
Thus, we regard the spectroscopic \logg\ as less reliable.

%
%
%
For \tauCet\ and \muCas,\ the mean \logg\ values from the parallax method are higher than the fundamental values by 0.1--0.2~dex (1--2$\sigma$), while the mean spectroscopic determinations agree within the uncertainties\footnote{However, for \muCas\ those of the spectroscopic determinations based on ionization balance of LTE Fe abundances could be too low by 0.2~dex due to non-LTE effects \citep[Table~8 in][]{1999ApJ...521..753T}.}.
These stars have very similar parameters (see e.g. \fig{fig:HRD2}), although \muCas\ is somewhat more metal poor. Both stars have direct measurements of angular diameter and bolometric flux.
However, the masses used for the fundamental \logg\ determination might be rather uncertain because they are derived from problematic evolutionary tracks predicting unreasonable ages (\sect{sect:mass}).
For \tauCet, using the mass or \logg\ from seismic data (see \tabs{tab:MassA} and \ref{tab:loggA}) results in gravities of 4.53 
and 4.58, respectively, which are compatible with the values by \citet{2004A&A...420..183A} and \citet{2007A&A...465..271R} from the parallax method.
For \muCas, using the dynamical mass (see \sect{sect:othermass}) results in a gravity of 4.51$\pm$0.04, which is closer to the values by \citet{2003A&A...404..187G}, \citet{2004A&A...420..183A}, \citet{2006MNRAS.367.1329R}, and \citet{2007A&A...465..271R} from the parallax method. In the first of these articles, a value of 4.46$\pm$0.10 was derived, which is fully compatible with our fundamental value. In that publication, a more recent version of the \emph{Padova} isochrones was used than in the other three (which quote \logg\ values of 4.6 to 4.7 with uncertainties of 0.04 to 0.2).

\subsubsection{The subgiant \epsFor}
\label{sect:subgiantslogg}
%
%
%
For the moderately metal-poor subgiant \epsFor,\ the mean spectroscopic \logg\ is 0.5~dex higher than the fundamental value (about 1$\sigma$). 
The spectroscopic \logg\ determinations show a large dispersion, compared to stars with similar \Teff.
The fundamental value is based on an angular diameter derived from a surface-brightness relation and has one of the largest uncertainties at \Teff$\gtrsim$5000~K, mainly due to the uncertainty of 17\% in mass (caused by the close spacing of evolutionary tracks, see \fig{fig:HRD2}).
The spectroscopic determination by \citet{2000AJ....120.1841F}, 
based on ionization equilibrium, agrees with the fundamental value. 
In all other publications, higher values have been presented, either using the same method (\citealt{2006A&A...458..609D} by 0.3~dex 
and \citealt{2006MNRAS.370..163B} by 0.9~dex), fitting the wings of strong Mg~I, Ca~I, and Fe~I lines (\citealt{2004A&A...425..187T}, by 0.7~dex), 
or using a hybrid method (\citealt{2005ApJS..159..141V}, by 0.4~dex). 
Including non-LTE effects in an Fe-line analysis would increase the derived \logg\ (by about 0.1~dex at the metallicity of \epsFor, \citealt{2013MNRAS.429..126R}).
Three of the five \logg\ values from the parallax method agree with the fundamental value \citep{2003A&A...404..187G,2003A&A...410..527B,2007A&A...465..271R}, 
while the other two are about 0.2~dex higher \citep{2006MNRAS.367.1329R,2011A&A...530A.138C}. 

\subsubsection{Four giants with discrepant gravities}
\label{sect:giantslogg}
%
%
For the K giants \betGem\ and \epsVir, the means of the spectroscopic \logg\ determinations are 0.2--0.3~dex (2$\sigma$) higher than the fundamental values.
In both cases, the dispersion of the three to four determinations is rather small (0.1~dex or less, see \tab{tab:lit_teff_logg}).
The uncertainties of the fundamental values (based on measured angular diameters) are below 0.1~dex as well.
The values from the parallax method agree better with the fundamental ones.
This comparison indicates that for solar-metallicity giants with \Teff$\approx$5000~K and \logg$\approx$2.8, the spectroscopic approach can lead to overestimated gravities.
On the other hand, for \ksiHya, a third giant in our sample with similar parameters as \betGem\ and \epsVir, the single spectroscopic \logg\ determination by \citet{2010MNRAS.405.1907B} agrees very well with the fundamental value.

%
%
For the metal-rich giant \muLeo\ and the metal-poor ($-0.7$~dex) giant HD~220009, the \logg\ values determined by \citet{2007AJ....133.2464L} 
with the parallax method differ from the fundamental one, by $-0.3$~dex (1$\sigma$) and by +0.6~dex (3$\sigma$), respectively.
For \muLeo\ the mean of five spectroscopic determinations agrees with the fundamental value within the uncertainties, while for HD~220009 the two spectroscopic determinations are close to the one from the parallax method. 
The parallax-\logg\ value by \citet{2007AJ....133.2464L} is based on isochrones from a previous version of the \emph{Padova} models \citep{1994A&AS..106..275B}. 
\citet{2007AJ....133.2464L}, \citet{2004A&A...420..183A}, and \citet{2007A&A...465..271R} applied the same models to the solar-metallicity giant \betGem, resulting in lower-than-fundamental \logg\ values.
In fact, a systematic trend of discrepancy with metallicity becomes apparent when comparing the parallax-\logg\ values determined by \citet{2007AJ....133.2464L} for six giants with the fundamental values (see \tab{tab:loggparLuck}).
On the other hand, \citet{2008A&A...484L..21M} 
used more recent model isochrones, similar to the ones used in the current work, for Arcturus and HD~107328, and derived \logg\ values closer to the fundamental ones.

\begin{table}
   \caption{Comparison of \logg\ values derived with the parallax method by \citet[LH07]{2007AJ....133.2464L} for six giants with the fundamental values.}
   \label{tab:loggparLuck}
   \centering
   \begin{tabular}{lrrrr}
      \hline\hline
Name & [Fe/H] & \logg(fund.) & LH07$-$fund. \\
      \hline                           
\muLeo    &    0.3 & 2.5 & $-$0.3 \\   
\epsVir   &    0.2 & 2.8 & $-$0.1 \\   
\betGem   &    0.1 & 2.9 & $-$0.1 \\   
HD~107328 & $-$0.3 & 2.1 &   +0.1 \\   
Arcturus  & $-$0.5 & 1.6 &   +0.1 \\   
HD~220009 & $-$0.7 & 1.4 &   +0.6 \\   
      \hline\hline
   \end{tabular}
\end{table}

\section{Discussion and conclusions}
\label{sect:lastsection}

\subsection{Atmospheric parameters quality for current sample}
\label{sect:Teffloggsummary}
In this section, we summarise our current knowledge of \Teff\ and \logg\ for each of the 29 FGK-type dwarfs, subgiants, and giants and refer to the discussions and comparisons in \sects{sect:teff} to \ref{sect:results} as appropriate. We consider two comparison values to ``agree well'' if their error bars overlap. (These are the cases that have not been explicitly discussed in \sects{sect:Teff_other} and \ref{sect:logg_other}.)
Based on these considerations, we assess for each star and each parameter whether they are suitable to be used as reference values for validation and/or calibration. 
Those of the fundamental values that we do not recommend as reference parameters are mentioned explicitly below and are indicated in \tab{tab:Teff_logg} by square brackets.

Concerning the status of the M~giants, we briefly note that all five of them have fundamental parameters based on directly measured input data (except for the bolometric flux of \betAra).
Their masses are highly uncertain for various reasons discussed in \sect{sect:mass}. Both
\alfTau\ and \alfCet\ have been discussed in \citet{2012A&A...547A.108L}.
For a discussion of the \Teff\ for \gamSge\ see \sect{sect:gamSgeTeff}.
For \betAra\ and \psiPhe\ the lack of comparison data prevents a comprehensive evaluation of their parameters.
%
\paragraph{Dwarfs}
\begin{description}
  \setlength{\itemsep}{3pt}      

   \item[Procyon:] 
      The quality of the available input data for the fundamental parameters is excellent. The interferometric diameter is based on 3D-RHD modelling of $K$-band visibilities. The mass is known to better than 10\% from stellar evolution models, dynamical orbit modelling, and asteroseismic data.
      The mean spectroscopic and photometric determinations of \Teff\ agree with the fundamental one to less than 1\%. We conclude that Procyon's \Teff\ is 6550~K with an uncertainty of 1\%.
      The fundamental surface gravity agrees well with several other parallax-based determinations (the difference is 3\%).
      Also in comparison to the spectroscopic \logg\ determinations, the error bars overlap. However, the latter show a large dispersion of 0.3~dex (factor 2).

   \item[HD~84937:] 
      The fundamental \Teff\ and \logg\ values are based on a calibrated angular diameter.
      Nevertheless, we recommend their use as reference values, since they show very good agreement in comparison with other determinations. For \Teff, the means of the spectroscopic and photometric values differ by less than 1\% from the fundamental one. For \logg, the means of the parallax-based values differ by 3\% and those of the spectroscopic ones by 9\%.
      In all cases the standard deviations of the comparison values are similar to the uncertainties of the fundamental value and larger than the differences quoted above.

   \item[HD~49933:] 
      The interferometric diameter for this sub-mas target was obtained from 3D-RHD modelling of optical visibilities. The resulting fundamental \Teff\ value agrees within 2\% with all but one of the individual comparison values from the literature.
      The evolutionary mass and the fundamental \logg\ values agree very well with the asteroseismic determinations (cf. \sects{sect:othermass} and \ref{sect:logg_seismic}), and there is a good agreement ($\approx$20\%) with spectroscopic \logg\ values.

   \item[\alfCen~A:] 
      Apart from the Sun, \alfCen~A has the most precisely determined fundamental \Teff\ value of the sample (0.3\% uncertainty), which agrees well with the literature spectroscopic measurements. Some of the photometric \Teff\ values are lower by 100~K to 300~K, but those by \citet{2005ApJ...626..446R} and \citet{2007A&A...465..271R} agree within 30~K (1\%).
      The evolutionary, dynamic, and seismic masses agree well, as do the fundamental \logg\ values with the asteroseismic, spectroscopic, and parallax-based \logg\ determinations.

   \item[HD~22879:] 
      As for HD~84937 the fundamental \Teff\ and \logg\ values based on a calibrated diameter agree well with all comparison values.

   \item[The Sun:] 
      Based on the measured total solar irradiance \citep[TSI,][]{2011GeoRL..38.1706K} and the measured solar radius \citep[][\sect{sect:sun}]{2014A&A...569A..60M}, the effective temperature of the Sun is 5771$\pm$1~K, corresponding to a precision of 0.01\%.
      A similar value has been derived by \citet[][their Sect.~2.4.2]{2006ApJS..165..618A} based on an earlier measurement of the TSI with the same instrumentation \citep{2005SoPh..230..129K}\footnote{The adopted TSI value is about 0.4\% or 5~Wm$^{-2}$ lower than even earlier TSI measurements with different instruments \citep[e.g.][]{2000SSRv...94...15F}, resulting in a downward revision of 0.1\% or about 6~K from the previously used canonical value of 5777~K for the solar \Teff.}.
      The solar surface gravity based on the same radius measurement and the measured solar mass parameter \citep[][\sect{sect:logg}]{2011Icar..211..401K} is \logg(\ulogg) = 4.4380 with an uncertainty of 0.05\%, which is dominated by the uncertainty in the radius.

   \item[\muCas:] 
      All \Teff\ determinations agree well.
      The fundamental \logg\ value based on the mass determined in \sect{sect:mass} and given in  \tab{tab:Teff_logg} might be about 0.1~dex too low, and we do not recommend to use it as a reference value (see discussion in \sect{sect:dwarfslogg}).

   \item[\tauCet:] 
      The fundamental \Teff\ value agrees well with the photometric determinations and with half of the spectroscopic determinations (see discussion in \sect{sect:tauCetmuAraTeff}).
      For the fundamental \logg\ value, the same case\ applies as for \muCas.

   \item[\alfCen~B:] 
      The interferometric diameter is based on 3D-RHD modelling of $K$-band visibilities. The fundamental \Teff\ value agrees well with the literature spectroscopic measurements. For the photometric \Teff\ values, the comparison is the same as for \alfCen~A.
      All mass and \logg\ values are consistent as in the case of \alfCen~A.

   \item[18~Sco:] 
      The fundamental \Teff\ value (based on the angular diameter measured at optical wavelengths by \citealt{2011A&A...526L...4B}) 
      agrees well with all comparison values (differences less than 1\%).
      The evolutionary and seismic masses are consistent, as are the fundamental \logg\ values with the asteroseismic, spectroscopic, and parallax-based \logg\ determinations.

   \item[\muAra:] 
      This star has a very uncertain \Teff, which should not be used as a reference value. The fundamental value is based on an indirect angular diameter, and the spectroscopic and photometric \Teff\ values are discrepant (see discussion in \sect{sect:tauCetmuAraTeff}).
      The evolutionary and seismic masses differ by more than 20\%, 
      and the fundamental gravity from the seismic one by 13\% (1.6$\sigma$).
      However, both the mean spectroscopic and the mean parallax-based \logg\ agree well with the fundamental one (with differences of 3\%\ and 9\%, respectively).

   \item[\betVir:] 
      The fundamental \Teff\ value (based on an angular diameter measured at optical wavelengths) agrees well with the mean spectroscopic and photometric values. (The systematic difference seen in \fig{fig:teff_comparison} is about 1\%.)
      The evolutionary and seismic masses agree well (with a difference of about 5\%). The fundamental \logg\ agrees well with the seismic, the mean spectroscopic, and the mean parallax-based ones (differences of 0, 6, and 11\%, respectively).

   \item[\epsEri:] 
      All \Teff\ and \logg\ determinations are consistent. The mean differences are around 1\%, except for the spectroscopic \logg\ values, which are on average lower by 20\%.

   \item[Gmb~1830:] 
      This star has a very uncertain effective temperature, which should not be used as a reference value. The interferometric diameter of this sub-mas object might be affected by calibration errors, and the spectroscopic and photometric determinations show a large dispersion (see discussion in \sect{sect:Gmb1830}).
      The mean spectroscopic and the mean parallax-based \logg\ are 40\% and 20\% higher than the fundamental one, respectively, although they agree within the uncertainties.

   \item[\Cyg~A and B:] 
      The adopted fundamental \Teff\ values for both components agree with the photometric ones.
      For \Cyg~B there are two different recent bolometric flux measurements (see \sect{sect:ang_fbol_comparison}), and we adopted the one by \citet{2013ApJ...779..188M} based on more realistic SEDs.
      The fundamental \logg\ of \Cyg~A is consistent with the single parallax-based comparison value.
      Using the dynamical mass by \citet[][see \sect{sect:othermass}]{2006Ap.....49..386G} instead of the evolutionary mass by \citet{2008A&A...488..667K} does not change the \logg\ value.
      For \Cyg~B the fundamental \logg\ value based on the dynamical mass is 0.1~dex lower than that based on the evolutionary mass and given in \tab{tab:Teff_logg}.
      Determination of \Teff\ and \logg\ based on equivalent width analysis seems to be unreliable in the temperature range of \Cyg~A and B (see discussion in \sects{sect:61CygTeff} and \ref{sect:dwarfslogg}).
\end{description}
\paragraph{Subgiants}
\begin{description}
  \setlength{\itemsep}{3pt}      

   \item[\delEri:] 
      The fundamental \Teff\ value is based on a precise (1\%) interferometric measurement of the angular diameter with a good coverage of the visibility curve. The fundamental \Teff\ is about 1\% lower than the mean photometric value and the closest spectroscopic determinations (see discussion in \sect{sect:delEriepsVirTeff}). 
      The fundamental \logg\ agrees well with the seismic \logg\ estimate (0.01~dex difference).
      The mean spectroscopic and parallax-based determinations are larger by around 40\% than the fundamental \logg, but the dispersions around the mean have similar magnitudes.

   \item[HD~140283:] 
      The fundamental \Teff\ of this sub-mas object is based on a direct angular diameter measured at optical wavelengths, but on a calibrated bolometric flux. The spectroscopic and photometric determinations are systematically higher, although they span a wide range of values (the lowest being within 1\%, and the highest differing by 7\% from the fundamental value, see discussion in \sect{sect:Teffphot}). We do not currently recommend using HD~140283 as a reference star for \Teff\ validation.

      The fundamental \logg\ is consistent with the mean spectroscopic and parallax-based determinations within its rather large uncertainty of 0.1~dex and the large dispersion of the spectroscopic values. It is also compatible with the value of 3.65$\pm$0.06 derived by \citet{2015A&A...575A..26C}.
      Two of the spectroscopic values (3.4 and 3.5~dex) are based on ionization balance of LTE Fe abundances \citep{2000AJ....120.1841F,2001A&A...370..951M} and could be too low by up to 0.5~dex owing to non-LTE effects \citep[Table~8 in][]{1999ApJ...521..753T}. The third spectroscopic value (3.68~dex) is based on a fit of the wings of the Mg~Ib triplet lines \citep{2000A&A...362.1077Z} and is close to the mean of the eight parallax-based values (3.69 with a dispersion of only 0.03~dex, see \tab{tab:lit_teff_logg}).

   \item[\epsFor:] 
      The fundamental \Teff\ is consistent with the mean spectroscopic and photometric determinations, although it is based on indirect input data. 
      The \logg\ determinations, including the fundamental one (\tab{tab:Teff_logg}), span a wide range of possible values from 3.5 to 4.4~dex (see discussion in \sect{sect:subgiantslogg}).
These values, together with the adopted effective temperature, are consistent with the subgiant classification, but we cannot consider any of them to be a reference value for the \logg\ of \epsFor.

   \item[\etaBoo\ and  \betHyi:] 
      The fundamental \Teff\ values agree well with the mean literature spectroscopic and photometric measurements with differences of less than 1\%.
      The evolutionary and seismic masses agree well (with differences of 2\%\ and 8\%, respectively).
      The fundamental \logg\ values agree well with the seismic and the mean parallax-based values ($<10$\% differences). 
      The mean spectroscopic \logg\ values are also consistent with the fundamental ones, although they are higher by about 40\%\ and 20\%, respectively.
\end{description}
\paragraph{Giants}
\begin{description}
  \setlength{\itemsep}{3pt}      

   \item[Arcturus:] 
      The fundamental \Teff\ agrees with the mean spectroscopic and photometric determinations of \Teff\ to less than 1\%, and agrees perfectly with the independent determination by \citet[][see \sect{sect:Teff_other}]{2011ApJ...743..135R}. We conclude that the effective temperature of Arcturus is 4290~K with an uncertainty of 1\%.

      The quality of the surface gravity is more uncertain. The seismic mass and \logg\ are 40\% lower than the evolutionary mass and the fundamental \logg\ (which is 1.6~dex). However, the nature of the oscillations and the applicability of the seismic scaling relations are being debated (see \sects{sect:othermass} and \ref{sect:logg_seismic}).
      The mean spectroscopic \logg\ value agrees well with the fundamental one with a 14\% difference. The individual determinations range from 1.4 to 1.9~dex. 
      The mean parallax-based \logg\ value is 50\% higher than the fundamental one with a narrower range from 1.7 to 2.0~dex.
      In conclusion, the uncertainty of the fundamental value (\tab{tab:Teff_logg}) seems underestimated, and we recommend to adopt \logg=1.6$\pm$0.2~dex, if Arcturus is to be used as a reference star. Exactly the same value was derived by \citet{1983UppOR..27.....E} using the wings of strong pressure-broadened metal lines as constraints.

   \item[HD~122563:] 
      The interferometric diameter is based on 3D-RHD modelling of $K$-band visibilities.
      The resulting fundamental \Teff\ agrees well with the mean spectroscopic one (to within 2\%). 
      The single photometric \Teff\ value shown in \fig{fig:teff_comparison} and given in \tab{tab:lit_teff_logg} deviates significantly from the fundamental one, but several other determinations agree within the uncertainties (see discussion in \sect{sect:Teffphot}).

      The fundamental \logg\ value agrees within 0.1~dex (30\%) with two parallax-based determinations and with the spectroscopic determination by \citet{2008A&A...478..529M} based on ionization equilibrium of non-LTE Ca abundances. All cases have uncertainties of similar magnitude.
      Three other spectroscopic determinations based on ionization equilibrium of LTE Fe abundances are 0.3 to 1.0~dex too low. \citet{2013MNRAS.429..126R} 
      show that an LTE analysis underestimates the surface gravity by at least 0.3~dex at the metallicity of HD~122563 and up to 1.5~dex in combination with an erroneous \Teff. This can easily explain the discrepancies.

   \item[\muLeo:] 
      The fundamental \Teff\ agrees well with the literature photometric determination \citep{2007AJ....133.2464L}, although it is based on indirect input data. 
      The mean of the four spectroscopic \Teff\ determinations is about 3\% higher, but is consistent with the fundamental \Teff\ within the uncertainties.
      The adopted \logg\ value (\tab{tab:Teff_logg}) may serve as a reasonable reference. The discrepancy of about 50\% with the single parallax-based comparison value (\fig{fig:logg_comparison}, \tab{tab:lit_teff_logg}) can be explained by inadequate isochrones (see discussion in \sect{sect:giantslogg}).

   \item[\betGem:] 
      The fundamental \Teff\ is based on direct input data, and agrees well with the mean of the photometric values. The spectroscopic \Teff\ determinations are on average about 2\% higher, but the difference has low significance.
      The evolutionary mass and the fundamental \logg\ values agree with the seismic determinations within the uncertainties. 
      The mean parallax-based and spectroscopic determinations are discrepant, in opposite directions, from the fundamental \logg\ value. As discussed in \sect{sect:giantslogg}, this may be explained by inadequate isochrones and a failure of the spectroscopic approach.

   \item[\epsVir:] 
      The fundamental \Teff\ and \logg\ are based on direct input data and are consistent with the means of the photometric and parallax-based values. 
      However, the mean spectroscopic determinations of \Teff\ and \logg\ are significantly higher, by about 3\% and 0.3~dex (90\%), respectively.
      This indicates an inadequate application of the spectroscopic approach for this star (see discussions in \sect{sect:delEriepsVirTeff} and \sect{sect:giantslogg}).

   \item[\ksiHya:] 
      The fundamental \Teff\ and \logg\ are based on direct input data. The fundamental \Teff\ value of 5040$\pm$40~K agrees well with the single available literature spectroscopic determination, and the photometric determination of 5010~K by \citet{1990ApJS...74.1075M}. 
      The spectroscopic determination of \logg\ deviates by 13\% from the fundamental value, which is less than the quoted uncertainty of 20\%. Also, the pre-Hipparcos parallax-based \logg\ by \citet[][2.93$\pm$0.3~dex]{1990ApJS...74.1075M} is within 15\% of the fundamental value.
      The seismic mass is consistent with the evolutionary one, and the seismic \logg\ is equal to the fundamental one.

   \item[HD~107328:] 
      The fundamental \Teff\ and \logg\ are based on indirect input data.
      However, the fundamental \Teff\ value agrees well ($<$2\% difference) with the individual spectroscopic and photometric determinations, and the same applies to the fundamental in comparison with the parallax-based \logg\ value. The spectroscopic \logg\ value deviates by 30\%, which is, however, within the uncertainties. Thus, we consider the \Teff\ and \logg\ in \tab{tab:Teff_logg} to be reasonable reference values.

   \item[HD~220009:] 
      The fundamental \Teff\ and \logg\ are based partly on preliminary direct and partly on indirect input data. All comparison values are inconsistent with the fundamental ones. The discrepancies may partly be explained by an erroneous photometric calibration in the case of \Teff\ and inadequate isochrones in the case of the parallax-based \logg\ (see discussions in \sect{sect:Teffphot} and \sect{sect:giantslogg}). In conclusion, the parameters remain uncertain for this star and are currently not recommended for using as reference values.
\end{description}

\subsection{Conclusions}
\label{sect:conclusions}

In the era of large Galactic stellar surveys, carefully calibrating and validating the data sets has become an important and integral part of the data analysis.
Successive generations of stellar atmosphere models need to be subjected to benchmark tests to assess progress in predicting stellar properties.
%
In this article we aimed at defining a sample of benchmark stars covering the range of F, G, and K spectral types at different metallicities.
A set of 34 \gbs\ was selected, based on the availability of data required to determine the effective temperature and the surface gravity independently from spectroscopy and atmospheric models as far as possible.
Most of these stars have been subject to frequent spectroscopic investigations in the past, and almost all of them have previously been used as reference, calibration, or test objects.
The stars are rather bright (V magnitudes ranging from 0 to 8), and about two thirds can be observed with telescopes on both hemispheres.

Fundamental values for \Teff\ and \logg\ were determined from their defining relations (the Stefan-Boltzmann law and Newton's law of gravitation), using a compilation of angular diameter measurements and bolometric fluxes, and from a homogeneous mass determination based on stellar evolution models.
Most of the available diameter measurements have formal uncertainties around 1\% (see \tab{tab:parameters}), 
which translate into uncertainties in effective temperature of about 0.5\%.
The measurements of bolometric flux seem to be accurate to 5\% or better, 
which translates into uncertainties in effective temperature of 1\% or less.
This would enable accuracies for late-type stars used in Galactic studies of 2\% \citep[$\sim$100~K, e.g.][]{2005ESASP.576..121N}. 

The derived parameters were compared to recent spectroscopic and photometric determinations and, in the case of gravity, to estimates based on seismic data.
The comparison with literature data is in general satisfactory. In a few cases, significant systematic deviations are seen, most notably for cool stars and metal-poor stars. Some of them can be explained by shortcomings in the methods used in the literature, while others call for further improvement of the fundamental values: that is to say, their input data (angular diameters, bolometric fluxes, masses).
The fundamental \Teff\ and \logg\ values of all \gbs\ are listed in \tab{tab:Teff_logg}, where those that need future adjustment are indicated by square brackets.
In summary, 21 of the 29 FGK-type stars in our sample (including the Sun) have parameters with good quality, which may be used for testing, validation, or calibration purposes.
There are four stars with good \Teff\ but uncertain \logg\ (including Arcturus), and three stars with good \logg\ but uncertain \Teff. For one star (HD~220009), both parameters remain uncertain.



Additional interferometric observations are needed for Gmb~1830 and \muAra\ (see \sects{sect:Gmb1830} and \ref{sect:tauCetmuAraTeff}, respectively).
Such observations are also desirable for the remaining stars with indirectly determined angular diameters (see \tab{tab:parameters}), in order to verify the good quality of their fundamental \Teff\ and \logg.
For most of these stars, we are planning observations with the CHARA array (Gmb~1830, \epsFor, HD~22879, \muLeo, HD~107328)\footnote{\muAra\ and HD~84937 are not accessible to current interferometers, see \sect{sect:sample}.}. 

For nine stars, the bolometric flux values are currently based on calibrations of broad-band photometry and bolometric corrections by \citet{1995AaA...297..197A,1999A&AS..140..261A}, see \tab{tab:parameters}.
More direct determinations should be obtained by compiling absolute flux measurements at numerous wavelength points distributed over a significant part of the spectrum for each star and integrating over the resulting SED (or over fitted model or template SEDs).
It is also worth considering redetermining the measurements for the whole sample with the same method in order to obtain homogeneous bolometric fluxes.

Concerning mass determinations, a few metal-poor stars with masses less than about 0.7~$M_\odot$ lie at the limits of the model grids that we employed (\muCas, \tauCet, HD~22879, HD~84937).
Their masses should be verified using stellar evolution models tailored to their properties, applying codes such as CESAM2k \citep{1997A&AS..124..597M,2008Ap&SS.316...61M} or the Dartmouth Stellar Evolution Program \citep{2008ApJS..178...89D,2011ApJ...740L..25F}.
The estimated mass uncertainties should be validated for the whole sample using statistical methods such as Monte Carlo simulations.
In the long run, seismic data should be obtained and seismic modelling applied to all stars for which solar-type oscillations are expected.

The initial sample of \gbs\ presented in this article should be extended in the future to improve the coverage of parameter space, in particular metallicity, and should be adapted to the needs of different surveys and studies.
Detailed suggestions for future candidate benchmark stars are given in Appendix~\ref{app:suggested}.

\begin{acknowledgements}
UH and AK acknowledge support from the Swedish National Space Board (SNSB/Rymdstyrelsen).
This work was partly supported by the European Union FP7 programme through ERC grant number 320360.
The authors acknowledge the role of the SAM collaboration (\url{http://www.astro.uu.se/~ulrike/GaiaSAM}) in stimulating this research through regular workshops.
We are thankful for valuable input from several members of the \emph{Gaia} Data Processing and Analysis Consortium and the \emph{Gaia}-ESO Public Spectroscopic Survey.
The results presented here benefitted from discussions held during \emph{Gaia}-ESO workshops and conferences supported by the ESF (European Science Foundation) through the GREAT (\emph{Gaia} Research for European Astronomy Training) Research Network Programme.
We thank the referee, G.~F. Porto de Mello, for carefully reading the manuscript.
This research has made use of the SIMBAD database, operated at the CDS, Strasbourg, France.
This publication makes use of data products from the Two Micron All Sky Survey, which is a joint project of the University of Massachusetts and the Infrared Processing and Analysis Center/California Institute of Technology, funded by the National Aeronautics and Space Administration and the National Science Foundation.
\end{acknowledgements}

\bibliographystyle{aa}
\bibliography{cbs,lit_teff_logg,Gaia,models,other}

\appendix

\section{HR-diagrams for individual stars}
\label{app:masses}
\begin{figure*}[ht]
   \begin{center}
      \resizebox{0.43\hsize}{!}{\includegraphics[trim=50 50 50 50,clip]{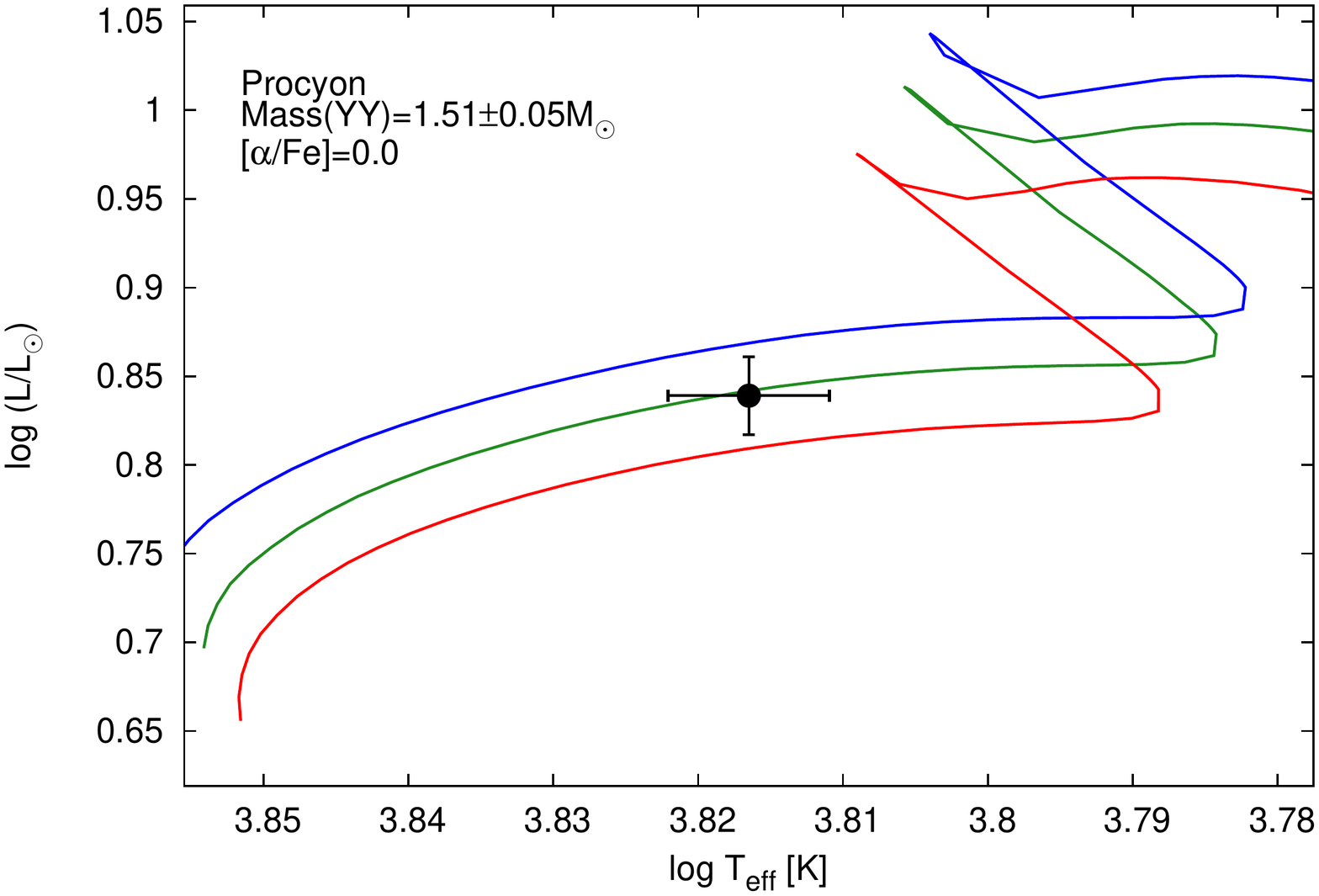}}
      \resizebox{0.43\hsize}{!}{\includegraphics[trim=50 50 50 50,clip]{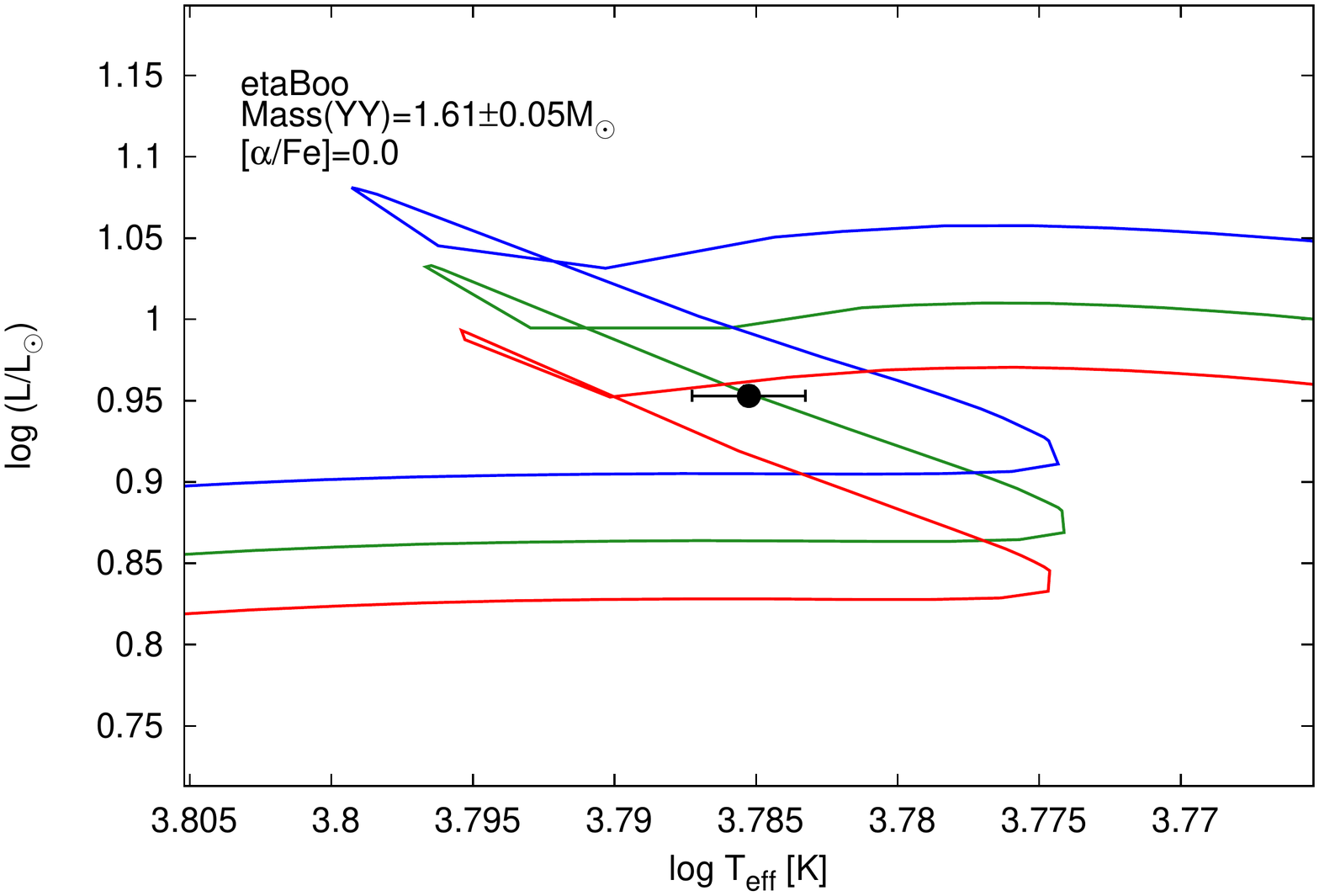}}
      \resizebox{0.43\hsize}{!}{\includegraphics[trim=50 50 50 50,clip]{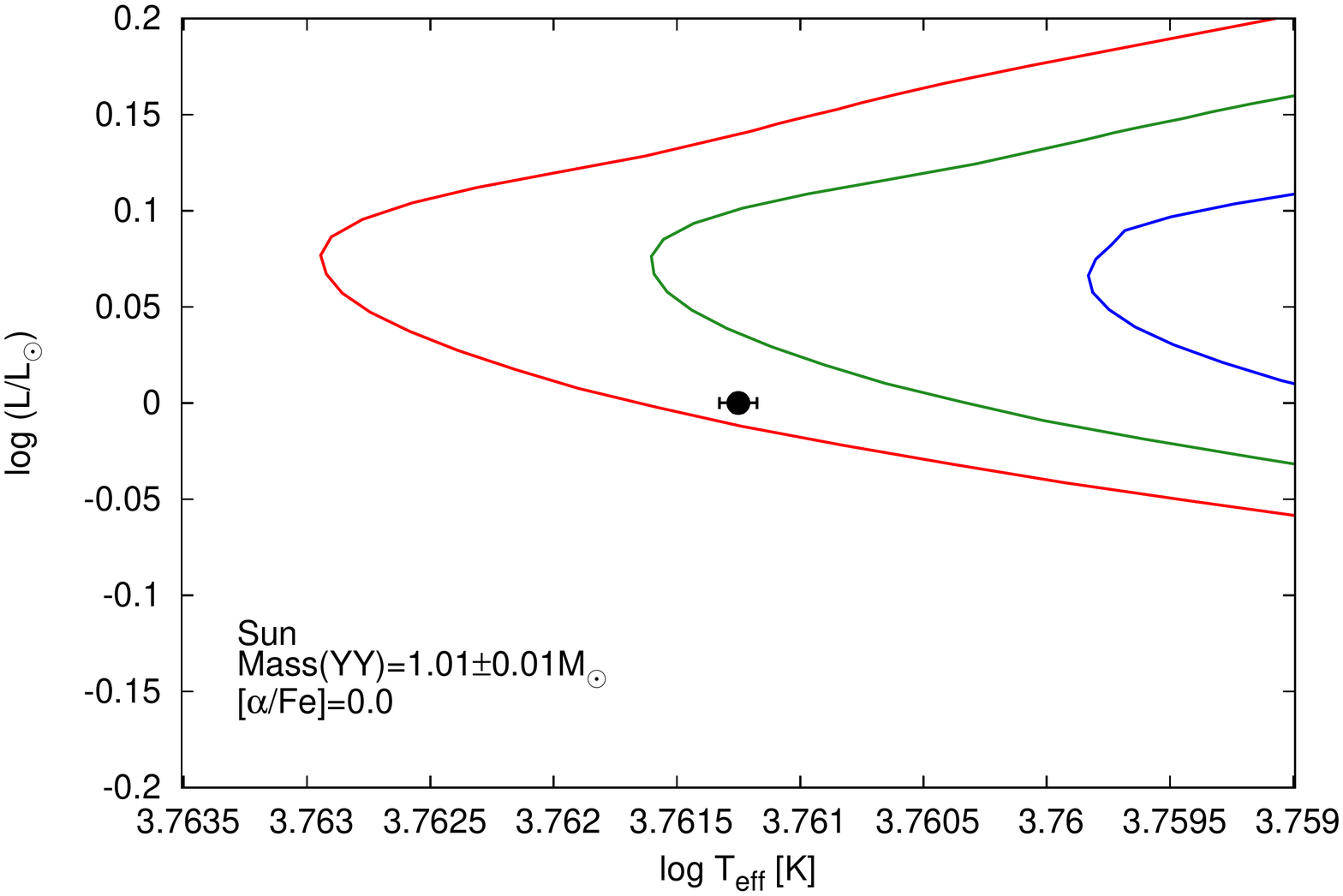}}
      \resizebox{0.43\hsize}{!}{\includegraphics[trim=50 50 50 50,clip]{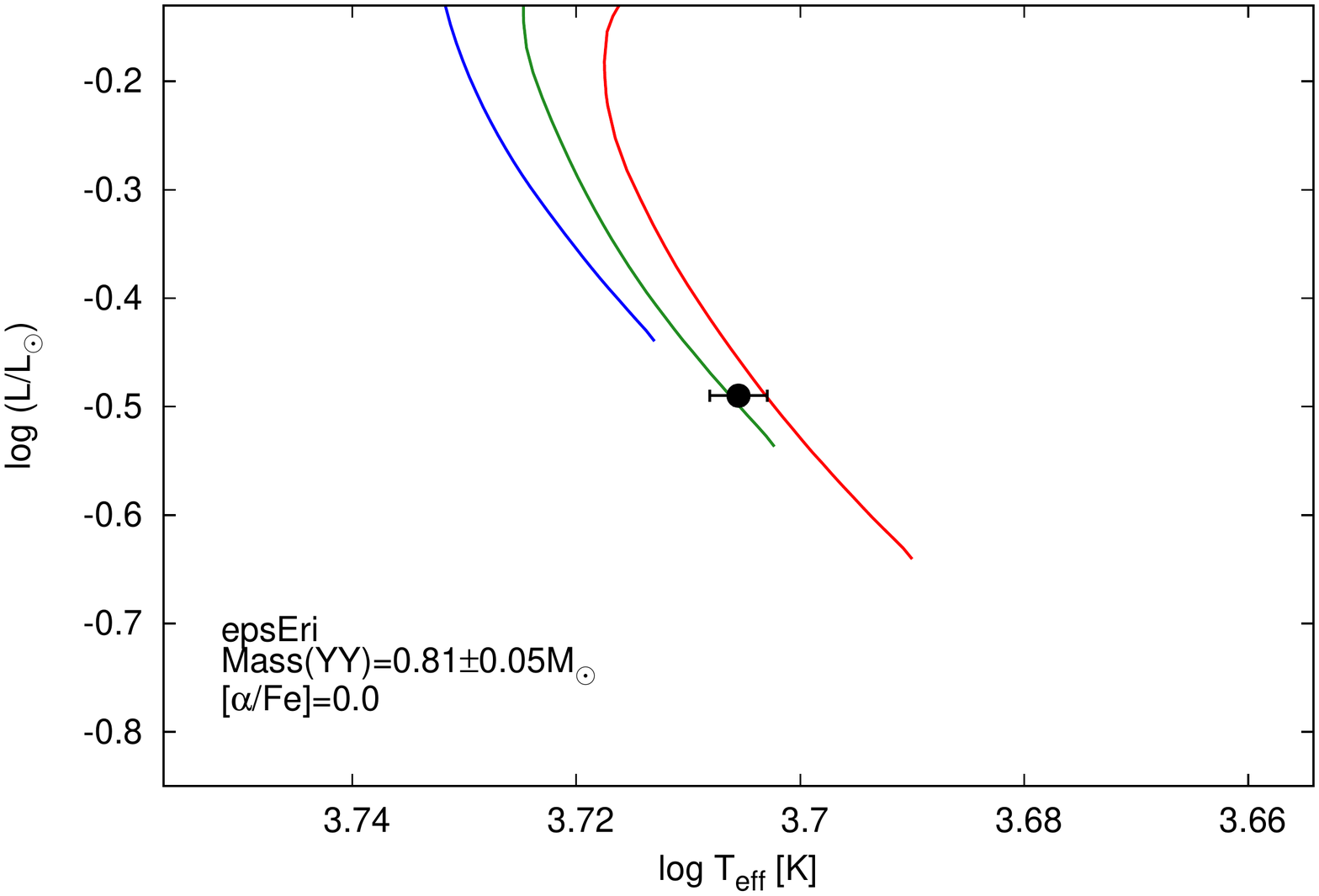}}
      \resizebox{0.43\hsize}{!}{\includegraphics[trim=50 50 50 50,clip]{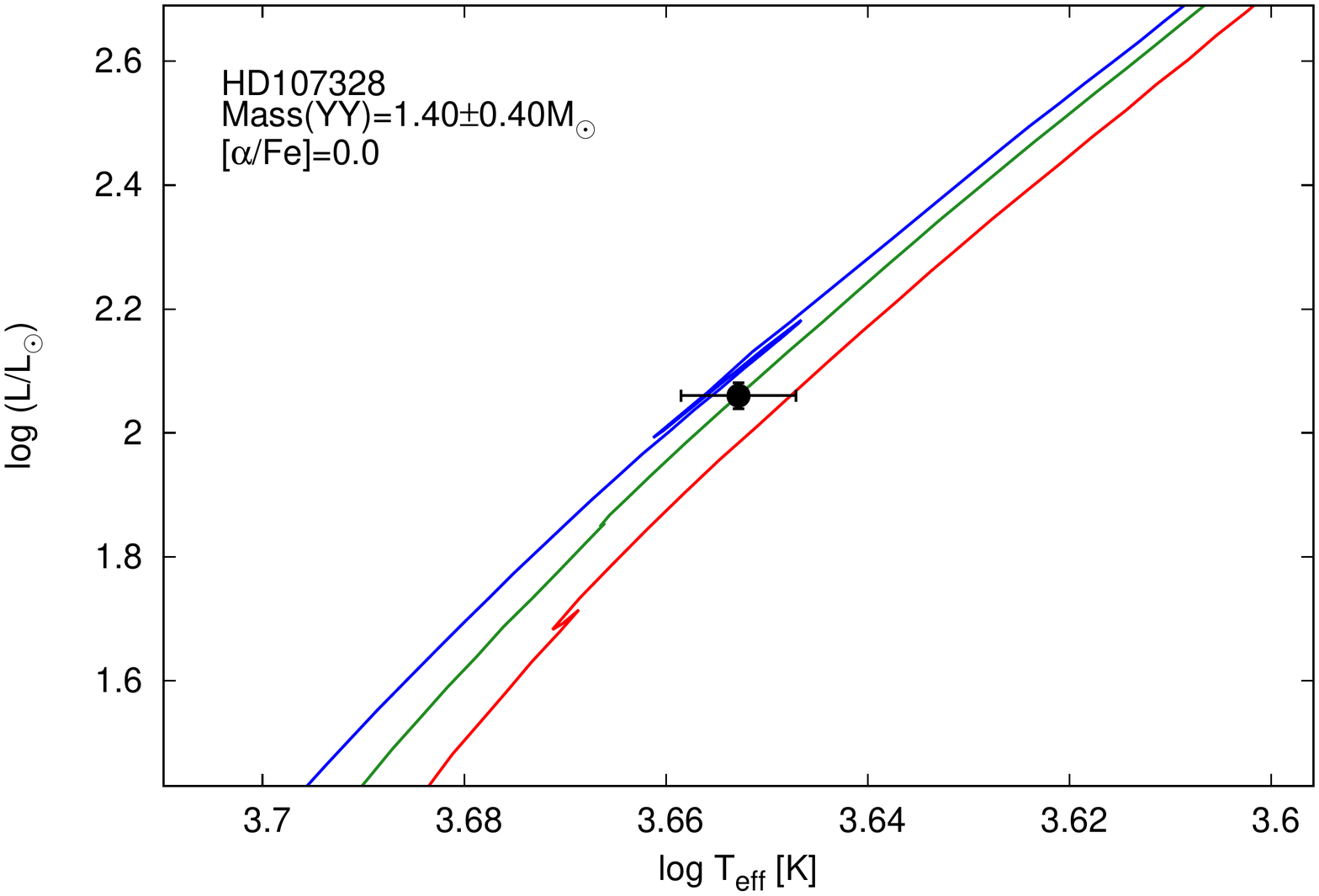}}
      \resizebox{0.43\hsize}{!}{\includegraphics[trim=50 50 50 50,clip]{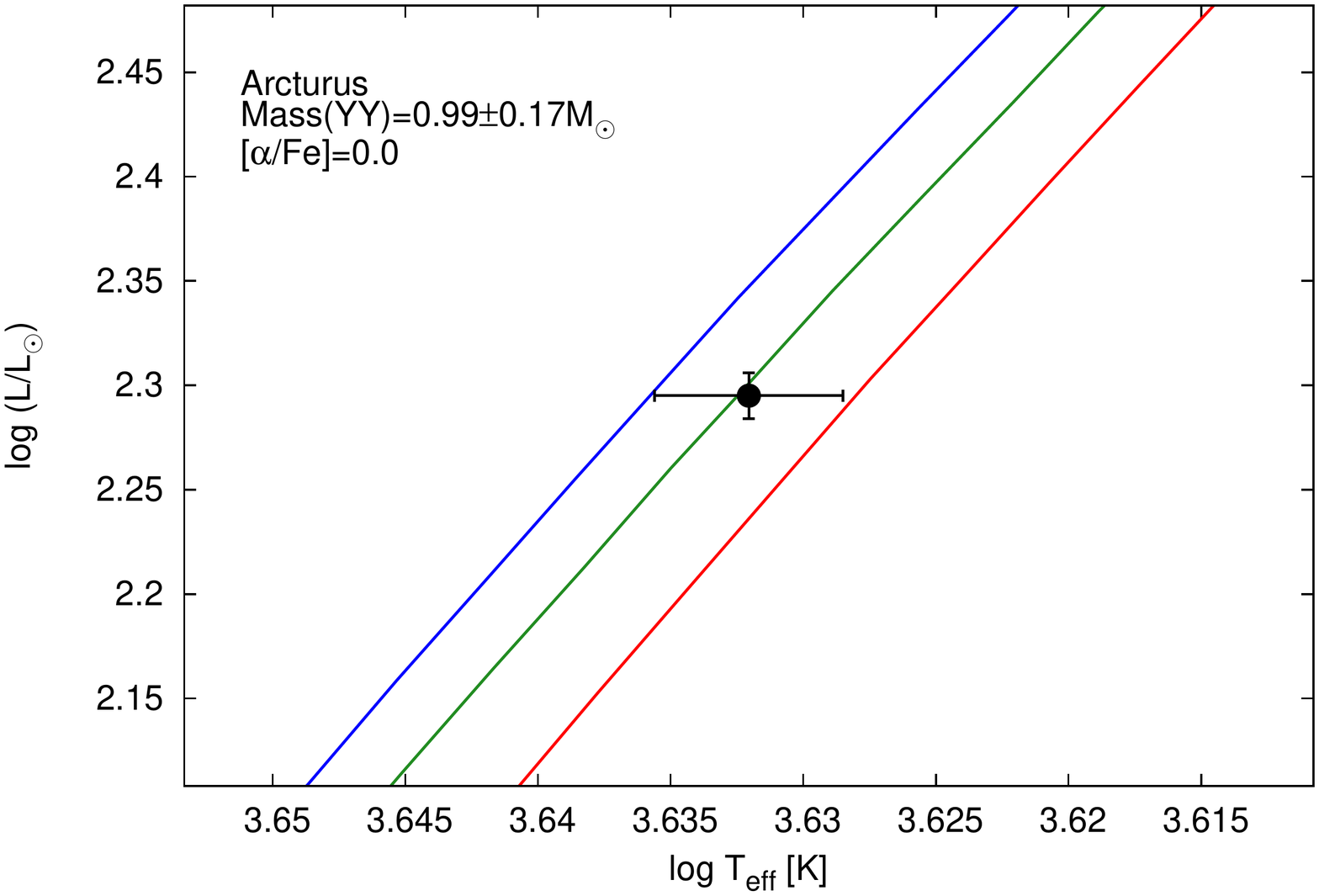}}
      \resizebox{0.43\hsize}{!}{\includegraphics[trim=50 50 50 50,clip]{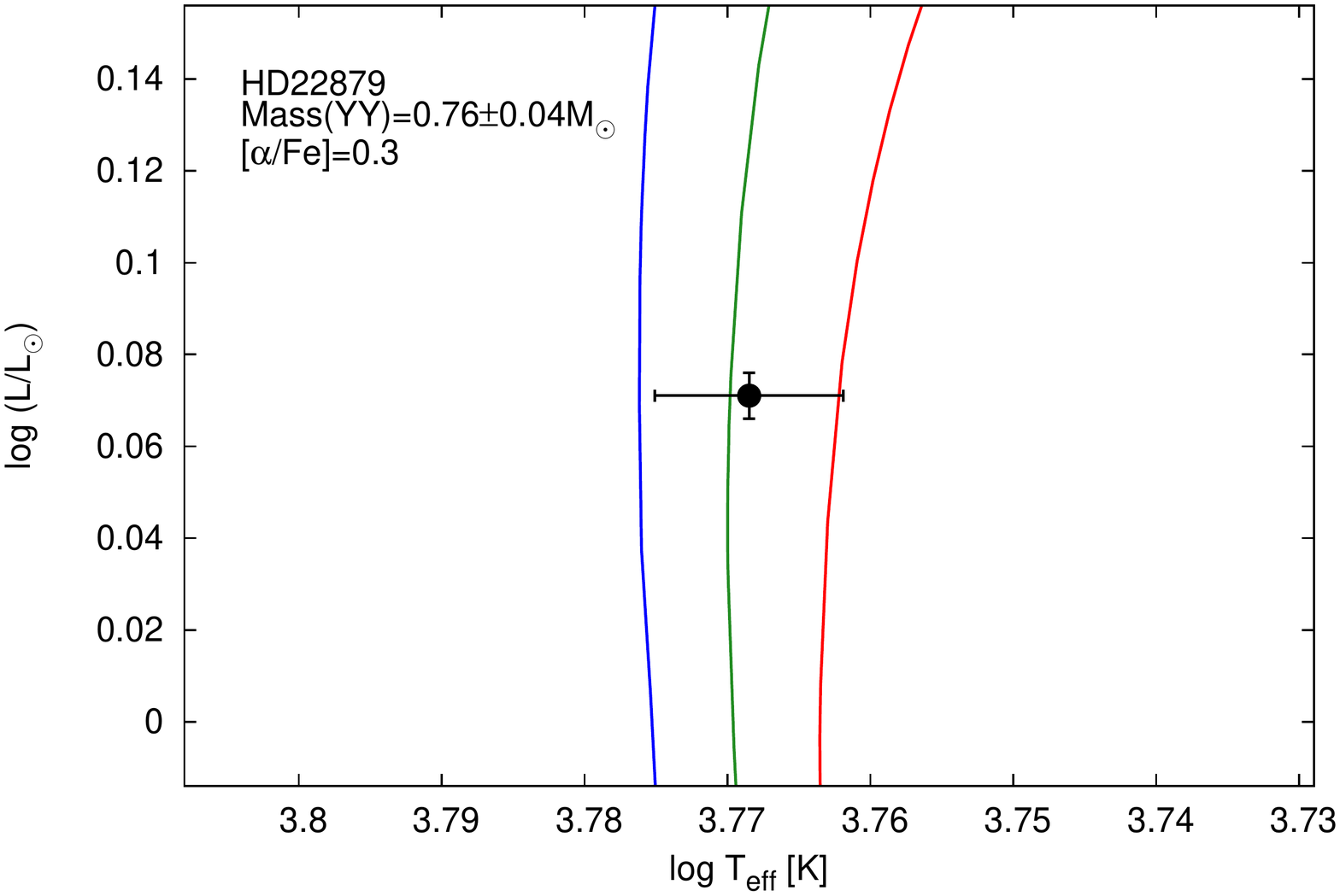}}
      \resizebox{0.43\hsize}{!}{\includegraphics[trim=50 50 50 50,clip]{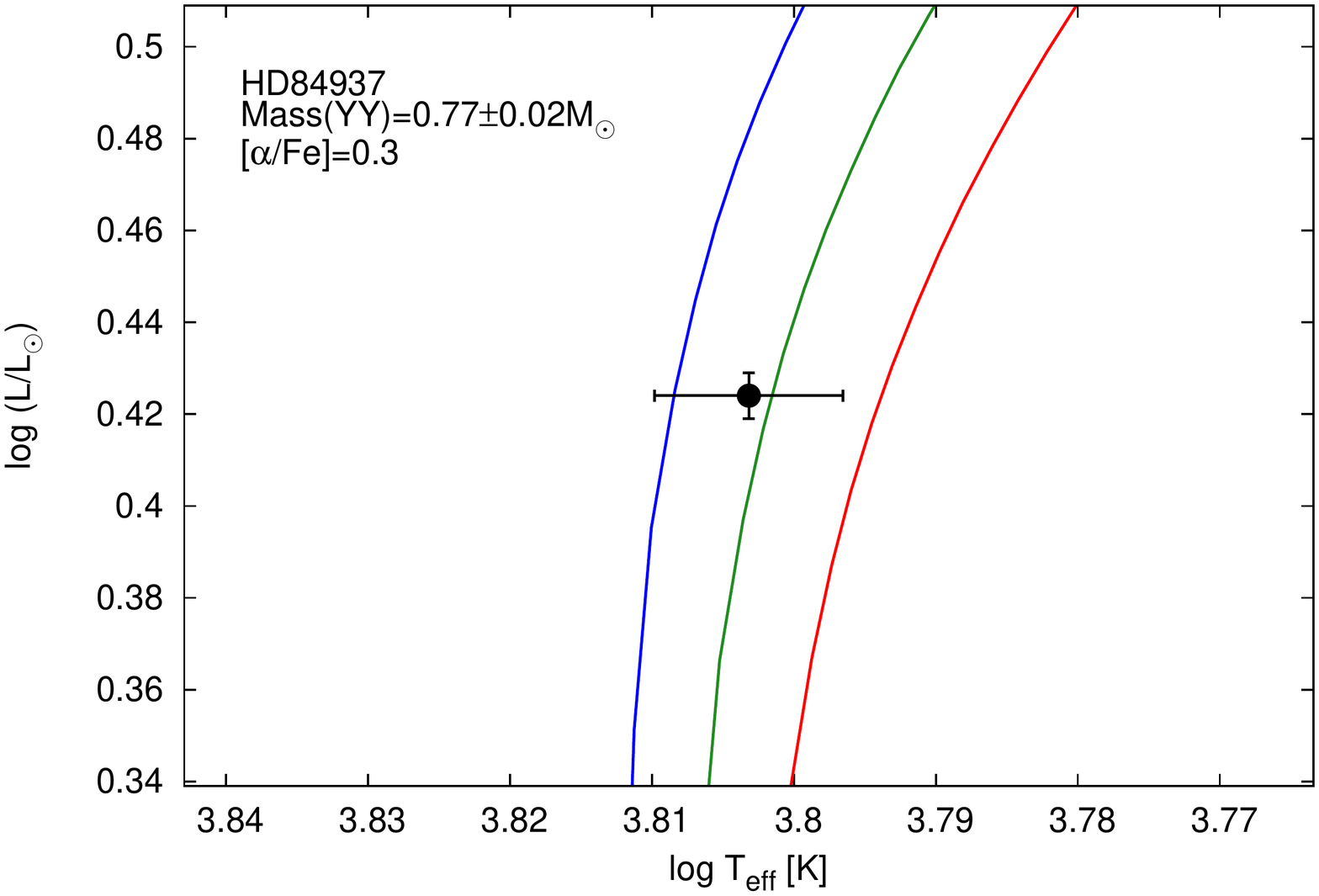}}
   \end{center}
   \caption{Evolutionary tracks for selected \gbs, interpolated in the \emph{Yonsei-Yale} grid
   for three different combinations of mass (given in each panel) and metallicity (taken from \tab{tab:general}), see text for meaning of line colours. Metallicities decrease from top to bottom. Effective temperature and luminosity determined in this paper are indicated by the black filled circles. The $\alpha$-element abundances [$\alpha$/Fe] of the grids are indicated as well.}
   \label{fig:HRDinterp}
\end{figure*}

In \fig{fig:HRDinterp}, we give examples for the results of mass
determination for individual stars. Effective temperature and luminosity of
each star are shown together with three evolutionary tracks interpolated in the
\emph{Yonsei-Yale} grid. The green lines correspond to a track for the derived
mass (indicated in the figure) and the adopted metallicity (\tab{tab:general}),
while blue lines are tracks for both mass and metallicity increased by their
uncertainties, and red lines are tracks for both mass and metallicity decreased by their
uncertainties. The mass uncertainties were adapted such that one of the tracks
with varied mass and metallicity was still consistent with the uncertainties
in \Teff\ and $L$.

\section{Future extension of the sample}
\label{app:suggested}
In this section, we describe different categories of possible additional stars, and give a list of suggested future benchmark stars in \tab{tab:suggested}.

%
%

Four metal-rich dwarfs with magnitudes and angular sizes similar to \muAra\ (cf. \sect{sect:sample}), but observable from the northern hemisphere, can be found in the PASTEL catalogue.
These are
HD~73752 \citep[HR 3430, \FeH=0.4,][]{2006AJ....131.3069L}, 
a close visual binary (5.4 and 6.8 mag, period of 123 yr, semi-major axis of 1.7~arcsec, \citealt{2012A&A...546A..69M}); 
%
HD~19994 (GJ~128) and HD~120136 ($\tau$~Boo, GJ~527), two planet-host visual binaries
(semi-major axes 10 and 5~arcsec, \citealt{2012A&A...546A..69M,1994RMxAA..28...43P}) 
with \FeH=0.3 \citep{2005PASJ...57...27T}, 
each of which has an angular diameter of 0.79$\pm$0.03~mas measured with the CHARA array \citep{2008ApJ...680..728B}; 
%
HD~161797 \citep[$\mu$~Her, GJ~695~A, \FeH=0.3,][]{2011A&A...526A..71D}, 
a star with asteroseismic data \citep{2008ApJ...676.1248B}, 
for which \citet{2014ApJ...781...90B} 
measured an angular diameter of 1.96$\pm$0.01~mas with the NPOI.

%
%

To find metal-poor candidate benchmark stars, we queried the PASTEL catalogue for stars with 4500~K $<$ \Teff\ $<$ 6500~K and $-2$~dex $<$ \FeH\ $<$ $-1$~dex. The resulting list includes 14 stars that are brighter than $V=9$ with at least four metallicity values published after 1990 with standard deviations less than 0.1~dex.
Apart from Gmb~1830, these are HD~21581, HD~23439A, HD~23439B, HD~63791, HD~83212, HD~102200, HD~175305, HD~196892, HD~199289, HD~201891, HD~204543, HD~206739, and HD~218857.

%
%

\citet{2012ApJ...757..112B,2013ApJ...771...40B} 
measured and compiled angular diameters with uncertainties of less than 5\% for 125 main-sequence stars.
Eighty-two of those are FGK stars not included in the \gbs\ sample.
\citet{2012ApJ...757..112B,2013ApJ...771...40B} also determined bolometric fluxes and effective temperatures for all of these stars.
Forty-seven of the 82 stars have \Teff\ determinations better than 1\% and have a metallicity determination published 1990 or later (in the PASTEL catalogue).
The mean literature metallicities of these stars range from $-0.5$ to +0.4~dex, with 80\% between $-0.3$ and +0.1~dex.
Candidate benchmark stars may be selected from 0.1~dex-wide metallicity bins as follows (ID, [Fe/H] bin centre, spectral type):
HD~158633, $-$0.45, K0V;
HD~128167, $-$0.35, F4V;
HD~69897, $-$0.25, F6V; HD~4628, $-$0.25, K3V;
HD~82328, $-$0.15, F7V; HD~16160, $-$0.15, K3V;
HD~173667, $-$0.05, F6V; HD~10476, $-$0.05, K1V;
HD~30652, +0.05, F6V; HD~149661, +0.05, K2V;
HD~217014, +0.15, G3IV;
HD~161797 ($\mu$~Her), +0.25, G5IV;
HD~217107, +0.35, G8IV.

Recently, \citet{2014ApJ...781...90B} 
have published new interferometric diameters for several stars with seismic data (see below): $\beta$~Aql, $\epsilon$~Oph, $\eta$~Ser, $\kappa$~Oph, $\xi$~Dra, $\zeta$~Her, and HR~7349, as well as the \gbs\ \tauCet\ and \etaBoo\ and the already mentioned $\mu$~Her.
For several additional candidate benchmark stars, interferometric data have been obtained and are being processed (O.~Creevey, priv. comm.).
These are
HD~148897, a metal-poor giant with \FeH=$-1.2$~dex,
and two solar-metallicity dwarfs with asteroseismic data: HD~52265 and HD~165341 (70~Oph).
Furthermore, an observing programme at ESO's VLTI has been started to measure about 20 giant and subgiant stars with predicted angular diameters larger than 3~mas (AMBER instrument, PI I.~Karovicova)\footnote{\url{http://archive.eso.org/wdb/wdb/eso/sched_rep_arc/query?progid=094.D-0572(A)}}.
The stars were selected to fill gaps in the parameter space covered by the current \gbs\ and have \FeH\ from $-$0.5 to +0.5~dex. They are part of a sample of about 100 stars identified as possible targets for future proposals.

%
%

The stars $\theta$~Cyg and 16~Cyg~A and B 
are the brightest dwarf stars 
with asteroseismic observations from the \emph{Kepler} mission.
\citet{2011arXiv1110.2120G} 
estimated a large frequency separation \Dnu=84~$\mu$Hz for $\theta$~Cyg.
\citet{2012ApJ...748L..10M} 
derived \Dnu=103.4~$\mu$Hz for 16~Cyg~A and \Dnu=117.0~$\mu$Hz for 16~Cyg~B and determined the stellar parameters for these stars by modelling over 40 individual oscillation frequencies detected for each star.
\citet{2013MNRAS.433.1262W} 
measured angular diameters of 0.753, 0.539, and 0.490~mas for $\theta$~Cyg, 16~Cyg~A, and 16~Cyg~B, respectively, with the CHARA array, each with a precision of 1\%. In combination with the asteroseismic and other data, they determined the effective temperatures with precisions better than 1\%.
\citet{2010MNRAS.405.1907B} 
list 12 
stars with seismic data in addition to those already included in the \gbs\ sample (cf. \sect{sect:sample}). These are 70~Oph~A, 171~Pup, $\alpha$~For, $\beta$~Aql, $\gamma$~Pav, $\gamma$~Ser, $\delta$~Pav, $\eta$~Ser, $\iota$~Hor, $\nu$~Ind, $\tau$~PsA, and HR~5803.
We note that 70~Oph~A was used as a standard star by \citet[][cf. \sect{sect:sample}]{2010ApJ...725.2176Q}. 
Additional stars with seismic data are listed in \citet{2012MNRAS.419L..34M}: 
$\beta$~Oph, $\beta$~Vol, $\epsilon$~Oph, $\kappa$~Oph, $\xi$~Dra, $\mu$~Her, $\zeta$~Her, HD~49385, HD~52265, HD~170987, HD~175726, HD~181420, and HD~181906.
%
Furthermore, \citet{2012ApJ...760...32H} 
measured the angular diameters of ten bright oscillating stars observed by \emph{Kepler} or \emph{CoRoT} with the CHARA array. The sample includes six dwarfs (HD~173701, HD~175726, HD~177153, HD~181420, HD~182736, HD~187637) and four giants (HD~175955, HD~177151, HD~181827, HD~189349).

%
%

%
%
Visual binary systems with well-defined orbits can be found in Table~2 of \citet{2012A&A...546A..69M}. 
Constraining the systems in that table to those with uncertainties in periods and semi-major axes of at most 5\%, primary magnitudes less than 5, and to those where the spectral types of both components are given results in six systems containing at least one FGK dwarf component (including \alfCen) and two systems with at least one FGK giant component, with periods between three months and 170~yr, and total dynamical system masses between 1.6 and 4.7~$M_\odot$.
The systems (in addition to \alfCen) are
$\gamma$~Vir (F0V+F0V),
$\psi$~Vel (F0IV+F3IV), 
$\alpha$~Equ (G0III+A5V),
10~UMa (F3V+K0V),
70~Oph (K0V+K4V),
113~Her (G4III+A6V), and
$\xi$~Boo (G8V+K5V).

%
%
Eclipsing binary systems with accurate mass and radius determinations (uncertainties lower than 3\%) were compiled by \citet{2010A&ARv..18...67T}. 
Their Table~2 contains 85 FGK stars in 46 systems.
Another catalogue of well-studied detached eclipsing binaries \citep{2014arXiv1411.1219S} 
is provided on-line\footnote{\url{http://www.astro.keele.ac.uk/~jkt/debcat/}}.
The version of 2014-07-04 contained 100 systems with one or two FGK components, including 39 of those listed in \citet{2010A&ARv..18...67T}.
This sample is dominated by F~dwarfs.
The brightest system with F-type components is
VV~Crv \citep[F5IV+F5V,][]{2013AJ....146..146F}, 
the brightest systems with a G-type giant or dwarf are
TZ~For \citep[G8III+F7IV,][]{1991A&A...246...99A} 
and
KX~Cnc \citep[G0V+G0V,][]{2012AJ....143....5S}, 
and the brightest system with a K-type component is
AI~Phe \citep[K0IV+F7V,][]{1988A&A...196..128A}. 

%
%

Several recent publications on automatic spectroscopic parameter determination have defined their own sets of FGK standard stars for testing their routines. All of these may be considered for inclusion in an extended sample of \gbs.
The solar-metallicity G-type subgiant 70~Vir 
with an interferometric angular diameter of 0.93~mas \citep{2009ApJ...694.1085V} 
was used for spectroscopic tests by \citet{2011MNRAS.411.2311F}. In their Table~2 
\citet[][cf. \sect{sect:sample}]{2011A&A...531A..83C} 
list four dwarfs between 5200 and 6100~K with interferometric angular diameters, which are not included in the \gbs\ sample: 
$\upsilon$~And, $\sigma$~Dra, $\zeta$~Her, and $\mu$~Her. (The last two have already been mentioned above.)
\citet[][cf. their Table~2]{2013MNRAS.429..126R} 
used four nearby giants between 4600 and 4900~K with interferometric \Teff\ 
to test effective temperatures from Balmer-line fitting, in addition to several \gbs\ and two stars from \citet{2011A&A...531A..83C}: 
HD~27697, HD~28305, HD~140573, and HD~215665.
The two solar-metallicity giants $\beta$~And and $\delta$~Oph (\Teff$\approx$3800~K) have been used as standard stars in the APOGEE survey \citep{2013ApJ...765...16S,2013AJ....146..133M} 
to test the pipeline for parameter determination and the line list for abundance measurements.
Two metal-poor stars were among the reference stars used by \citet{2014A&A...564A.109S} 
for pipeline validation (cf. \sect{sect:sample}): the dwarf HD~126681, and the giant HD~26297.
\citet{2014A&A...564A.119M} 
used three solar-metallicity giants with interferometric angular diameters to validate their abundance analysis 
of red giants in the \emph{CoRoT} asteroseismology fields, in addition to Arcturus and \ksiHya:
$\beta$~Aql, 
$\epsilon$~Oph, 
and $\eta$~Ser. 
All of these have already been mentioned above.
See their Table~A.1 for a compilation of literature atmospheric parameters and abundances.

Finally, it could be worth considering the FGKM dwarfs and giants with P$\delta$ and H$\alpha$ observations contained in the catalogue of hydrogen line profiles by \citet{2012A&A...547A..62H} 
These include HD~2665, HD~2796, HD~4306, HD~6755, HD~6833, HD~6860 ($\beta$~And), HD~25329, HD~62345, HD~71369, HD~165195, HD~187111, HD~216143, HD~217906, HD~221170, HD~232078, and BD~+44~493.

Basic information for the 102 stars explicitly mentioned in this section is given in \tab{tab:suggested}.
Furthermore, we extracted parallax measurements from the SIMBAD database for all stars except one.
About half of these have uncertainties below 1.5\%, and the uncertainties are below 10\% for 80\% of the stars.
We also list the mean and standard deviations of metallicity determinations extracted from the PASTEL catalogue, except for a few stars for which metallicity references are given in the table notes.
Twenty-eight of these stars have \FeH$<-1$~dex.

\onecolumn
\begin{longtab}
\centering
\begin{longtable}{rllllrrr}
   \caption{\label{tab:suggested} Basic information for stars suggested for future extension of the \gbs\ sample.} \\
      \hline\hline
       HD & Name   & RA (J2000) & DEC (J2000) & Spectral type & $V$mag & \FeH & $\sigma$(\FeH) \\
      \hline
      \endfirsthead
      \caption{ continued.} \\
      \hline\hline
       HD & Name   & RA (J2000) & DEC (J2000) & Spectral type & $V$mag & \FeH & $\sigma$(\FeH) \\
      \hline
      \endhead
      \hline
      \endfoot
      \hline\hline
      \multicolumn{8}{l}{\parbox{0.80\textwidth}{Coordinates, $V$ magnitudes, and spectral types are extracted from the SIMBAD database. Spectral types for binary stars were taken from the respective catalogues. \FeH\ and $\sigma$ are mean and standard deviations of metallicity determinations contained the PASTEL catalogue (publications from 1990 and later), except for the following stars, for which individual measurements and uncertainties were taken from the sources in parentheses:
  HD~6980  \citep{1988A&A...196..128A}, 
 HD~20301  \citep{1991A&A...246...99A}, 
 HD~49385  \citep{2010A&A...515A..87D}, 
 HD~74057  \citep{2012AJ....143....5S}, 
 HD~82434  \citep{2012ApJS..203...30F}, 
HD~146051  \citep{2013ApJ...765...16S}, 
HD~170987  \citep{2010A&A...518A..53M}, 
HD~181907  \citep{2014A&A...564A.119M}, 
HD~189349  \citep{2012ApJ...760...32H}, 
HD~217906  \citep{1985ApJ...294..326S}, 
BD~+44~493 \citep{2009ApJ...698L..37I}. 
}}\\
      \endlastfoot
        2665&         HD 2665 & 00 30 45.446 & +57 03 53.63 &          G5IIIw & 7.7 & -2.00 & 0.09\\
  2796&         HD 2796 & 00 31 16.915 & -16 47 40.80 &              Fw & 8.5 & -2.32 & 0.13\\
  4306&         HD 4306 & 00 45 27.163 & -09 32 39.79 &           KIIvw & 9.0 & -2.70 & 0.19\\
  4628&         HD 4628 & 00 48 22.977 & +05 16 50.21 &           K2.5V & 5.7 & -0.26 & 0.05\\
  6980&          AI Phe & 01 09 34.195 & -46 15 56.09 &        K0IV+F7V & 8.6 & -0.14 & 0.10\\
  6755&         HD 6755 & 01 09 43.065 & +61 32 50.19 &             F8V & 7.7 & -1.55 & 0.05\\
  6860&         bet And & 01 09 43.924 & +35 37 14.01 &           M0III & 2.0 & -0.04 &     \\
  6833&         HD 6833 & 01 09 52.265 & +54 44 20.28 &           G9III & 6.7 & -0.88 & 0.11\\
  9826&         ups And & 01 36 47.842 & +41 24 19.64 &             F9V & 4.1 &  0.08 & 0.05\\
 10476&        HD 10476 & 01 42 29.762 & +20 16 06.60 &             K1V & 5.2 & -0.04 & 0.04\\
      &      BD +44 493 & 02 26 49.738 & +44 57 46.52 &           G5IV  & 9.1 & -3.68 & 0.11\\
 16160&        HD 16160 & 02 36 04.895 & +06 53 12.75 &             K3V & 5.8 & -0.12 & 0.06\\
 17051&         iot Hor & 02 42 33.466 & -50 48 01.06 &             F8V & 5.4 &  0.13 & 0.10\\
 20010&         alf For & 03 12 04.527 & -28 59 15.43 &         F6V+G7V & 3.9 & -0.28 & 0.06\\
 19994&        HD 19994 & 03 12 46.437 & -01 11 45.96 &             F8V & 5.1 &  0.19 & 0.07\\
 20301&          TZ For & 03 14 40.093 & -35 33 27.60 &      G8III+F7IV & 6.9 &  0.10 & 0.15\\
 21581&        HD 21581 & 03 28 54.486 & -00 25 03.11 &              G0 & 8.7 & -1.69 & 0.09\\
 23439&       HD 23439A & 03 47 02.113 & +41 25 38.06 &             K1V & 8.1 & -1.06 & 0.09\\
 23439&       HD 23439B & 03 47 02.636 & +41 25 42.56 &             K2V & 8.8 & -1.03 & 0.13\\
 25329&        HD 25329 & 04 03 14.999 & +35 16 23.79 &             K1V & 8.5 & -1.79 & 0.06\\
 26297&        HD 26297 & 04 09 03.418 & -15 53 27.06 &        G5/G6IVw & 7.5 & -1.79 & 0.09\\
 27697&        HD 27697 & 04 22 56.093 & +17 32 33.05 &           K0III & 3.8 &  0.05 & 0.09\\
 28305&        HD 28305 & 04 28 36.999 & +19 10 49.54 &           K0III & 3.5 &  0.11 & 0.09\\
 30652&        HD 30652 & 04 49 50.411 & +06 57 40.59 &             F6V & 3.2 &  0.00 & 0.03\\
 49385&        HD 49385 & 06 48 11.503 & +00 18 17.90 &             G0V & 7.4 &  0.09 & 0.05\\
 52265&        HD 52265 & 07 00 18.036 & -05 22 01.78 &             G0V & 6.3 &  0.21 & 0.03\\
 62345&        HD 62345 & 07 44 26.854 & +24 23 52.79 &           G8III & 3.6 & -0.02 & 0.17\\
 63077&         171 Pup & 07 45 35.022 & -34 10 20.51 &             F9V & 5.4 & -0.81 & 0.13\\
 63791&        HD 63791 & 07 54 28.724 & +62 08 10.76 &              G0 & 7.9 & -1.70 & 0.06\\
 69897&        HD 69897 & 08 20 03.862 & +27 13 03.74 &             F6V & 5.1 & -0.27 & 0.06\\
 71878&         bet Vol & 08 25 44.195 & -66 08 12.80 &           K2III & 3.8 & -0.01 &     \\
 71369&        HD 71369 & 08 30 15.871 & +60 43 05.41 &           G5III & 3.4 & -0.10 & 0.14\\
 73752&        HD 73752 & 08 39 07.901 & -22 39 42.81 &            G5IV & 5.0 &  0.39 &     \\
 74057&          KX Cnc & 08 42 46.211 & +31 51 45.37 &         G0V+G0V & 7.2 &  0.07 & 0.10\\
 76943&          10 UMa & 09 00 38.381 & +41 46 58.61 &         F3V+K0V & 4.0 &  0.25 &     \\
 82434&         psi Vel & 09 30 42.000 & -40 28 00.26 &       F0IV+F3IV & 3.6 &  0.00 & 0.20\\
 82328&        HD 82328 & 09 32 51.434 & +51 40 38.28 &             F7V & 3.2 & -0.18 & 0.06\\
 83212&        HD 83212 & 09 36 19.952 & -20 53 14.76 &       G6/K0IIIw & 8.3 & -1.44 & 0.06\\
102200&       HD 102200 & 11 45 34.235 & -46 03 46.39 &             F2V & 8.8 & -1.22 & 0.06\\
110317&          VV Crv & 12 41 15.951 & -13 00 50.03 &        F5IV+F5V & 5.8 &  0.00 &     \\
110379&         gam Vir & 12 41 39.643 & -01 26 57.74 &         F0V+F0V & 2.7 & -0.05 & 0.07\\
117176&          70 Vir & 13 28 25.809 & +13 46 43.64 &          G5IV-V & 5.0 & -0.06 & 0.03\\
120136&       HD 120136 & 13 47 15.743 & +17 27 24.86 &         F6IV+M2 & 4.5 &  0.29 & 0.08\\
126681&       HD 126681 & 14 27 24.911 & -18 24 40.44 &             G3V & 9.3 & -1.21 & 0.12\\
128167&       HD 128167 & 14 34 40.817 & +29 44 42.46 &             F4V & 4.5 & -0.37 & 0.10\\
131156&         ksi Boo & 14 51 23.380 & +19 06 01.70 &         G8V+K5V & 4.6 & -0.14 & 0.06\\
139211&         HR 5803 & 15 39 56.543 & -59 54 30.02 &            F6IV & 6.0 &  0.00 & 0.06\\
140573&       HD 140573 & 15 44 16.074 & +06 25 32.26 &           K2III & 2.6 &  0.05 & 0.11\\
142860&         gam Ser & 15 56 27.183 & +15 39 41.82 &            F6IV & 3.8 & -0.20 & 0.04\\
146051&         del Oph & 16 14 20.739 & -03 41 39.56 &         M0.5III & 2.8 & -0.01 & 0.09\\
146791&         eps Oph & 16 18 19.290 & -04 41 33.03 &           G8III & 3.2 & -0.10 & 0.10\\
148897&       HD 148897 & 16 30 33.549 & +20 28 45.07 &           G8III & 5.2 & -1.16 &     \\
149661&       HD 149661 & 16 36 21.450 & -02 19 28.52 &             K2V & 5.8 &  0.04 & 0.04\\
150680&         zet Her & 16 41 17.161 & +31 36 09.79 &            G0IV & 2.8 &  0.02 & 0.06\\
153210&         kap Oph & 16 57 40.098 & +09 22 30.11 &           K2III & 3.2 &  0.02 & 0.07\\
158633&       HD 158633 & 17 25 00.099 & +67 18 24.15 &             K0V & 6.4 & -0.49 & 0.07\\
161096&         bet Oph & 17 43 28.353 & +04 34 02.30 &           K2III & 2.8 &  0.05 & 0.07\\
161797&          mu Her & 17 46 27.527 & +27 43 14.44 &            G5IV & 3.4 &  0.22 & 0.08\\
163588&         ksi Dra & 17 53 31.730 & +56 52 21.51 &           K2III & 3.8 & -0.09 &     \\
165195&       HD 165195 & 18 04 40.071 & +03 46 44.73 &             K3p & 7.3 & -2.12 & 0.13\\
165341&          70 Oph & 18 05 27.285 & +02 30 00.36 &         K0V+K4V & 4.0 & -0.02 & 0.12\\
168723&         eta Ser & 18 21 18.601 & -02 53 55.78 &        K0III-IV & 3.2 & -0.18 & 0.12\\
170987&       HD 170987 & 18 32 01.731 & +06 46 48.36 &              F5 & 7.5 & -0.15 & 0.06\\
173701&       HD 173701 & 18 44 35.119 & +43 49 59.80 &              K0 & 7.5 &  0.28 & 0.09\\
173667&       HD 173667 & 18 45 39.726 & +20 32 46.72 &             F6V & 4.2 & -0.05 & 0.05\\
175305&       HD 175305 & 18 47 06.440 & +74 43 31.45 &           G5III & 7.2 & -1.41 & 0.08\\
175492&         113 Her & 18 54 44.872 & +22 38 42.10 &       G4III+A6V & 4.6 & -0.16 &     \\
175955&       HD 175955 & 18 55 33.327 & +47 26 26.79 &              K0 & 7.0 &  0.12 & 0.08\\
175726&       HD 175726 & 18 56 37.173 & +04 15 54.47 &             G0V & 6.7 & -0.04 & 0.06\\
177151&       HD 177151 & 19 01 06.548 & +48 02 08.01 &              K0 & 7.0 & -0.10 & 0.01\\
177153&       HD 177153 & 19 01 39.678 & +41 29 24.33 &              G0 & 7.2 & -0.07 & 0.03\\
181827&       HD 181827 & 19 19 55.068 & +45 01 55.53 &              K0 & 7.2 &  0.14 &     \\
181420&       HD 181420 & 19 20 27.072 & -01 18 35.13 &             F6V & 6.5 &  0.00 &     \\
181906&       HD 181906 & 19 22 21.340 & +00 22 58.74 &              F8 & 7.7 & -0.11 &     \\
181907&         HR 7349 & 19 22 21.545 & -00 15 08.43 &           G8III & 5.8 & -0.15 & 0.12\\
182736&       HD 182736 & 19 24 03.379 & +44 56 00.74 &              G0 & 7.0 & -0.10 & 0.06\\
185144&         sig Dra & 19 32 21.590 & +69 39 40.24 &             G9V & 4.7 & -0.23 & 0.05\\
185395&         tet Cyg & 19 36 26.534 & +50 13 15.96 &            F3+V & 4.5 & -0.03 & 0.05\\
232078&       HD 232078 & 19 38 12.070 & +16 48 25.64 &           K3IIp & 8.7 & -1.56 & 0.03\\
186408&        16 Cyg A & 19 41 48.953 & +50 31 30.22 &             G2V & 6.0 &  0.08 & 0.03\\
186427&        16 Cyg B & 19 41 51.972 & +50 31 03.08 &             G3V & 6.2 &  0.06 & 0.02\\
187111&       HD 187111 & 19 48 39.574 & -12 07 19.74 &       G8III/IV  & 7.8 & -1.71 & 0.09\\
187637&       HD 187637 & 19 49 18.139 & +41 34 56.86 &              F5 & 7.5 & -0.14 & 0.05\\
188512&         bet Aql & 19 55 18.793 & +06 24 24.34 &          G9.5IV & 3.7 & -0.19 & 0.07\\
189349&       HD 189349 & 19 58 02.382 & +40 55 36.64 &              G5 & 7.3 & -0.56 & 0.16\\
190248&         del Pav & 20 08 43.610 & -66 10 55.44 &            G8IV & 3.6 &  0.29 & 0.12\\
196892&       HD 196892 & 20 40 49.380 & -18 47 33.28 &             F6V & 8.2 & -1.03 & 0.10\\
199289&       HD 199289 & 20 58 08.522 & -48 12 13.46 &             F5V & 8.3 & -1.01 & 0.08\\
201891&       HD 201891 & 21 11 59.032 & +17 43 39.89 &       G5VFe-2.5 & 7.4 & -1.05 & 0.08\\
202448&         alf Equ & 21 15 49.432 & +05 14 52.24 &       G0III+A5V & 3.9 &  0.09 &     \\
203608&         gam Pav & 21 26 26.605 & -65 21 58.31 &             F9V & 4.2 & -0.69 & 0.07\\
204543&       HD 204543 & 21 29 28.213 & -03 30 55.37 &              G0 & 8.3 & -1.80 & 0.09\\
206739&       HD 206739 & 21 44 23.945 & -11 46 22.85 &             G5V & 8.6 & -1.57 & 0.02\\
210302&         tau Psa & 22 10 08.780 & -32 32 54.27 &             F6V & 4.9 &  0.05 & 0.04\\
211998&          nu Ind & 22 24 36.884 & -72 15 19.49 & G9VFe-3.1CH-1.5 & 5.3 & -1.50 & 0.13\\
215665&       HD 215665 & 22 46 31.878 & +23 33 56.36 &        G8II-III & 3.9 & -0.12 & 0.08\\
216143&       HD 216143 & 22 50 31.089 & -06 54 49.56 &              G5 & 7.8 & -2.20 & 0.06\\
217014&       HD 217014 & 22 57 27.980 & +20 46 07.79 &          G2.5IV & 5.5 &  0.17 & 0.08\\
217107&       HD 217107 & 22 58 15.541 & -02 23 43.38 &            G8IV & 6.2 &  0.36 & 0.04\\
217906&       HD 217906 & 23 03 46.457 & +28 04 58.03 &      M2.5II-III & 2.4 & -0.11 & 0.13\\
218857&       HD 218857 & 23 11 24.596 & -16 15 04.02 &             G6w & 8.9 & -1.88 & 0.04\\
221170&       HD 221170 & 23 29 28.809 & +30 25 57.85 &            G2IV & 7.7 & -2.13 & 0.05\\

\end{longtable}
\end{longtab}

\end{document}